\documentclass[prd,twocolumn,nofootinbib,showpacs,superscriptaddress]{revtex4-1}

\usepackage{amsfonts}
\usepackage{amsmath}
\usepackage{amssymb}
\usepackage{bm}
\usepackage{dcolumn}
\usepackage{graphicx}
\usepackage{graphics}
\usepackage[latin1]{inputenc}
\usepackage{latexsym}
\usepackage{rotating}
\usepackage{hyperref}
\usepackage[all]{hypcap} 
\usepackage{xspace} 
\usepackage[usenames]{color}
\usepackage{mathrsfs}

\usepackage{ulem}
\definecolor {darkgreen}{rgb}{0.2,0.7,0.2}

\newcommand\be{\begin{equation}}
\newcommand\ba{\begin{eqnarray}}
\newcommand\ee{\end{equation}}
\newcommand\ea{\end{eqnarray}}

\newcommand\bw{\begin{widetext}}
\newcommand\ew{\end{widetext}}

\newcommand{\nn}{\nonumber}

\newcommand{\orb}{{\mbox{\tiny orb}}}

\newcommand{\pd}{\partial}

\newcommand{\Part}{{\mbox{\tiny P}}}
\newcommand{\Hom}{{\mbox{\tiny H}}}

\newcommand{\ext}{\mathrm{ext}}
\newcommand{\inter}{\mathrm{int}}

\newcommand{\mrm}{\mathrm}

\begin{document}
\title{Effective No-Hair Relations for Neutron Stars and Quark Stars: \\ Relativistic Results}

\author{Kent Yagi}
\affiliation{Department of Physics, Montana State University, Bozeman, MT 59717, USA.}

\author{Koutarou Kyutoku}
\affiliation{Department of Physics, University of Wisconsin-Milwaukee,
P.O. Box 413, Milwaukee, Wisconsin 53201, USA}

\author{George Pappas}
\affiliation{School of Mathematical Sciences, The University of Nottingham,
University Park, Nottingham NG7 2GD, United Kingdom}
\affiliation{SISSA, Via Bonomea 265, 34136 Trieste, Italy}

\author{Nicol\'as Yunes}
\affiliation{Department of Physics, Montana State University, Bozeman, MT 59717, USA.}

\author{Theocharis A. Apostolatos}
\affiliation{Section of Astrophysics, Astronomy and Mechanics, Department of Physics, University of Athens, Panepistimiopolis Zografos GR15783, Athens, Greece}

\date{\today}

\begin{abstract} 

Astrophysical charge-free black holes are known to satisfy no-hair relations through which all multipole moments can be specified in terms of just their mass and spin angular momentum. 
We here investigate the possible existence of no-hair-like relations among multipole moments for neutron stars and quark stars that are independent of their equation of state. 
We calculate the multipole moments of these stars up to hexadecapole order by constructing uniformly-rotating and unmagnetized stellar solutions to the Einstein equations.
For slowly-rotating stars, we construct stellar solutions to quartic order in spin in a slow-rotation expansion, while for rapidly-rotating stars, we solve the Einstein equations numerically with the LORENE and RNS codes. 
We find that the multipole moments extracted from these numerical solutions are consistent with each other, and agree with the quartic-order slow-rotation approximation for spin frequencies below roughly $500$ Hz. 
We also confirm that the current-dipole is related to the mass-quadrupole in an approximately equation of state independent fashion, which does not break for rapidly rotating neutron stars or quark stars.  
We further find that the current-octupole and the mass-hexadecapole moments are related to the mass-quadrupole in an approximately equation of state independent way to roughly $\mathcal{O}(10\%)$, worsening in the hexadecapole case.
All of our findings are in good agreement with previous work that considered stellar solutions to leading-order in a weak-field, Newtonian expansion.  
In fact, the hexadecapole-quadrupole relation agrees with the Newtonian one quite well even in moderately relativistic regimes.
The quartic in spin, slowly-rotating solutions found here allow us to estimate the systematic errors in the measurement of the neutron star's mass and radius with future X-ray observations, such as NICER and LOFT.
We find that the effect of these quartic-in-spin terms on the quadrupole and hexadecapole moments and stellar eccentricity may dominate the error budget for very rapidly-rotating neutron stars.
The new universal relations found here should help to reduce such systematic errors.

\end{abstract}



\maketitle

\section{Introduction}

Stationary and axisymmetric black holes (BHs) satisfy certain no-hair relations~\cite{MTW,robinson,israel,israel2,hawking-uniqueness0,hawking-uniqueness,carter-uniqueness}, which allow us to completely describe them in terms of their mass, spin angular momentum and charge. Such relations imply that the exterior gravitational field of BHs can be expressed as an infinite series of multipole moments that only depend on the first two (in the absence of charge): the mass-monopole (the mass) and the current-dipole (the spin angular momentum)~\cite{geroch,hansen}. 

The multipole moments used to describe the gravitational potential far from a source are analogous to those used in electromagnetism to describe the electric and vector potential of a distribution of charge and current. As in electromagnetism, in General Relativity (GR) these multipoles come in two flavors: mass moments and current moments~\cite{thorne-MM}. The former is sourced by the energy density (or the time-time component and the trace of the spatial part of the matter stress-energy tensor) while the latter is sourced by the energy current density (or the time-space component of the stress-energy tensor)~\cite{kidder}. Multipole moments are important not only because they allow us to describe the exterior gravitational field~\cite{Backdahl:2005uz,Backdahl:2006ed}, but also because they are directly related to astrophysical observables~\cite{Ryan:1995wh,Ryan:1997hg,Pappas:2012nt}.

The BH no-hair relations do not apply to neutron stars (NSs) and quark stars (QSs) since these are not vacuum solutions to the Einstein equations. One may then expect that the multipole moments of NSs and QSs would depend strongly on their internal structure, or more precisely, on their equation of state (EoS), which relates their internal pressure to the energy density. The goal of this paper is to investigate whether such NSs and QSs satisfy approximate no-hair relations in full GR. If this were the case, one would then be able to describe their exterior gravitational field by measuring their first few multipole moments, which could have interesting applications to X-ray NS observations~\cite{Baubock:2013gna,Psaltis:2013fha}.

Some evidence already exists to support the existence of such no-hair relations for NSs and QSs. A universal relation between the moment of inertia (directly related to the current dipole moment) and the mass quadrupole moment was found in~\cite{I-Love-Q-Science,I-Love-Q-PRD}, using an unmagnetized, uniform- and slow-rotation approximation. This result was immediately confirmed by~\cite{lattimer-lim} using different EoSs. Haskell \textit{et al.}~\cite{I-Love-Q-B} extended the analysis of~\cite{I-Love-Q-Science,I-Love-Q-PRD} to magnetized NSs and found that the universality still holds, provided that stars spin moderately fast (spin period less than 0.1s) and the magnetic fields are not too large (less than $10^{12}$G). Such properties are precisely those one expects millisecond pulsars to have.

Several studies have relaxed the slow-rotation approximation~\cite{doneva-rapid,Pappas:2013naa,Chakrabarti:2013tca}, leading to an apparent initial disagreement on whether rapid rotation can destroy the EoS-universality \emph{in general}. Doneva \textit{et al.}~\cite{doneva-rapid} constructed NS and QS sequences by varying the dimensional spin frequency and found, as anticipated in~\cite{I-Love-Q-Science,I-Love-Q-PRD}, that the relation between the moment of inertia and the quadrupole moment depend on the spin frequency. By constructing stellar sequences by varying the dimensional spin-frequency, they found that EoS-universality breaks down and from that, they concluded that \emph{in general} the EoS-universality is broken for rapidly rotating stars. However, shortly after, Pappas and Apostolatos~\cite{Pappas:2013naa} and Chakrabarti \textit{et al.}~\cite{Chakrabarti:2013tca} constructed stellar sequences by varying dimensionless combinations of the spin angular momentum and found that the relations remain EoS-universal. Although the conclusions of~\cite{doneva-rapid} and~\cite{Pappas:2013naa,Chakrabarti:2013tca} are contradictory, their calculations are not actually in disagreement; instead, they show that whether the EoS-universality holds for rapidly-rotating relativistic stars depends on the choice of spin parametrization. More recently, Stein \textit{et al.}~\cite{Stein:2013ofa} proved analytically that universality is preserved regardless of rotation, provided one works with dimensionless spin parameterizations.

Recent studies have also considered whether approximately EoS independent relations exist between higher-$\ell$ multipole moments. Reference~\cite{Pappas:2013naa} in fact found one such relation between the current octupole and the mass quadrupole moments of NSs. This relation was not only approximately EoS-universal but also approximately spin-insensitive. The Newtonian results of~\cite{Stein:2013ofa} analytically confirmed this result. The latter, in fact, proved that higher-$\ell$ multipole moments in the non-relativistic Newtonian limit can be expressed in terms of just the mass monopole, spin current dipole and mass quadrupole moments through relations that are approximately EoS-universal and spin-independent. This universality, however, was found to deteriorate with increasing $\ell$ multipole number.  

The existence of approximately universal relations is not only of academic interest, but it also has practical applications. For example, if one could measure any two quantities in a given relation independently, one could perform an EoS-independent test of GR in the strong-field regime~\cite{I-Love-Q-Science,I-Love-Q-PRD}. Moreover, these relations may play a critical role when attempting to measure the mass and radius of NSs with future X-ray telescopes, such as NICER~\cite{2012SPIE.8443E..13G} and LOFT~\cite{2012AAS...21924906R,2012SPIE.8443E..2DF}. The pulse and atomic line profiles of such stars depend not only on the stellar mass and radius, but also on the moment of inertia, the quadrupole moment and the stellar eccentricity~\cite{Morsink:2007tv,Baubock:2012bj,Psaltis:2013zja}. Universal relations between these quantities~\cite{I-Love-Q-Science,I-Love-Q-PRD,Baubock:2013gna} allow one to break parameter degeneracies and measure the mass and radius~\cite{Psaltis:2013fha}. Such measurements, in turn, would allow for exquisite constraints on the EoS in the high density regime~\cite{2006Natur.441.1115O}. 

In this paper, we study whether approximately EoS-independent relations among multipole moments exist up to hexadecapole order in full GR for both NSs and QSs. To do so, we construct unmagnetized, uniformly-rotating NS and QS solutions to the Einstein equations. For rapidly-rotating stars, we extract multipole moments by numerically constructing stellar solutions with the LORENE~\cite{bonazzola_gsm1993,bonazzola_gm1998} and RNS~\cite{stergioulas_friedman1995} codes. For slowly-rotating stars, we extract multipole moments by solving the Einstein equations in a slow-rotation expansion to quartic order in spin, extending previously-found quadratic~\cite{hartle1967,Hartle:1968ht} and cubic~\cite{benhar} solutions. Validity of the quadratic solution is discussed in~\cite{berti-white}. Such an extension allows us to estimate the importance of quartic-order-in-spin terms in X-ray observations of millisecond pulsars, which were neglected in~\cite{Psaltis:2013zja,Psaltis:2013fha}.   

\begin{figure}[htb]
\centering
\includegraphics[width=\columnwidth,clip=true]{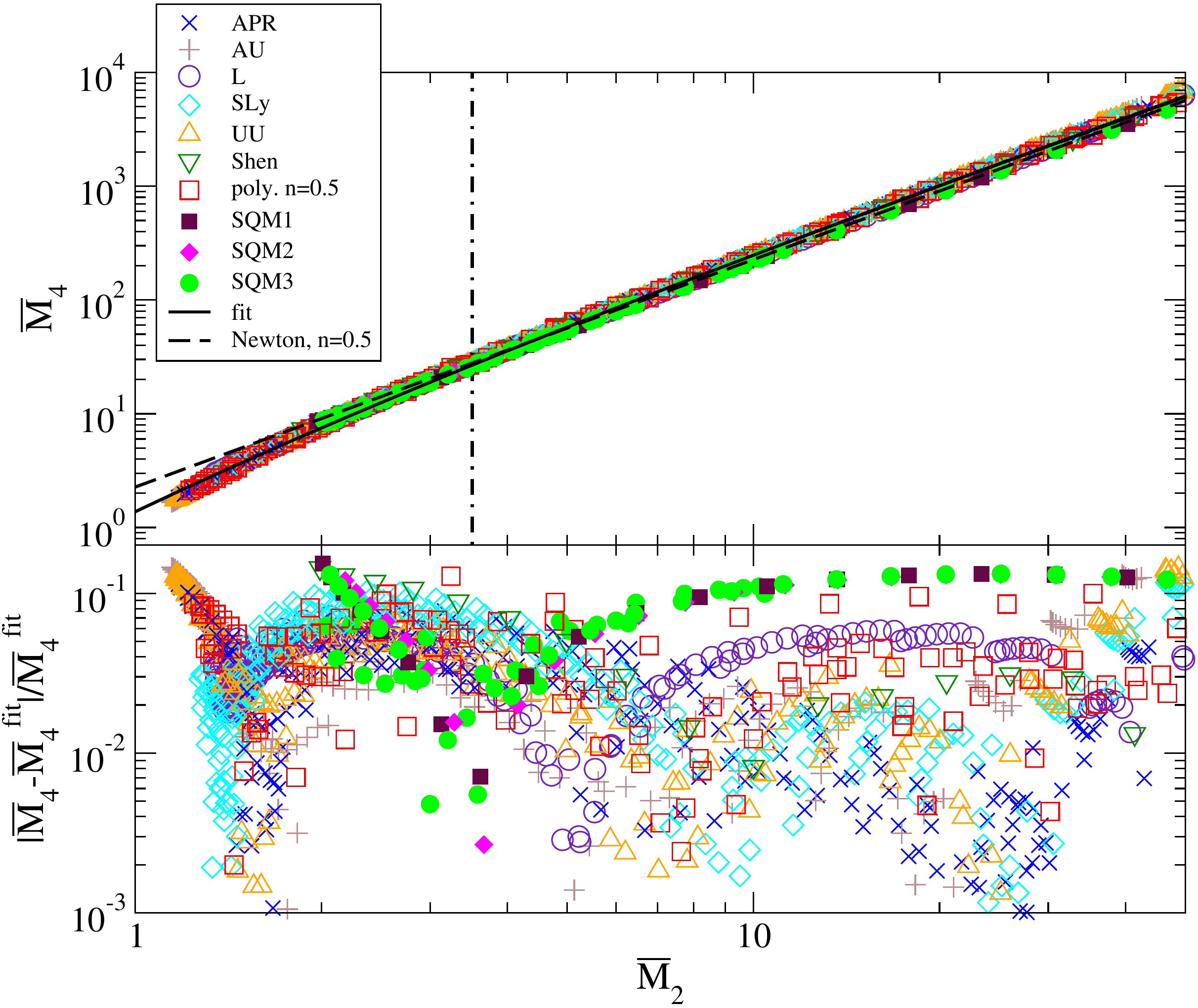}
\caption{
(Color online) (Top) The (reduced dimensionless) hexadecapole $(\bar{M}_4)$--quadrupole $(\bar{M}_2)$ moments relation with various NS and QS EoSs and spins, together with the fit given by Eq.~\eqref{eq:fit} and the Newtonian relation found in~\cite{Stein:2013ofa}. Observe the relation approaches the Newtonian one as one increases $\bar{M}_2$. The Newtonian relation for an $n=0.5$ polytrope agrees with the relativistic fit for various realistic EoSs within 10\% accuracy above the critical $\bar{M}_2$ denoted by the dotted-dashed, vertical line. (Bottom) Fractional difference between the data and the fit. Observe the relation is universal to roughly 20\%. This means that the hexadecapole moment can be approximately expressed in terms of just the stellar mass, spin and quadrupole moment.
\label{fig:M4-M2-intro}
}
\end{figure}

\subsection{Executive Summary}

Given the length of the paper, let us here present a brief summary of the main results. First, we confirm that the LORENE and RNS codes lead to numerically extracted multipole moments up to hexadecapole order that are not only consistent with each other, but also consistent with the slow-rotation approximation. We find that the numerical codes become inaccurate in the extraction of the mass hexadecapole moment for $\chi \lesssim 0.1$--$0.2$, while the slow-rotation expansion becomes inaccurate for $\chi \gtrsim 0.3$, where $\chi \equiv S_1/M_0^{2}$, with $S_1$ the spin angular momentum and $M_0$ is the stellar mass. The latter suggest that spin corrections to the moment of inertia and the quadrupole moment become important only for NSs with spin frequency larger than 100-450Hz, depending on the stellar compactness.

Second, we confirm the results of~\cite{Pappas:2013naa,Chakrabarti:2013tca}, who found that the moment of inertia and the mass quadrupole moment obey approximately EoS-independent relations for arbitrarily rapidly spinning NSs, when one parametrized the stellar sequence in terms of $\chi$. We further confirm that the current octupole and the mass quadrupole moments also satisfy approximately EoS-independent relations to roughly 10\% variability~\cite{Pappas:2013naa}. We find that these relations hold not only for NSs, but also for QSs.

Third, we investigate for the first time whether approximately EoS-independent relations exist in full GR between the mass hexadecapole and the mass quadrupole moment. The top panel of Fig.~\ref{fig:M4-M2-intro} shows the reduced (dimensionless) mass hexadecapole as a function of the mass quadrupole (see Eq.~\eqref{eq:def-dimensionless} for a definition), for various realistic NS and QS EoSs and spin frequencies. For comparison, the top panel also shows an analytic fit to all the data and the Newtonian relations of~\cite{Stein:2013ofa} for an $n=0.5$ polytropic EoS. The bottom panel shows the fractional difference between all the data and the fit. Observe that the relation is both approximately EoS-independent and also spin-independent, with 20\% variability at most. 
The variability though drops to only 10\% if the sequence of NSs and the sequence of QSs were considered seperately. 

Fourth, we find that the approximate EoS-universality between the hexadecapole and the quadrupole moments is worse than that found between the octupole and the quadrupole moments. This suggests that the universality becomes worse as one considers higher-$\ell$ multipoles, which is consistent with the Newtonian limit~\cite{Stein:2013ofa}. The top panel of Fig.~\ref{fig:M4-M2-intro} shows that the full GR $\bar{M}_{4}$--$\bar{M}_{2}$ relation approaches the Newtonian result rapidly as $\bar{M}_{2}$ increases, i.e.~as the compactness decreases. 

Fifth, we find that the Newtonian-limit relations of~\cite{Stein:2013ofa} are quite good at approximating the full GR relations, even for NSs with moderately large compactness. The vertical line in the top panel of Fig.~\ref{fig:M4-M2-intro} shows the value of $\bar{M}_{2}$ at which the Newtonian expression differs from the full GR one by 10\%. This occurs at $\bar{M}_2 = 3.5$, which corresponds to a compactness of approximately $0.2$. When considering the current octupole-mass quadrupole relation, the threshold value of $\bar{M}_2$ increases by an order of magnitude. This suggests that the Newtonian-limit of the universal relations become better approximations to the  full GR relations as one considers relations between quadrupole and higher-$\ell$ multipole moments.

Sixth, we confirm that the NS quadrupole and octupole moments scale with the spin-squared and the spin-cubed, respectively, and the coefficients roughly depend only on the mass and EoS~\cite{pappas-apostolatos}. We find that such a property is preserved also for QSs. We also find that the NS and QS hexadecapole moment scales with the fourth power of spin, and again, the coefficient only depends on the star's mass and EoS, extending the results found in~\cite{pappas-apostolatos}. Due to
the universality mentioned in the paragraphs above, these coefficients are
intimately related with each other. Such scaling is similar to that found for Kerr BHs, and may help as a guide to construct an exact analytic solution with a small number of parameters, by using one of the generating techniques to describe a NS and QS exterior spacetime. For example, the two-soliton solution of~\cite{1995JMP....36.3063M} was proposed as a possible candidate to describe the exterior spacetime of relativistic stars analytically~\cite{Pappas:2012nt}. However, such a solution has a hexadecapole moment whose spin dependence starts at quadratic order and not at quartic order. The scaling results suggest that a better analytic solutions would be one that respects the scaling of multipole moments described above.

Seventh, we study the implications of our results to future X-ray observations of NSs, including NICER and LOFT. Reference~\cite{Baubock:2012bj} showed that the quadrupole moment affects the atomic line profiles significantly. Psaltis et al.~\cite{Psaltis:2013fha} found that the quadrupole moment and the stellar eccentricity contribute to the X-ray pulse profiles by 1--5\% and 10--30\% respectively, for pulsars with mass $\sim 1.8M_\odot$ and a spin frequency of 600Hz.  Since the goal of NICER and LOFT is to measure the mass and radius independently within 5\% accuracy, both quantities must be included when analyzing pulse profiles. Reference~\cite{Psaltis:2013fha}, however, carried out this analysis in the slow-rotation approximation, neglecting cubic and higher order terms in the spin. Our construction of NS and QS solutions to quartic order in spin allows us to estimate the systematic errors in the analysis of~\cite{Psaltis:2013fha} due to the quadratic order in spin truncation. 

Quartic order in spin terms should lead to order $\chi^2$ correction to quadratic order in spin effects. The fastest spinning pulsar~\cite{2006Sci...311.1901H} discovered could have $\chi \sim 0.53$, which means that quartic in spin terms could lead to corrections of $\sim 30\%$ on quadratic in spin effects. Given that the leading-order-in-spin contribution of the quadrupole moment affects the X-ray pulse profiles by 1--5\%, the spin correction to the quadrupole moment can be neglected in future X-ray observations; one expects similar results to hold for the effect of the hexadecapole moment on the pulse profile. The quartic in spin corrections to the stellar eccentricity, however, can lead to modifications on the pulse profile of $\sim 6\%$ for rapidly rotating pulsars. Such spin corrections as a function of the stellar compactness are EoS-insensitive if one restricts the star to be a NS and not a QS, with stellar compactness larger than 0.15.

We also found that although the (leading-order in spin) eccentricity-compactness relation is approximately EoS independent within $\sim 1\%$ for NSs, as found in~\cite{Baubock:2013gna} and used in~\cite{Psaltis:2013fha}, this independence breaks for QSs. For the latter, different QS EoSs lead to a variability of order 10\%. Such relatively large variability originates from the relation between the moment of inertia and the compactness. This means that if one wants to use the eccentricity-compactness or the moment-of-inertia-compactness relation to achieve 5\% accuracy in the measurement of the mass and radius, one needs to assume that the pulsar is a NS. The new universal relations found in this paper should allow us to reduce the number of parameters in X-ray observations, breaking degeneracies and improving parameter estimation.

\subsection{Organization and Convention}

The organization of the paper is as follows. In Sec.~\ref{sec:slow-rot}, we explain how to construct perturbative NS and QS solutions to quartic order in spin in the slow-rotation approximation, extending the results of~\cite{hartle1967,benhar}. We present the perturbed Einstein equations, the asymptotic behavior of the solution near the stellar center, the exterior solutions, and the matching conditions at the stellar surface and explain how to extract multipole moments. In Sec.~\ref{sec:rapid-rot}, we explain how to construct rapidly-rotating NS and QS solutions using the LORENE and RNS codes. In Sec.~\ref{sec:results}, we present our numerical results. We show the universal relations among the multipole moments and compare them with the Newtonian relations. In Sec.~\ref{sec:X}, we explain how our work may help to estimate the pulse profile of X-ray observations. We conclude in Sec.~\ref{sec:conclusions} and point to future work. 

Henceforth, we use geometric units with $c=1=G$. The $\mathcal{O}(N)$ symbol will represent a term of order $N$ or a term that is $\sim N$ in magnitude. Given the length of the paper, we find it convent to summarize the following conventions here. We use the following masses:
\begin{itemize}

\item $M_0$ is the Geroch-Hansen mass monopole moment~\cite{geroch, hansen}.  

\item $M_*$ is the stellar mass for a non-rotating star, which agrees with the Geroch-Hansen mass monopole moment in the slow-rotation limit.

\item $M_\mrm{Komar}$ is the Komar mass, which agrees with $M_0$.

\end{itemize}
We use the following radial quantities:
\begin{itemize}

\item $R_\mrm{eq}$ is the gauge-invariant circumferential stellar radius at the equator.  

\item $R_\mrm{pol}$ is the gauge-invariant circumferential stellar radius at the poles.  

\item $R_*$ is the stellar radius for a non-rotating star. 

\item $r_\mrm{eq}$ is the coordinate equatorial radius in quasi-isotropic coordinates [Eq.~\eqref{eq:quasi-isotropic}]  (not to be confused with that in the Hartle-Thorne coordinates [Eq.~\eqref{Eq:metric-slow-rot}]).  

\end{itemize}
We use the following spin-related quantities:
\begin{itemize}

\item $f$ is the stellar (linear) spin frequency.

\item $\Omega$ is the stellar (angular) spin velocity, which is related to the linear spin frequency by $\Omega = 2 \pi f$.  

\item $S_1$ is the Geroch-Hansen current dipole moment~\cite{geroch, hansen}.  

\item $J$ is the magnitude of the spin angular momentum to leading order in spin frequency.

\item $J_\mrm{Komar}$ is the Komar angular momentum, which agrees with $S_1$.

\item $j$ is the dimensionless spin parameter, defined by $j = J/M_*^2$ and valid to leading order in spin frequency.

\item $\chi$ is the dimensionless spin parameter, defined by $\chi = S_1/M_0^2$ and valid to all orders in spin frequency.

\end{itemize}

\section{Slowly-Rotating Stars}
\label{sec:slow-rot}

In this section, we construct slowly-rotating stars for a fixed value of the central energy density to quartic order in spin, i.e.~to ${\cal{O}}(J^{4}/M_{*}^{8})$, where $J$ is the magnitude of the star's spin angular momentum to leading order in spin frequency and $M_{*}$ is its mass in the non-rotating limit. Slowly-rotating stars to quadratic order in spin have been constructed in~\cite{hartle1967}, and extended to third order in~\cite{benhar}. We here focus on the equations relevant to quartic order only, whose solutions allow us to calculate the star's multipole moments to quadratic order. 

\subsection{Metric Ansatz, Stress Energy Tensor and Differential Equations of Structure}

Following and extending~\cite{hartle1967,benhar}, we impose the following metric ansatz:
\bw
\begin{align}
\label{Eq:metric-slow-rot}
ds^2 &= - e^{\nu (r)} \left[ 1 + 2 \epsilon^2 h_2(r,\theta) + 2 \epsilon^4 h_4(r,\theta) \right] dt^2
+  e^{\lambda (r)} \left[ 1 + \frac{2 \epsilon^2 m_2(r,\theta)+ 2 \epsilon^4 m_4(r,\theta)}{r - 2 M(r)} \right] dr^2 
\nn \\
&+  r^2 \left[ 1 + 2 \epsilon^2 k_2(r,\theta) + 2 \epsilon^4 k_4(r,\theta) \right] 
\left( d\theta^2 + \sin^2 \theta \left\{ d\phi 
- \epsilon \left[ \Omega - \omega_1 (r,\theta)  + \epsilon^2 \omega_3 (r,\theta) \right] dt \right\}^2 \right) + \mathcal{O}(\epsilon^5)\,,
\end{align}
\ew
where $\Omega$ is the spin angular velocity of the star, $\epsilon$ is a book-keeping parameter that counts powers of $j=J/M_{*}^{2}$, and we define $M(r)$ via
\be
e^{- \lambda(r)}  \equiv 1 - \frac{2 M(r)}{r}\,.
\ee
We further decompose the metric perturbations in terms of Legendre polynomials as
\allowdisplaybreaks
\ba
\omega_1(r,\theta) &=& \omega_{11}(r) P'_1 (\cos \theta)\,, \\
h_2 (r,\theta) &=& h_{20}(r) + h_{22}(r) P_2(\cos\theta)\,, \\
m_2 (r,\theta) &=& m_{20}(r) + m_{22}(r) P_2(\cos\theta)\,, \\
k_2 (r,\theta) &=& k_{22}(r) P_2(\cos\theta)\,, \\
\omega_3(r,\theta) &=& \omega_{31}(r) P'_1 (\cos \theta) + \omega_{33}(r) P'_3 (\cos \theta)\,, \\ 
h_4 (r,\theta) &=& h_{40}(r) + h_{42}(r) P_2(\cos\theta) + h_{44}(r) P_4(\cos\theta)\,, \nn \\
\\
m_4 (r,\theta) &=& m_{40}(r) + m_{42}(r) P_2(\cos\theta) + m_{44}(r) P_4(\cos\theta)\,, \nn \\
\\
k_4 (r,\theta) &=& k_{42}(r) P_2(\cos\theta) + k_{44}(r) P_4(\cos\theta)\,, 
\ea
with $P_\ell' (\cos\theta) = d P_\ell (\cos \theta) /d \cos \theta$. The function $k_\ell (r,\theta)$ does not have an $\ell=0$ mode since this can be eliminated by performing a radial coordinate transformation.

As pointed out in~\cite{hartle1967}, one needs to be careful when considering perturbation near the surface. We perform a coordinate transformation such that the stellar density $\rho[r(R,\theta),\theta]$ evaluated at the new radial coordinate $R$ is the same as $\rho^{(0)} (r)$ in the static configuration:
\be
\rho[r(R,\theta),\theta] = \rho(R) = \rho^{(0)}(R)\,.
\ee
By construction, the stellar density and pressure only contain the static part (i.e.~no spin perturbation). The old and new radial coordinates are related by
\be
r(R, \theta) = R + \epsilon^2 \xi_2(R,\theta) +  \epsilon^4 \xi_4(R,\theta) + \mathcal{O}(\epsilon^6)\,,
\ee
where we will decompose the gauge functions through Legendre polynomials:
\ba
\xi_2(R,\theta) &=& \xi_{20}(R) + \xi_{22}(R) P_2 (\cos\theta)\,, \\
\xi_4(R,\theta) &=& \xi_{40}(R) + \xi_{42}(R) P_2 (\cos\theta) + \xi_{44}(R) P_4 (\cos\theta)\,. \nn \\
\ea
Note that $\xi_\ell (R)$ is well-defined only inside a star and we let them take a constant value outside.

The matter stress-energy tensor will be chosen to be that of a perfect fluid with energy density $\rho$ and pressure $p$:
\be
\label{Eq:matter-slow-rot}
T_{\mu \nu} = (\rho + p) u_\mu u_\nu + p g_{\mu\nu}\,,
\ee
where $u^{\mu}$ is the fluid 4-velocity. For a stationary and axisymmetric star, the latter is simply
\be
\label{eq:umu}
u^\mu = \left( u^0, 0, 0, \Omega u^0 \right)\,,
\ee
with $u^{0}$ determined by the normalization condition $u_\mu u^\mu = -1$.

Inserting Eqs.~\eqref{Eq:metric-slow-rot} and~\eqref{Eq:matter-slow-rot} into the Einstein equations and expanding to quartic order in spin, one obtains differential equations for the NS structure functions. At $\mathcal{O}(\epsilon^4)$, one finds the constraints
\allowdisplaybreaks
\begin{align}
\label{eq:xi4-slow-rot}
\xi_{4i} &= -\frac{R (R-2 M)}{M+4 \pi R^3 p} h_{4i} + S_{\xi 4i}\,, \\
\label{eq:m4-slow-rot}
m_{4i} &=  -(R-2 M) h_{4i} + S_{m 4i}\,, 
\end{align}
where $i$ equals 2 or $4$ and $S_{\xi 4i}$ and $S_{m 4i}$ are source terms that depend only on quantities of $\mathcal{O}(\epsilon^3)$ (see Appendix~\ref{app:exterior}). To ${\cal{O}}(\epsilon^{4})$, one finds the differential equations
\begin{align}
\label{eq:h42-slow-rot}
\frac{d h_{42}}{dR} &= {\frac {2+R
\nu '  }{R
\nu '
  }} \frac{d v_{42}}{d R} -\,{\frac {2}{R \left( R-2\,M   \right) \nu' }} h_{42} \nn \\
& - \,{\frac {4}{R \left( R-2\,M   \right) \nu '  }} v_{42} - {\frac {2\,R
\nu '  +2}{{R}^{2} \left( R-
2\,M   \right) 
\nu '  }}
 m_{42} \nn \\
& + \frac{1}{{R}^{3} \left( R- 2\,M \right) \nu'} \left[ {R}^{2} \left( R-2\,M \right) \nu'' \right. \nn \\
& \left. + \, \left( 4M  - R - M' R \right) R \nu' -2 M'  R+6M   \right]
 \xi_{42} + S_{h 42}\,,  \\
\label{eq:v42-slow-rot}
\frac{d v_{42}}{dR} &= {\frac {-8\,\pi \,  {R}^{3} p-2\,M  }{R
 \left( R-2\,M   \right) }} h_{42} + S_{v 42}\,,  \\
\label{eq:h4-slow-rot}
\frac{d h_{44}}{d R} &= \frac{h_{44}}{(M+4 \pi R^3 p) R (R-2 M)} 
\nn \\
& \times \left[2 M^2-2 R (1 +12 \pi R^2 p  +4 \pi R^2 \rho) M
\right. 
\nn \\
&\left.-4 \pi R^4 (8 \pi R^2 p^2- p- \rho)\right]  -\frac{9 v_{44}}{M+4 \pi R^3 p}  +S_{h 44}\,, \\
\label{eq:v4-slow-rot}
\frac{d v_{44}}{d R} &= -\frac{8 \pi R^3 p +2 M}{R (R-2M)} h_{44} +S_{v 44}\,,
\end{align}
where $v_{4i} \equiv h_{4i} + k_{4i}$ and $S_{h 4i}$ and $S_{v 4i}$ are also source functions (see Appendix~\ref{app:exterior}). Since the star's multipole moments will be calculated to quadratic order in spin, we do not need to solve the $\ell=0$ mode equations at $\mathcal{O}(\epsilon^4)$, except for their exterior solution, which we do need as we will explain in Sec.~\ref{sec:ryan}.

\subsection{Boundary Conditions}

In order to solve the differential equations presented in the previous subsection, we need to prescribe boundary conditions. One set can be obtained by finding the solution asymptotically about the stellar center. The other set can be derived from the exterior solution to the differential equations. 

\subsubsection{Asymptotic Behavior near the Stellar Center}

The asymptotic behavior about the stellar center at ${\cal{O}}(\epsilon^{4})$ of the $\ell=4$ mode of the structure functions is
\begin{align}
h_{44}(R) &= C_{44}^\inter R^4 + \mathcal{O}(R^6/M_*^{6})\,, \\
v_{44}(R) &= \frac{24}{35} (C_{22}^\inter{})^2 R^4 + \mathcal{O}(R^6/M_*^{6})\,, \\
m_{44}(R) &= \frac{R^{5}}{105 p_c+35 \rho_c} \left[-105 p_c C_{44}^\inter +72 p_c (C_{22}^\inter{})^2 
\right. 
\nn \\
&\left. + 132 \rho_c (C_{22}^\inter{})^2 + 36 \rho_c \omega_{11,c}^2 C_{22}^\inter e^{-\nu_c} - 35 \rho_c C_{44}^\inter\right] 
\nn \\
&+ \mathcal{O}(R^7/M_*^{7})\,, \\
\xi_{44}(R) &= \frac{162 \omega_{11,c}^2 C_{22}^\inter e^{-\nu_c} + 27 \omega_{11,c}^4 e^{-2 \nu_c}+243 (C_{22}^\inter{})^2}{560 \pi^2 (3 p_c+ \rho_c)^2}R 
\nn \\
&+ \mathcal{O}(R^3/M_*^{3})\,, 
\end{align}
where $p_c = p(R=0)$, $\rho_c = \rho(R=0)$, $\nu_c = \nu(R=0)$, and $\omega_{11,c} = \omega_{11}(R=0)$. The constant $C_{44}^\inter$ is to be determined by matching interior and exterior solutions at the stellar surface. The constant $C_{22}^\inter$ first appears via $h_{22} (R) = C_{22}^\inter R^2 + \mathcal{O}(R^4/M_*^4)$ and is determined by the matching of interior and exterior solutions for the $\ell=2$ mode functions at $\mathcal{O}(\epsilon^2)$.

The asymptotic behavior about the stellar center at ${\cal{O}}(\epsilon^{4})$ of the $\ell=2$ mode of the structure functions is
\begin{align}
h_{42}(R) &= C_{42}^\inter R^2 + \mathcal{O}(R^4/M_*^4)\,, \\
v_{42}(R) &= 2 C_{22}^\inter h_{20,c} R^2 + \mathcal{O}(R^4/M_*^4)\,, \\
m_{42}(R) &= (2 C_{22}^\inter h_{20,c} - C_{42}^\inter) R^3 + \mathcal{O}(R^5/M_*^5)\,, \\
\xi_{42}(R) &= \frac{1}{112(3 p_c + \rho_c )^2 \pi^2} \left\{  \omega_{11,c} \; e^{-\nu_c} 
\right. \nn \\ & \left.
\left[ 56 (h_{20,c} \omega_{11,c} + \omega_{31,c} ) \left(3 p_c+ \rho_c \right) \pi - 45 C_{22}^\inter \omega_{11,c} \right] 
\right. \nn \\ & \left.
- 18 \omega_{11,c}^4 e^{-2 \nu_c} + 84 \left( 2 C_{22}^\inter h_{20,c} -  C_{42}^\inter \right) \left( 3 p_c+  \rho_c \right) \pi 
\right. \nn \\ & \left.
+ 27 (C_{22}^\inter)^2 \right\} R + \mathcal{O}(R^3/M_*^3)\,, 
\end{align}
where $h_{20,c} = h_{20}(R=0)$ and $\omega_{31,c} = \omega_{31}(R=0)$. The constant $C_{42}^\inter$ is again determined through matching. 

\subsubsection{Exterior Solution}

One could derive the asymptotic behavior of the solution to the structure equations near spatial infinity, but in fact, one can find exact solutions in the exterior. The exterior solutions for $\nu$, $M$, $\omega_{11}$, $h_{20}$, $h_{22}$, $k_{22}$, $\omega_{31}$ and $\omega_{33}$ were first obtained in~\cite{hartle1967,benhar} (see Appendix~\ref{app:ext-sol-O3}). The exterior solution for $h_{40}$, $h_{4i}$ and $v_{4i}$, with $i=2,4$, are
\ba
h_{40}^\ext (R) = h_{40}^{\ext, \Part} + C_{40}^\ext h_{40}^{\ext, \Hom}\,, \\
h_{4i}^\ext (R) = h_{4i}^{\ext, \Part} + C_{4i}^\ext h_{4i}^{\ext, \Hom}\,, \\
v_{4i}^\ext (R) = v_{4i}^{\ext, \Part} + C_{4i}^\ext v_{4i}^{\ext, \Hom}\,,
\ea
where $C_{40}^\ext$ and $C_{4i}^\ext$ are integration constants.
The homogeneous and particular solutions for $h_{40}$ are given by
\bw
\ba
h_{40}^{\ext, \Hom} &=& -\frac{M_*}{R-2 M_*}\,, \\
h_{40}^{\ext, \Part} &=& \frac{1}{10 M_*^3 R^8 (R-2 M_*)^2} \{24 (C_{22}^\ext)^2 R^6 M_*^7 - [40 C_{31}^\ext J R^5 -48 C_{22}^\ext R^3 J^2 + 40 (C_{22}^\ext)^2 R^7] M_*^6 \nn \\
& & - [16 C_{22}^\ext R^4 J^2 + 40 (C_{22}^\ext)^2 R^8 - 20 C_{31}^\ext R^6 J - 24 J^4] M_*^5 +[180 (C_{22}^\ext)^2  R^9- 124 C_{22}^\ext J^2 R^5 + 24 J^4 R] M_*^4 \nn \\ 
& & - [6 (C_{22}^\ext)^2 R^{10} + 176 C_{22}^\ext R^6 J^2 + 24 J^4 R^2] M_*^3 - [72 (C_{22}^\ext)^2 R^{11} + R^3 J^4 - 382 C_{22}^\ext R^7 J^2] M_*^2 \nn \\
& & - 2 R^4 [96 C_{22}^\ext R^4 J^2 - J^4 - 9 (C_{22}^\ext)^2 R^8] M_* + 30 C_{22}^\ext R^9 J^2\} \nn \\
& & + \frac{3 C_{22}^\ext}{10 R^4 (R-2 M_*) M_*^4} ( 6 C_{22}^\ext M_* R^8 - 18 C_{22}^\ext M_*^2 R^7 - 16 C_{22}^\ext M_*^3 R^6 + 56 C_{22}^\ext M_*^4 R^5 + 5 R^5 J^2 - 12 C_{22}^\ext M_*^5 R^4 \nn \\
& &  - 27 J^2 M_* R^4 - 8 C_{22}^\ext M_*^6 R^3 + 40 J^2 M_*^2 R^3 - 4 M_*^3 R^2 J^2 - 12 M_*^4 R J^2 - 8 M_*^5 J^2 ) \ln \left( 1-\frac{2 M_*}{R} \right) \nn \\
& & + \frac{9 (C_{22}^\ext)^2 R (R + 2 M_*) (R-2 M_*)^2}{20 M_*^4}  \left[ \ln \left( 1-\frac{2 M_*}{R} \right) \right]^2\,,
\ea
\ew
where recall that $M_*$ is the stellar mass and $J$ is the spin angular momentum in the slow-rotation limit, as we will see in Sec.~\ref{sec:ryan}. $C_{22}^\ext$ and $C_{31}^\ext$ are integration constants that enter the exterior solutions at $\ell=2$, $\mathcal{O}(\epsilon^2)$ and $\ell=1$, $\mathcal{O}(\epsilon^3)$ respectively. The particular and homogeneous solutions for $h_{42}$, $v_{42}$, $h_{44}$ and $v_{44}$ are rather lengthy, and we thus present them in Appendix~\ref{app:ext-sol-O4}.

\subsection{Numerical Algorithm}

The differential equations at $\mathcal{O}(\epsilon^4)$ will be numerically solved as follows. First, we choose an arbitrary trial value for $C_{4i}^\inter$ and solve both homogeneous and inhomogeneous differential equations at $\mathcal{O}(\epsilon^4)$ using a 4th-order Runge-Kutta method. This solution leads to trial homogeneous solutions ($h_{4i}^{\inter,\Hom}(R)$, $v_{4i}^{\inter,\Hom}(R)$) and trial particular solutions ($h_{4i}^{\inter,\Part}(R)$, $v_{4i}^{\inter,\Part}(R)$) in the interior region. The interior solutions for $h_{4i}(R)$ and $v_{4i}(R)$ with the correct boundary conditions are given by
\ba
h_{4i}^\inter (R) &=& h_{4i}^{\inter,\Part}(R) + C_{4i}^\inter{}' h_{4i}^{\inter,\Hom}(R)\,, \\
v_{4i}^\inter (R) &=& v_{4i}^{\inter,\Part}(R) + C_{4i}^\inter{}' v_{4i}^{\inter,\Hom}(R)\,, 
\ea
where $C_{4i}^\inter{}'$ are constants that are to be determined by matching $h_{4i}^\inter (R)$ and $v_{4i}^\inter (R)$ with $h_{4i}^\ext (R)$ and $v_{4i}^\ext (R)$ at the stellar surface $R=R_*$. 

The matching procedure described above must be carried out with care when the stellar density profile is discontinuous at the stellar surface, for example when considering a constant density NS or a QS~\cite{damour-nagar,hinderer-lackey-lang-read}. This is because the differential equations for $h_{4i}$ and $v_{4i}$ [Eqs.~\eqref{eq:h42-slow-rot}--\eqref{eq:v4-slow-rot}] contain terms proportional to $d\rho/dR$ in the source terms. Let us peel the discontinuous density derivatives and denote these terms as $F_{4i}^{\rho', h_{4i}}(R) \; (d\rho/dR)$ and $F_{4i}^{\rho', v_{4i}}(R) \; (d\rho/dR)$. When the energy density is discontinuous at the surface, $(d\rho/dR) = - \rho^{-} (R) \delta (R-R_*)$, where $\rho^{-}(R_*)$ is the finite part of the interior energy density at the surface. Then, to find the matching conditions, we analytically integrate $dh_{4i}/dR$ and $dv_{4i}/dR$ from $R_{*} - \epsilon_{R}$ to $R_{*} + \epsilon_{R}$ for $\epsilon_{R}/R_{*} \ll 1$ and constant, and then take the $\epsilon_R \to 0$ limit:
\ba
h_{4i}^\inter (R_*) -\rho^{-} (R_*) F_{4i}^{\rho', h_{4i}}(R_*) \ &=& h_{4i}^\ext (R_*)\,, \\
v_{4i}^\inter (R_*) -\rho^{-} (R_*) F_{4i}^{\rho', v_{4i}}(R_*) &=& v_{4i}^\ext (R_*)\,. 
\ea
These conditions are also valid for ordinary stars in which the energy density is analytic at the stellar surface, and thus, $\rho^{-}(R_{*}) = 0$. 

From these two matching conditions, one can solve for the constants $C_{4i}^\ext$ and $C_{4i}^\inter{}'$. The former is given by
\ba
C_{4i}^\ext &=& \frac{1}{h_{4i}^{\inter, \Hom}(R_*) v_{4i}^{\ext, \Hom}(R_*) - v_{4i}^{\inter, \Hom}(R_*) h_{4i}^{\ext, \Hom}(R_*)} \nn \\
& & \times \left[ h_{4i}^{\inter, \Hom}(R_*) \tilde{v}_{4i}^{\inter, \Part}(R_*) +v_{4i}^{\inter, \Hom}(R_*) h_{4i}^{\ext, \Part}(R_*) \right. \nn \\
& & \left. -h_{4i}^{\inter, \Hom}(R_*) v_{4i}^{\ext, \Part}(R_*) -v_{4i}^{\inter, \Hom}(R_*) \tilde{h}_{4i}^{\inter, \Part}(R_*) \right]\,, \nn \\
\ea
with 
\ba
\tilde{h}_{4i}^{\inter, \Part}(R_*) &=& h_{4i}^{\inter, \Part}(R_*) -\rho^{-} (R_*) F_{4i}^{\rho', h_{4i}}(R_*)\,, \\
\tilde{v}_{4i}^{\inter, \Part}(R_*) &=& v_{4i}^{\inter, \Part}(R_*) -\rho^{-} (R_*) F_{4i}^{\rho', v_{4i}}(R_*)\,.
\ea
We do not present the expression for $C_{4i}^\inter{}'$ because, as we will see, it will not be needed to compute the multipole moments in the next section. However, we did use $C_{4i}^\inter{}'$ to check the numerical matching at the surface for $h_{4i}(R)$, $v_{4i}(R)$ and $m_{4i}(R)$.  

\subsection{Extracting Multipole Moments}
\label{sec:ryan}

Multipole moments are the coefficients in a multipole expansion of a given field. Such expansions are common when studying electromagnetic or gravitational fields far from the source. In electromagnetism, for example, the potential $V$ for a source confined to a small region near the origin can be expanded as
\be
V(R,\theta,\phi) = \sum_{\ell,m} \frac{V_{\ell m}}{R^{\ell +1}} Y^{\ell m}(\theta,\phi)\,,
\ee
where $Y^{\ell m}$ are scalar spherical harmonics and $V_{\ell m}$ are the constant multipole moments. Clearly then, the multipole moments are the angle-independent coefficients of the $1/R^{\ell+1}$ term in a series expansion. The $\ell=0$ term (the monopole) corresponds to the electric charge (or the mass in the gravitational case). 

In GR, there are two types of multipole moments that must be distinguished: the \emph{mass-moments} and the \emph{current-moments}. The former are the gravitational analog to the $V_{\ell m}$ coefficients of the expansion of the potential in electromagnetism. The latter are the gravitational analog to the coefficients in the expansion of the vector potential in electromagnetism. 

In order to extract the multipole moments of a slowly-rotating NS to quartic order in spin, we must understand how these depend on the integration constants in the exterior solution ($M_*$, $J$, $C_{20}^\ext$, $C_{22}^\ext$, $C_{31}^\ext$, $C_{33}^\ext$, $C_{40}^\ext$, $C_{42}^\ext$, $C_{44}^\ext$). We will here concentrate on the so-called Geroch-Hansen (GH) multipole moments~\cite{geroch, hansen}, which can be thought of as GR generalizations of the multipole moments in electrodynamics.  The mapping between GH moments and the integration constants can be found through Ryan's method~\cite{Ryan:1995wh}, which we now explain. 

Let us first consider a test-particle in an equatorial, circular orbit around a stationary and axisymmetric spacetime, such as that of a slowly-rotating star. The orbital frequency $\Omega_{\orb}$ is given by
\be
\Omega_{\orb} = \frac{-g_{t\phi,R} + \sqrt{(g_{t\phi,R})^2-g_{tt,R} g_{\phi\phi,R}}}{g_{\phi\phi,R}}\,, 
\ee
where commas in index lists stand for partial derivatives. This expression for $\Omega_{\orb}$ can be derived by studying the conserved quantity (the orbital angular momentum) associated with the Killing vector related to axisymmetry.  For the spacetime prescribed in Eq.~\eqref{Eq:metric-slow-rot}, and using the exterior solution derived in the previous subsection, the above equation can be inverted to obtain 
\bw
\ba
\label{x}
\sqrt{\frac{M_0}{R}} &=& {\frac {m^{1/6}}{72}}\, \left[ 72-12\,C_{20}^\ext\,\epsilon^{2}+ \left( 7\,{C_{20}^\ext
}^{2}-12\,C_{40}^\ext \right) \epsilon^{4} \right] v + \frac{\epsilon}{18 {m}^{5/6}}\,{
 \left[ 6\,j - (7\,j\,C_{20}^\ext\, -6\,C_{31}^\ext)\,\epsilon^{
2} \right] {v}^{4}} \nn \\
& & -{\frac {\epsilon^2}{120{m}
^{7/6}}}\,{ \left[ 48
\,C_{22}^\ext +30\,{j}^{2} - (88\,C_{22}^\ext\,C_{20}^\ext\,-30\,C_{42}^\ext \,+55\,{j}^{2}C_{20}^\ext )\,\epsilon^{2} \right] {v}^{5}} \nn \\
& & -{\frac {7 \epsilon^2}{9 {m}^{{11/6}}}}\, \left[ \frac{1}{28}\,{j}^{2} +\frac{6}{7}\,C_{22}^\ext -
 \left( C_{22}^\ext\,C_{20}^\ext-{\frac {15}{28}}\,C_{42}^\ext+C_{31}^\ext\,j
-{\frac {11}{24}}\,{j}^{2}C_{20}^\ext \right) \epsilon^{2} \right] {
v}^{7} \nn \\
& & -{\frac {\epsilon^{3}}{60 {m}^{{13/6}}}}\, \left( 30\,C_{33}^\ext+88
\,j\,C_{22}^\ext+55\,{j}^{3} \right) {v}^{8} \nn \\
& & +\frac{5 \epsilon^2}{7 {m}^{5/2}}\, \left\{ {\frac {7}{10}}\,{j}^{2}-2\,C_{22}^\ext+ \left[ {
\frac {101}{480}}\,{j}^{4}+ \left( -{\frac {295}{84}}\,C_{22}^\ext-{
\frac {23}{10}}\,C_{20}^\ext \right) {j}^{2}+ \left( {\frac {39}{10}}
\,C_{31}^\ext+\frac{1}{4}\,C_{33}^\ext \right) j-{\frac {7}{16}}\,C_{44}^\ext \right. \right. \nn \\
& & \left. \left. -{
\frac {1377}{350}}\,(C_{22}^\ext)^2-\frac{5}{4}\,C_{42}^\ext+C_{22}^\ext\,C_{20}^\ext
 \right] \epsilon^{2} \right\} {v}^{9} -{\frac {\epsilon
^{3}}{324 {m}^{{17/6}}}}\,
 \left( 673\,{j}^{3}+1368\,j\,C_{22}^\ext+540\,C_{33}^\ext \right) {v}^{10} \nn \\
& & -{\frac {8 \epsilon^{2}}{15 {m}^{{19/6}}}}\, \left\{ 6\,C_{22}^\ext + \left[ {\frac {367}{96}}\,{j}^{4}+ \left( {\frac {18805}{672}}\,
C_{22}^\ext+{\frac {15}{4}}\,C_{20}^\ext \right) {j}^{2}+ \left( -\frac{15}{2}\,C_{31}^\ext +{\frac {65}{32}}\,C_{33}^\ext \right) j+{\frac {12589}{560}}
\,(C_{22}^\ext)^2 \right. \right. \nn \\
& & \left. \left. +{\frac {15}{4}}\,C_{42}^\ext +{\frac {315}{128}}\,C_{44}^\ext +C_{22}^\ext\,C_{20}^\ext \right] \epsilon^{2} \right\} {v}^{11} +\mathcal{O} \left( \epsilon^5, v^{12} \right)\,,
\ea
\ew
where we have defined the orbital velocity $v = (M_0 \Omega_{\orb})^{1/3}$, with $M_0$ the GH mass moment, $m\equiv M_0/M_*$ and $j = J/M_{*}^{2}$ related to the current dipole moment.

Let us now calculate the energy per unit mass of a particle in such an orbit, i.e.~the conserved quantity associated with the Killing vector related to stationarity:
\be
E = - \frac{g_{tt}+g_{t\phi} \Omega_{\orb}}{\sqrt{-g_{tt}-2 g_{t\phi} \Omega_{\orb} - g_{\phi\phi} \Omega_{\orb}^2}}\,.
\ee
As before, using the spacetime in Eq.~\eqref{Eq:metric-slow-rot}, the exterior solution derived in the previous subsection, and Eq.~\eqref{x}, one can re-express the energy change per logarithmic interval of the orbital frequency, a gauge invariant quantity, in terms of $v$ as
\bw
\ba
\Delta E &\equiv & - \frac{d E}{d \ln \Omega_{\orb}} = \frac{1}{27 {m}^{2/3}}\,{ \left[ 9 + 6\,C_{20}^\ext\,\epsilon^{2} + \left( 6\,C_{40}^\ext -{
C_{20}^\ext}^{2} \right) \epsilon^{4} \right] {v}^{2}} - \frac{1}{18 {m}^{4/3}}\,{
 \left[ 9 + 12 \,C_{20}^\ext\,\epsilon^{2}+ 2\left( 6\,C_{40}^\ext+{C_{20}^\ext
}^{2} \right) \epsilon^{4} \right] {v}^{4}} \nn \\
& &  +{\frac {20 \epsilon}{27 {m}^{5/3}}}\,{
 \left[ 3\,j + \left( 3\,C_{31}^\ext -j\,C_{20}^\ext \right) {
\epsilon}^{2} \right]  {v}^{5}} \nn \\
& & +\frac{1}{40 {m}^{2}}\,{ \left\{ -135+
 \left( -64\,C_{22}^\ext-270\,C_{20}^\ext-40\,{j}^{2} \right) \epsilon^{2}+
 \left[ -135\,(C_{20}^\ext)^2+ \left( 64\,C_{22}^\ext+40\,{j}^{2}
 \right) C_{20}^\ext-270\,C_{40}^\ext-40\,C_{42}^\ext \right] \epsilon^{4}
 \right\} {v}^{6}} \nn \\
& & +{\frac {28 \epsilon}{9 {m}^{7/3}}}\,{ \left[ 3\,j+
 \left( 3\,C_{31}^\ext+j\,C_{20}^\ext \right] \epsilon^{2} \right) {v}^{7}
} \nn \\
& & -{\frac {125}{4 {m}^{8/3}}}\, \left\{ {\frac {9}{20}}+ \left( {\frac {
448}{1125}}\,C_{22}^\ext+\frac{6}{5}\,C_{20}^\ext+{\frac {104}{675}}\,{j}^{2}
 \right) \epsilon^{2}+ \left[ (C_{20}^\ext)^2+ \left( -{\frac {448}{3375}}\,
C_{22}^\ext +{\frac {88}{2025}}\,{j}^{2} \right) C_{20}^\ext+{\frac {56}{
225}}\,C_{42}^\ext+\frac{6}{5}\,C_{40}^\ext \right. \right. \nn \\
& & \left. \left. -{\frac {128}{675}}\,C_{31}^\ext\,j
 \right] \epsilon^{4} \right\} {v}^{8} +{\frac {81 \epsilon}{2 {m}^{3}}}\, \left\{ 
j+ \left[ -{\frac {4}{27}}\,{j}^{3}+ \left( C_{20}^\ext-{\frac {32}
{135}}\,C_{22}^\ext \right) j+C_{31}^\ext-{\frac {4}{27}}\,C_{33}^\ext
 \right] \epsilon^{2} \right\} {v}^{9} \nn \\
& & -\frac{1}{18 {m}^{10/3}}\, \left\{ {\frac {59535}
{64}}+ \left( {\frac {99225}{32}}\,C_{20}^\ext+1122\,C_{22}^\ext+{\frac {
2345}{4}}\,{j}^{2} \right) \epsilon^{2}+ \left[ {j}^{4}+ \left( {
\frac {3538}{7}}\,C_{22}^\ext+{\frac {3725}{12}}\,C_{20}^\ext \right) {j}
^{2} \right. \right. \nn \\
& & \left. \left. + \left( -230\,C_{31}^\ext-30\,C_{33}^\ext \right) j+374\,C_{22}^\ext\,
C_{20}^\ext+{\frac {231525}{64}}\,(C_{20}^\ext)^2+{\frac {105}{2}}\,C_{40}^\ext +{\frac {18876}{35}}\,(C_{22}^\ext)^2+{\frac {99225}{32}}\,C_{40}^\ext \right. \right. \nn \\
& & \left. \left. +{\frac {2805}{4}}\,C_{42}^\ext \right] \epsilon^{4} \right\} {v}^{10} 
+\frac{275\epsilon}{{m}^{11/3}}\, \left\{ \frac{3}{5}\,j+ \left[ -{\frac {664}{6075}}\,{j}^{3}+
 \left( -{\frac {704}{3375}}\,C_{22}^\ext+C_{20}^\ext \right) j+\frac{3}{5}\,C_{31}^\ext -{\frac {32}{225}}\,C_{33}^\ext \right] \epsilon^{2} \right\} {v}^{11}
 \nn \\
& & +{\frac {1}{313600 {m}^
{4}}}\, \left\{ -56260575+ \left( -
81957120\,C_{22}^\ext-225042300\,C_{20}^\ext-53978400\,{j}^{2} \right) {\epsilon
}^{2} \right. \nn \\
& & \left. + \left[ -8144640\,{j}^{4}+ \left( -88270080\,C_{22}^\ext-
51223200\,C_{20}^\ext \right) {j}^{2}+ \left( -3091200\,C_{33}^\ext-
5510400\,C_{31}^\ext \right) j \right. \right. \nn \\
& & \left. \left. -225042300\,C_{40}^\ext-80936448\,(C_{22}^\ext)^{2}-7761600\,C_{44}^\ext-51223200\,C_{42}^\ext-337563450\,({C_{20}^\ext})^{
2}-81957120\,C_{22}^\ext\,C_{20}^\ext \right] \epsilon^{4} \right\} {v}^{12} \nn \\
& &   +\mathcal{O} \left( \epsilon^5, v^{13} \right)\,. 
\label{deltaE-us}
\ea
%

The energy change per logarithmic interval of the orbital frequency can also be computed assuming a spacetime expanded in multipole moments~\cite{Ryan:1995wh} 
\ba
\Delta E &=& \frac{1}{3}v^2 - \frac{1}{2} v^4 + \frac{20}{9} \frac{S_1}{M_0^2}v^5 - \left( \frac{27}{8} - \frac{M_2}{M_0^3}  \right) v^6 + \frac{28}{3} \frac{S_1}{M_0^2}v^7 - \left( \frac{225}{16} - \frac{80}{27} \frac{S_1^2}{M_0^4} - \frac{70}{9} \frac{M_2}{M_0^3} \right) v^8 \nn \\
& & + \left( \frac{81}{2} \frac{S_1}{M_0^2} + 6 \frac{S_1 M_2}{M_0^5} - 6 \frac{S_3}{M_0^4} \right) v^9 - \left( \frac{6615}{128} - \frac{115}{18} \frac{S_1^2}{M_0^4} - \frac{935}{24} \frac{M_2}{M_0^3} - \frac{35}{12} \frac{M_2^2}{M_0^6} + \frac{35}{12} \frac{M_4}{M_0^5} \right) v^{10} \nn \\
& & + \left( 165 \frac{S_1}{M_0^2}+ \frac{1408}{243} \frac{S_1^3}{M_0^6} + \frac{968}{27} \frac{S_1 M_2}{M_0^5} - \frac{352}{9}\frac{S_3}{M_0^4} \right) v^{11} \nn \\
& &  - \left( \frac{45927}{256} + \frac{123}{14} \frac{S_1^2}{M_0^4} - \frac{9147}{56} \frac{M_2}{M_0^3} - \frac{93}{4} \frac{M_2^2}{M_0^6} -24 \frac{S_1^2 M_2}{M_0^7} + 24 \frac{S_1 S_3}{M_0^6} + \frac{99}{4} \frac{M_4}{M_0^5} \right) v^{12} 
 + \mathcal{O} \left( v^{13} \right)\,,
\label{deltaE-ryan}
\ea
\ew
where $M_0$, $S_1$ $M_2$, $S_3$, $M_4$ are the GH mass monopole, current dipole, mass quadrupole, current octupole, and mass hexadecapole moments. 

Comparing Eqs.~\eqref{deltaE-us} and~\eqref{deltaE-ryan} term by term, i.e.~equating coefficients of terms proportional to $v^2$, $v^5$, $v^6$, $v^9$ and $v^{10}$, one can derive expressions for the GH multipole moments in terms of integration constants:
\begin{align}
\label{eq:M0}
M_0 &= M_{*} \left(1 + \epsilon^2 C_{20}^\ext +  \epsilon^4 C_{40}^\ext \right) + \mathcal{O}(\epsilon^6)\,,\\
S_1 &= \epsilon M_*^2 \left(j + \epsilon^2 C_{31}^\ext \right)  + \mathcal{O}(\epsilon^5)\,, \\
\label{eq:M2}
M_2 &= - \epsilon^2 M_*^3 \left( j^2 + \frac{8}{5} C_{22}^\ext + \epsilon^2 C_{42}^\ext \right) + \mathcal{O}(\epsilon^6)\,, \\
\label{eq:S3}
S_3 &= \epsilon^3 M_*^4 C_{33}^\ext   + \mathcal{O}(\epsilon^5)\,, \\
\label{eq:M4}
M_4 &= \frac{\epsilon^4 M_*^5}{735} \left[ 9432 (C_{22}^\ext)^2 + 9428 j^2 C_{22}^\ext + 749 j^4 \right. \nn \\
&  \left. + 735 C_{44}^\ext - 420 j C_{33}^\ext \right] + \mathcal{O}(\epsilon^6)\,.
\end{align}
$M_4$ can be rewritten in terms of previous multipole moments as
\begin{align}
\label{M4}
M_4 &= \frac{\epsilon^4 M_0^5}{5880} (29475 q^2 + 11810 \chi^2 q - 11673 \chi^4 \nn \\
& + 5880 C_{44}^\ext - 3360 \chi s_3) + \mathcal{O}(\epsilon^6)\,,
\end{align}
where $\chi \equiv S_1/M_0^2$, $q \equiv M_2/M_0^3$ and $s_3 \equiv S_3/M_0^4$. We have checked that the coefficients of terms proportional to $v^{4}$, $v^{7}$, $v^{8}$, $v^{11}$ and $v^{12}$ in Eqs.~\eqref{deltaE-us} and~\eqref{deltaE-ryan} are consistent with those in Eqs.~\eqref{eq:M0}--\eqref{eq:M4}, given the expressions for the multipole moments found above. This serves as an important consistency check of our calculations. Equations~\eqref{eq:M0}--\eqref{eq:M2} agree with~\cite{hartle1967}, while Eqs.~\eqref{eq:S3}--\eqref{M4} are new results. Once the multipole moments have been found, the moment of inertia $I$ is defined via $I \equiv S_{1}/\Omega$. Notice that $I$ is non-vanishing for a non-rotating configuration, while all the multipole moments except for the mass monopole vanish for such configuration.

\section{Rapidly-Rotating Stars}
\label{sec:rapid-rot}

The line element of a stationary, axisymmetric, and circular spacetime
is given in quasi-isotropic coordinates as
\begin{align}
\label{eq:quasi-isotropic}
 ds^2 & = - e^{2 \tilde{\nu}} dt^2 + e^{2(\zeta - \tilde{\nu})} ( dr^2 + r^2 d \theta^2
 ) \notag \\
 & + e^{-2\tilde{\nu}} B^2 r^2 \sin^2 \theta ( d \varphi - \omega dt )^2 ,
\end{align}
without loss of generality (see reviews for
\cite{stergioulas2003,gourgoulhon2010,friedman_stergioulas}). Here, $\tilde{\nu}
= \tilde{\nu} (r,\theta)$, $\zeta = \zeta (r,\theta)$, $\omega = \omega
(r,\theta)$, and $B = B (r,\theta)$. We consider uniformly rotating
configurations of a perfect fluid, and thus the stress-energy tensor
is given by Eq.~\eqref{Eq:matter-slow-rot} with $4$-velocity given in
Eq.~\eqref{eq:umu}. The Einstein equations and the conservation of 
stress-energy then become differential equations for the gravitational fields 
$(\tilde{\nu},\zeta,\omega,B)$ and an algebraic equation for the specific enthalpy, which we will later
solve numerically in the fully nonlinear regime.

Global quantities are computed for each stellar equilibrium model once
the gravitational and hydrostationary fields are solved. The mass and
angular momentum are computed by the Komar integral for compact sources
\cite{komar1959}:
\begin{align}
\label{eq:M0-komar}
 M_\mrm{Komar} &= - \int_\mathcal{V} ( 2 T^\mu{}_\nu - T \delta^\mu{}_\nu ) t^\nu
  dS_\mu\,,
\\
\label{eq:S1-komar}
 J_\mrm{Komar} &= \int_\mathcal{V} \left( T^\mu{}_\nu - \frac{1}{2} T
		       \delta^\mu{}_\nu \right) \phi^\nu dS_\mu\,,
\end{align}
where $\mathcal{V}$ is a constant time hypersurface and $dS_\mu$ is the
associated volume element. The circumferential equatorial
radius is given by $R_\mrm{eq} = e^{-\tilde{\nu}} B r_\mathrm{eq}$, where $r_\mathrm{eq}$ is the
coordinate equatorial radius of the star in quasi-isotropic coordinates defined by $p=0$ at $\theta = \pi / 2$. 

The GH multipole moments in terms of the metric functions in quasi-isotropic coordinates are given by
 \begin{align}
  M_0 & = M_\mrm{Komar} , \\
  S_1 & = J_\mrm{Komar} \equiv \chi M_0^2, \\
  M_2 & = - \frac{1 + 4 b_0 + 3 q_2}{3} M_0^3 , \\
  S_3 & = - \frac{3 ( 2\chi + 8\chi b_0 - 5 w_2 )}{10} M_0^4 , \\
  M_4 & = \left(\frac{19}{105} - \frac{6}{35} \chi^2 + \frac{32}{21} b_0 + \frac{8}{7} q_2 + \frac{16}{5} b_0^2 
\right.  \nn \\ & \left.
  + \frac{24}{7} b_0 q_2
  - q_4 - \frac{64}{35} b_2  \right) M_0^5\,,
 \end{align}
extending the results of~\cite{pappas-apostolatos} (see also~\cite{friedman_stergioulas}). The constants $(b_{0},b_{2},q_{2},q_{4}, w_{2})$ in this expression are defined
by the expansion of the metric components at
spatial infinity \cite{butterworth_ipser1976}:
\begin{widetext}
 \begin{align}
  \tilde{\nu} =& - \frac{M_0}{r} + \frac{b_0}{3} \left( \frac{M_0}{r} \right)^3
  + \chi^2 \left( \frac{M_0}{r} \right)^4 - \frac{3b_0^2 - b_2 + 36\chi^2}{15}
  \left( \frac{M_0}{r} \right)^5 + \frac{4 ( 12 - 7 b_0 ) \chi^2}{15} \left(
  \frac{M_0}{r} \right)^6 \notag \\
  & + \left[ q_2 \left( \frac{M_0}{r} \right)^3 - 2 \chi^2 \left( \frac{M_0}{r}
  \right)^4 - \frac{9 b_0 q_2 - 16 b_2 - 72 \chi^2}{21} \left( \frac{M_0}{r}
  \right)^5 + \frac{(45 w_2 + 56 \chi b_0 - 84\chi) \chi}{21} \left( \frac{M_0}{r}
  \right)^6 \right] P_2 ( \mu ) \notag \\
  & + \left[ q_4 \left( \frac{M_0}{r} \right)^5 - \frac{36 \chi w_2}{7}
  \left( \frac{M_0}{r} \right)^6 \right] P_4 ( \mu ) + \mathcal{O} \left[ \left( \frac{M_0}{r} \right)^7 \right] ,
\\
 M \omega = & \biggl[ 2\chi \left( \frac{M_0}{r} \right)^3 - 6\chi \left(
 \frac{M_0}{r} \right)^4 + \frac{6( 8-3 b_0 )}{5} \left( \frac{M_0}{r}
 \right)^5 - \frac{4( 40 - 50 b_0 + 3 q_2 )}{15} \left( \frac{M_0}{r}
 \right)^6 \notag \\
 & \, + \frac{2 \{ 160 - 10 b_0 ( 44 - 9 b_0 ) + 9 b_2 + 84 \chi^2 + 48 q_2
 \}}{35} \left( \frac{M_0}{r} \right)^7 \notag \\
 & \, - \frac{4\chi ( 56 - 280 b_0 + 189 b_0^2 + 18 b_2 + 228 \chi^2 + 42 q_2
 - 18 b_0 q_2 ) - 45 q_2 w_2 }{35} \left( \frac{M_0}{r} \right)^8 \biggr]
 \frac{dP_1 (\mu)}{d\mu} \notag \\
 + & \biggl[ w_2 \left( \frac{M_0}{r} \right)^5 + \frac{18 q_2 \chi - 25
 w_2}{10} \left( \frac{M_0}{r} \right)^6 - \frac{4\chi ( 8 b_2 + 8 \chi^2 + 21
 q_2 ) - 25 ( 2 - b_0 ) w_2}{15} \left( \frac{M_0}{r} \right)^7 \notag \\
 & \, + \frac{2\chi ( 1216 b_2 + 1552 \chi^2 + 1428 q_2 - 621 b_0 q_2 - 70 q_4
 ) + 35 ( - 28 + 47 b_0 + 6 q_2 ) w_2}{315} \left( \frac{M_0}{r} \right)^8
 \biggr] \frac{dP_3 (\mu)}{d\mu} \notag \\
 + & \left[ w_4 \left( \frac{M_0}{r} \right)^7 + \frac{70 \chi q_4 + 66 q_2
 w_2 - 147 w_4}{63} \left( \frac{M_0}{r} \right)^8 \right] \frac{dP_5
 (\mu)}{d\mu} + \mathcal{O} \left[ \left( \frac{M_0}{r} \right)^9 \right] .
\\ 
B &= 1 + \sqrt{\frac{\pi}{2}} \sum_\ell b_{2\ell} \left( \frac{M_0}{r}
						  \right)^{2(\ell+1)}
 T_{2\ell}^{1/2} ( \mu ) ,
\end{align}
\end{widetext}
where $T_\ell^{1/2} ( \mu )$ are Gegenbauer polynomials and we recall that
$P_\ell (\mu)$ are Legendre polynomials, with $\mu \equiv \cos \theta$. 

\subsection{Numerical Methods}
\label{sec:LORENE-RNS}

Equilibrium configurations of rotating stars are computed using two
public numerical codes to double-check results. One is
LORENE/rotstar \cite{bonazzola_gsm1993,bonazzola_gm1998}, and
the other is RNS \cite{stergioulas_friedman1995}. The consistency
of these two codes has been extensively studied in \cite{nozawa_sge1998}. We here compare these codes by studying multipole moments for the first time. Hereafter, we briefly summarize the computational methods used in each
code.

\subsubsection{LORENE/rotstar}

In LORENE/rotstar, an equilibrium configuration is computed
once the central specific enthalpy [$h \equiv ( \rho + p ) / \rho_0$, where $\rho_0$ is the baryon rest-mass density] and the angular velocity are given, along with an EoS. Four elliptic-type equations are solved with multi-domain spectral methods to obtain the four metric functions $\tilde{\nu}$ and $\zeta$, as well as the combination $B(r) r \sin \theta$, and $\omega(r) r \sin \theta$. 

The four main equations can be obtained from the Einstein equations
\cite{gourgoulhon2010}; they are
\begin{align}
 \Delta_3 \tilde{\nu} & = \sigma_{\tilde{\nu}} , \\
 \tilde{\Delta}_3 ( \omega r \sin \theta )& = \sigma_\omega , \\
 \Delta_2 ( B r \sin \theta ) & = \sigma_B , \\
 \Delta_2 \zeta & = \sigma_\zeta ,
\end{align}
where we have defined the differential operators
\begin{align}
 \Delta_3 & \equiv \frac{\partial^2}{\partial r^2} + \frac{2}{r}
  \frac{\partial}{\partial r} + \frac{1}{r^2} \frac{\partial^2}{\partial
  \theta^2} + \frac{\cos \theta}{r^2 \sin \theta}
  \frac{\partial}{\partial \theta} , \\
 \tilde{\Delta}_3 & \equiv \Delta_3 - \frac{1}{r^2 \sin^2 \theta} , \\
 \Delta_2 & \equiv \frac{\partial^2}{\partial r^2} + \frac{1}{r}
  \frac{\partial}{\partial r} + \frac{1}{r^2} \frac{\partial^2}{\partial
  \theta^2}\,,
\end{align}
and the source functions are 
\begin{align}
 \sigma_{\tilde{\nu}} & = 4 \pi e^{2( \zeta - \tilde{\nu} )} \left\{ \left[ 2 e^{2\tilde{\nu}}
 \left( u^0 \right)^2 - 1 \right] ( \rho + p ) + 2 p \right\} \notag \\
 & + \frac{1}{2} e^{-4\tilde{\nu}} B^2 r^2 \sin^2 \theta \partial \omega
 \partial \omega - \partial \tilde{\nu} \partial ( \ln B ) , \\
 \sigma_\omega & = - 16 \pi e^{2\zeta} \left( u^0 \right)^2 ( \rho + p )
 ( \Omega - \omega) r \sin \theta \notag \\
 & + r \sin \theta \partial \omega \partial ( 4 \tilde{\nu} - 3 \ln B ) , \\
 \sigma_B & = 16 \pi e^{2( \zeta - \tilde{\nu} )} B p r \sin \theta , \\
 \sigma_\zeta & = 8 \pi e^{2( \zeta - \tilde{\nu} )} \left\{ \left[ e^{2\tilde{\nu}}
 \left( u^0 \right)^2 - 1 \right] ( \rho + p )+ p \right\} \notag \\
 & + \frac{3}{4} e^{-4\tilde{\nu}} B^2 r^2 \sin^2 \theta \partial \omega
 \partial \omega - \partial \tilde{\nu} \partial \tilde{\nu} ,
\end{align}
where
\begin{equation}
 \partial f \partial g \equiv \frac{\partial f}{\partial r}
  \frac{\partial g}{\partial r} + \frac{1}{r^2} \frac{\partial
  f}{\partial \theta} \frac{\partial g}{\partial \theta} .
\end{equation}
Boundary conditions are chosen such that the metric remains asymptotically flat. 

The only nontrivial hydrostationary equation is the
relativistic Euler equation, which can be obtained from the spatial sector of the local energy-momentum conservation equation. Stationarity and axisymmetry allows us to rewrite this equation in first-integral form:
\begin{equation}
 \frac{h}{u^0} = \mathrm{const} .
\end{equation}
where recall $u^0$ is a function of the metric through the normalization
of the fluid's $4$-velocity. This equation gives the specific enthalpy field,
given its central value. The gravitational and hydrostationary equations are solved iteratively, until a convergent solution is obtained.

The metric coefficients $q_{2\ell}$, $w_{2\ell}$, and $b_{2\ell}$
are computed through certain integrals. 
Following \cite{butterworth_ipser1976,salgado_bgh1994}, these coefficients are obtained from
\begin{align}
 q_{2\ell} &= - \frac{1}{4\pi M_0^{2\ell + 1} } \int_\mathcal{V} \sigma_{\tilde{\nu}}
  P_{2\ell} r^{2\ell + 2} dr d\mu d\phi , \label{eq:lormass}
\\
w_{2\ell} &= - \frac{1}{8\pi (\ell + 1) (2\ell + 1)  M_0^{2\ell + 2} }
 \nn \\
 & \times \int_\mathcal{V} \sigma_\omega \frac{dP_{2\ell + 1}}{d\mu}
 r^{2\ell + 3} \sin \theta dr d\mu d\phi , \label{eq:lorcurrent}
\\
 b_{2\ell} &= - \frac{1}{4 \pi (2\ell + 1) M_0^{2\ell+2}} \label{eq:lorb}
  \sqrt{\frac{2}{\pi}} \int_\mathcal{V} \sigma_B r^{2\ell+2}
  T_{2\ell}^{1/2} dr d\mu d\phi .
\end{align}

The computational domain of LORENE/rotstar is composed of three
regions. The first region, the so-called \emph{nucleus}, is a spheroidal domain, whose surface is adapted to the stellar surface. The second region is a shell region surrounding the nucleus. The inner boundary of this shell is the same as the outer boundary of the nucleus, while the outer boundary of the shell is a sphere with twice the radius of the nucleus at the equator. The third region is a
compactified external domain that extends from the outer boundary of the
shell to spatial infinity. The compactified external
domain allows us to impose exact boundary conditions at spatial
infinity. 

The elliptic equations are solved in each computational
domain, and matching conditions are imposed so that values of the metric
functions and their derivatives agree on both sides of each domain. 
In LORENE, functions of $r$ and $\theta$ are expanded in Chebyshev polynomials
and trigonometric functions, respectively, and the latter are re-expanded
in Legendre polynomials (or spherical harmonics in general situations)
when it is advantageous. We usually use $49 \times 3$ and $25$
collocation points in the $r$ and $\theta$ directions, respectively, to
compute all quantities except for $q_4$. We compute
$q_4$ with a smaller number of collocation points, such as $33 \times 3$
and $17$ in the $r$ and $\theta$ directions. This is because a large number of collocation
points introduce substantial contamination, due to the numerical
noise associated with the high-order coefficients of the spectral
expansions. We carry out convergence tests and present results only when
all multipole moments extracted (and in particular $M_4$) are accurate to 
$\sim 1\%$. Since the numerical error for higher multipolar coefficients increases as the NS spin decreases, this implies we cannot extract multipoles when $\chi \lesssim 0.1$.

\subsubsection{RNS}

As in LORENE/rotstar, in RNS an equilibrium configuration is calculated once the rotation law and the EoS are specified, together with two more parameters. One of them is the central energy density, while the other can be any of the following: the angular velocity, the mass, the rest mass, the angular momentum, or the ratio of the polar coordinate radius to the equatorial coordinate radius in quasi-isotropic coordinates. In our case, we choose to construct uniformly rotating models by varying the central density and the ratio of the polar to the equatorial coordinate radius.  

In RNS, we solve for the metric functions $(\tilde{\rho},\omega,\gamma,\alpha)$, which are related to the metric functions of the line element in quasi-isotropic coordinates via $\tilde{\nu}=(\gamma+\tilde{\rho})/2$, $B=e^{\gamma}$, and $\zeta-\tilde{\nu}=\alpha$. Three of the four gravitational field equations are solved through a Green's function approach, as developed in~\cite{komatsu_eh1989}, with the modifications introduced in~\cite{cook_st1992}, so that the infinite radial domain is compactified. The fourth gravitational equation reduces to a first order differential equation and is integrated straightforwardly. The first three metric functions are given by the integrals,
\begin{widetext}
 \begin{align}
 \tilde{\rho}(s,\mu) =& - e^{-\gamma/2} \sum_{n=0}^{\infty} P_{2n}(\mu)
                           \left[\left(\frac{1-s}{s}\right)^{2n+1} \right. \int_0^s\frac{ds's'^{2n}}{(1-s')^{2n+2}}\int_0^1 d\mu'P_{2n}(\mu')\tilde{S}_{\tilde{\rho}}(s',\mu')\nn\\
                                        &+\left.  \left(\frac{s}{1-s}\right)^{2n}\int_s^1\frac{ds'(1-s')^{2n-1}}{s'^{2n+1}}\int_0^1 d\mu'P_{2n}(\mu')\tilde{S}_{\tilde{\rho}}(s',\mu')\right], \label{metricrho}\\
\omega(s,\mu) =& -\frac{1}{r_\mrm{eq}} e^{(2\tilde{\rho}-\gamma)/2} \sum_{n=1}^{\infty} \frac{1}{2n(2n-1)} \frac{d P_{2n-1} (\mu)}{d\mu}
                           \left[\left(\frac{1-s}{s}\right)^{2n+1}\int_0^s\frac{ds's'^{2n}}{(1-s')^{2n+2}}\int_0^1 d\mu' (1-\mu'{}^2) \frac{d P_{2n-1}(\mu')}{d\mu'}\tilde{S}_{\hat{\omega}}(s',\mu')\right.\nn\\
                                        &           +\left.  \left(\frac{s}{1-s}\right)^{2n-2}\int_s^1\frac{ds'(1-s')^{2n-3}}{s'^{2n-1}}\int_0^1 d\mu' (1-\mu'{}^2) \frac{d P_{2n-1}(\mu')}{d\mu'}\tilde{S}_{\hat{\omega}}(s',\mu')\right], \label{metricomega}\\                                        
\gamma(s,\mu) =& -\frac{2e^{-\gamma/2}}{\pi}  \sum_{n=1}^{\infty} \frac{\sin[(2n-1)\theta]}{(2n-1)\sin\theta}
                           \left[\left(\frac{1-s}{s}\right)^{2n}\int_0^s\frac{ds's'^{2n-1}}{(1-s')^{2n+1}}\int_0^1 d\mu' \sin[(2n-1)\theta']\tilde{S}_{\gamma}(s',\mu')\right.\nn\\
                                        &           +\left.  \left(\frac{s}{1-s}\right)^{2n-2}\int_s^1\frac{ds'(1-s')^{2n-3}}{s'^{2n-1}}\int_0^1 d\mu' \sin[(2n-1)\theta']\tilde{S}_{\gamma}(s',\mu')\right],\label{metricgama}
\end{align}
\end{widetext}
where recall that $\mu=\cos\theta$ and $s$ is the radial domain parameter, defined through $r=r_\mrm{eq}\left(\frac{s}{1-s}\right)$ with $s \in [0,1]$. Note that $r = 1/2$ at the coordinate equatorial radius of the star in quasi-isotropic coordinates, $r_\mrm{eq}$. These integrals are defined in terms of source functions $\tilde{S}_{\tilde{\rho}}(s,\mu),\tilde{S}_{\hat{\omega}}(s,\mu),\tilde{S}_{\gamma}(s,\mu)$, which depend on the energy density, the pressure, the angular velocity and the metric functions (see~\cite{cook_st1992} and Appendix~\ref{RNSfield}). These integrals, together with the integration of the metric function $\alpha$ and the equation of hydrostatic equilibrium, are evaluated iteratively until a convergent solution is obtained to a given accuracy. 

As in LORENE/rotstar, in RNS the multipole moments are evaluated through the following integrals:
\ba M_0^{2\ell+1}q_{2\ell}&=&-\frac{r_\mrm{eq}^{2\ell+1}}{2}\int_0^1 \frac{ds' s'^{2\ell}}{(1-s')^{2\ell+2}}\nn\\
                                  &&\times\int_0^1 d\mu' P_{2\ell}(\mu')\tilde{S}_{\tilde{\rho}}(s',\mu'),\label{RNSq}\\
M_0^{2\ell}w_{2\ell-2}&=& \frac{r_\mrm{eq}^{2\ell}}{4\ell}\int_0^1 \frac{ds' s'^{2\ell}}{(1-s')^{2\ell+2}}\nn\\
                                   &&\times\int_0^1 d\mu' (1 - \mu'{}^2) \frac{d P_{2\ell-1} (\mu')}{d \mu'} \tilde{S}_{\hat{\omega}}(s',\mu'),\label{RNSw} \nn \\
\\
M_0^{2\ell+2} b_{2\ell}&=&-\frac{16 \sqrt{2\pi} r_\mrm{eq}^{2\ell+4}}{2\ell+1}\int_0^{1/2} \frac{ds' s'^{2\ell+3}}{(1-s')^{2\ell+5}}\nn\\
                                                            &&\times\int_0^1 d\mu' \sqrt{1-\mu'{}^2} p(s',\mu') e^{\gamma+2\alpha} T_{2\ell}^{1/2}(\mu'), \nn \\ \label{RNSb}
\ea
where in the second integral $\ell\geq 1$, while in the other two integrals $\ell\geq 0$,  and the last integral is evaluated inside the source only. Note that in the last integral the pressure $p$ has units of inverse length squared in geometric units, so the equations are dimensionally correct.       

The computational domain of RNS is a grid in $(r,\theta)$ space, with extent $r \in [0,\infty]$ and $\theta \in [0,\pi/2]$. In practice, RNS uses the angular coordinate $\mu=\cos\theta \in [0,1]$ and the compactified radial coordinate $s=\frac{r}{r+r_\mrm{eq}} \in[0,1]$. If one chooses the size of the angular domain to be \textit{MDIV}, then the angular variable takes the values, $\mu[k]=\frac{k-1}{MDIV-1}$, where $k \in [1,MDIV]$. Similarly, if one chooses the size of the radial domain to be \textit{SDIV}, then the radial variable takes the values, $s[i]=0.9999\frac{i-1}{SDIV-1}$, where $i \in [1,SDIV]$. In this paper, we choose $MDIV\times SDIV=301\times601$.

\begin{figure*}[htb]
\centering
\includegraphics[width=\columnwidth,clip=true]{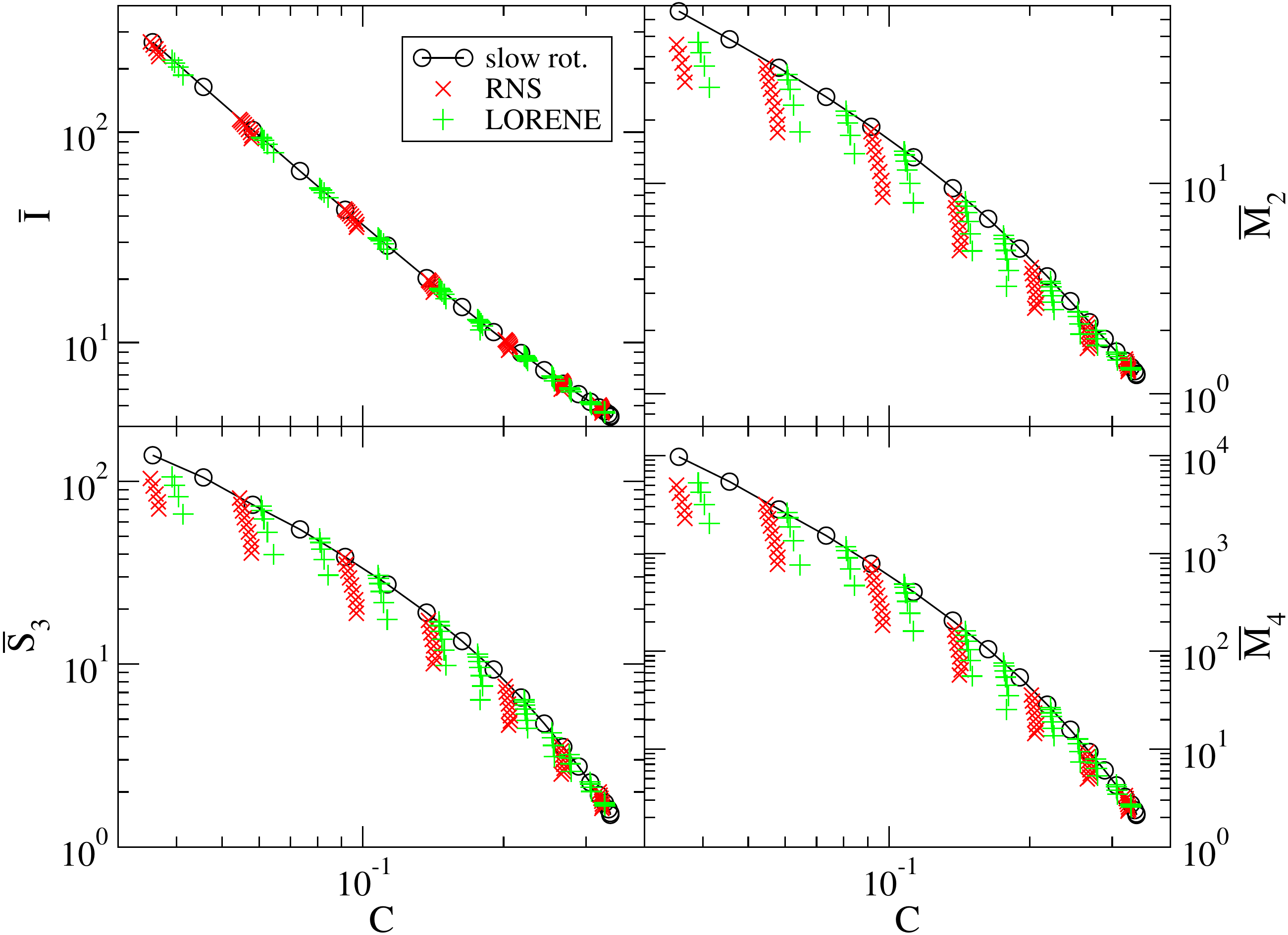}
\includegraphics[width=\columnwidth,clip=true]{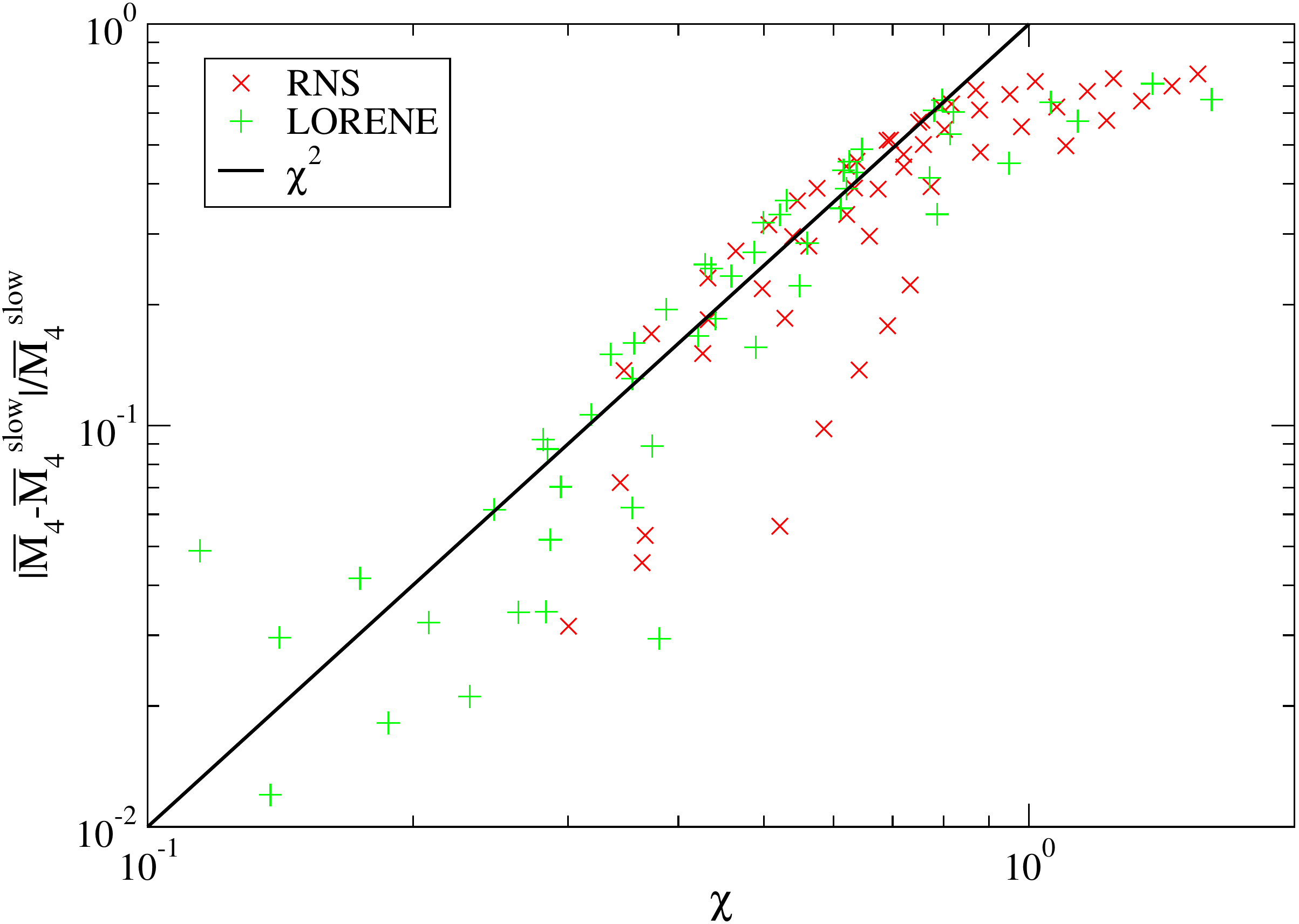}
\caption{
(Color online) (Left) $\bar{I}$--$C$ (top left), $\bar{M}_2$--$C$ (top right), $\bar{S}_3$--$C$ (bottom left) and $\bar{M}_4$--$C$ (bottom right) relations for an $n=0.5$ polytropic EoS, where $C$ is the stellar compactness. The black circles are calculated to leading-order in slow-rotation, while the green plusses and red crosses are computed with the LORENE and RNS codes respectively for a sequence of spins (increasing from top to bottom). Observe that the $\bar{I}$--$C$ relation is almost insensitive to spin, while the other three relations depend clearly on spins. (Right) Fractional difference,  as a function of dimensionless spin, of $\bar{M}_4$ computed with the LORENE or RNS codes and in the slow-rotation approximation. Observe that the fractional difference scales as $\chi^2$ as expected. The scattering is because (i) $\bar{M}_4$ depends on $C$ for any given $\chi$ and (ii) $\bar{M}_4$ computed with the LORENE and RNS codes contains spin corrections higher than $\mathcal{O}(\chi^2)$.
\label{fig:C-dep}
}
\end{figure*}

The main difficulty of the numerical calculation comes in the evaluation of the $q_4$ coefficient for the models with small angular velocity. The quantities $q_{2 \ell}$ are expected to be finite and become smaller as one considers models with smaller angular velocity. Therefore, one expects the result of the angular integration in Eq.~(\ref{RNSq}) to have such a radial dependence that, when multiplied by the appropriate power of $r$ and integrated to infinity, to converge to a finite result. That means that the integrand of the radial integration should have the appropriate asymptotic behavior so as to be integrable. Moreover, because of the asymptotic behavior of the metric potential one expects the integral in the neighborhood of spatial infinity to have a sub-leading contribution to the final result, with the main contribution coming from smaller radii. Unfortunately, numerical errors and the finite accuracy of the computation introduce a contribution to the radial integrand that results in singular behavior near $s=1$ ($r \to \infty$), as the multipole order increases. To avoid this singularity, one can exclude some of the last points from the integration, which have a negligible effect in the final result due to the asymptotic behavior of the metric functions. By varying the size of the grid used and testing the consistency of the integration as a function of the number of excluded points, we found that excluding the last three points gives a relatively stable result when computing $q_4$. Errors in $q_{4}$ are estimated to be of ${\cal{O}}(5\%)$ or less, for spin parameters $\chi\gtrsim0.2$ (see Appendix \ref{q4truncation}).   

\section{Numerical Results}
\label{sec:results}

We here present the multipole moments of NSs and QSs computed with assuming a slow-rotation expansion [Sec.~\ref{sec:slow-rot}], the LORENE and RNS codes [Sec.~\ref{sec:LORENE-RNS}]. We work in dimensionless multipole moments defined by
\be
\label{eq:def-dimensionless}
\bar{M}_{\ell} = (-)^{\frac{\ell}{2}} \frac{M_{\ell}}{M_0^{\ell + 1} \chi^{\ell}}\,,
\quad
\bar{S}_{\ell} = (-)^{\frac{\ell-1}{2}} \frac{S_{\ell}}{M_0^{\ell + 1} \chi^{\ell}}\,,
\ee
where recall that $\chi \equiv S_1/M_0^2$. For $\bar{M}_4$, we use RNS data only when $\chi > 0.3$ and LORENE data when $\chi>0.1$, since numerical errors are relatively large for smaller spin values.

We consider various EoSs. For NSs, we use the APR~\cite{APR}, AU~\cite{Arnett:1976dh}, L~\cite{Wiringa:1988tp}, SLy~\cite{SLy}, UU~\cite{Arnett:1976dh} and Shen~\cite{Shen1,Shen2} EoSs. These have a maximum NS mass above 1.97$M_\odot$ for a non-rotating configuration, the lower bound of the recently-found massive pulsar PSR J0348+0432~\cite{2.01NS}. Details of the first five EoSs are explained in~\cite{Pappas:2013naa}. We adopt a neutrino-less and beta-equilibrium configuration for the Shen EoS. Strange QS EoSs are based on the MIT bag model~\cite{farhi_jaffe1984}. Specifically, they are constructed according to the linear interpolation developed in~\cite{zdunik2000}. Values of the bag constant, QCD coupling constant and strange quark mass are adopted from the ``SQM'' family, provided in~\cite{lattimer_prakash2001}. We also consider a polytropic EoS of the form
\be
p = K \rho_0^{1+1/n}\,.
\ee
Using the first law of thermodynamics, the energy density $\rho$ is related to the rest mass density $\rho_0$ by $\rho = \rho_0 + n p$. NSs can be approximately modeled with a polytropic index in the range $n=0.5-1$~\cite{lattimer_prakash2001,flanagan-hinderer-love}.

\subsection{Consistency of the LORENE, RNS and Slow-Rotation Calculations}

The left panel of Fig.~\ref{fig:C-dep} compares $\bar{I} (\equiv I/M_0^3)$, $\bar{M}_2$, $\bar{S}_3$ and $\bar{M}_4$ as a function of compactness $(C \equiv M_0/R_\mrm{eq})$ for an $n=0.5$ polytropic EoS, computed with the LORENE code, the RNS code and in the slow-rotation approximation. We have checked that the slow-rotation numerical calculations for $\bar{M}_4$ have a numerical error smaller than $\mathcal{O}(10^{-5})$ for $\bar{M}_2<20$, $\mathcal{O}(10^{-3})$ for $20< \bar{M}_2<40$ and $\mathcal{O}(10^{-2})$ for $40< \bar{M}_2<50$. As the Newtonian regime is approached (as $\bar{M}_2$ is increased) it becomes increasingly more difficult to construct NS and QS solutions with a fully relativistic code. On the other hand, the LORENE data has a numerical error smaller than $\sim 1\%$, while the RNS data has an error smaller than $\sim 5\%$ for slowly-rotating stars, with the error decreasing for rapidly-rotating stars as explained in detail in Appendix~\ref{q4truncation}. Observe that all results are consistent with each other, in spite of the stellar sequences being constructed by fixing different quantities, and thus the moments being  computed at slightly different compactnesses. The vertical scatter of LORENE and RNS results arises because the dimensionless moments depend on $\chi$ when computed to all orders in spin, but they are $\chi$-independent to leading order in the slow-rotation calculation. We thus expect the LORENE and RNS results to disagree with the slow-rotation calculation at ${\cal{O}}(\chi^{2})$ in a $\chi \ll 1$ expansion, as verified in the right panel of Fig.~\ref{fig:C-dep}. Observe, however, that the $\bar{I}$--$C$ relation is almost insensitive to spin, which can be explained analytically in the Newtonian limit [see discussion in Sec.~\ref{sec:Newton}].   

The right panel of Fig.~\ref{fig:C-dep} shows the fractional difference between $\bar{M}_{4}$ computed with the LORENE or RNS codes, and the slow-rotation result. As already explained, this fractional difference should scale as ${\cal{O}}(\chi^{2})$ to leading order in a $\chi \ll 1$ expansion, a feature borne out by the figure. Observe, however, that there is still scattering in the fractional difference calculation about the $\chi^2$ line. One reason for this is that $\bar{M}_{4}$ depends on $C$ for any given value of $\chi$, and we have included all LORENE and RNS results, irrespective of the star's compactness. Another reason is that the LORENE and RNS codes contain spin corrections higher than $\mathcal{O}(\chi^2)$, since these codes compute the moments to all orders in spin. Higher-order in $\chi$ corrections become more important as $\chi$ increases.

\begin{figure}[htb]
\centering
\includegraphics[width=\columnwidth,clip=true]{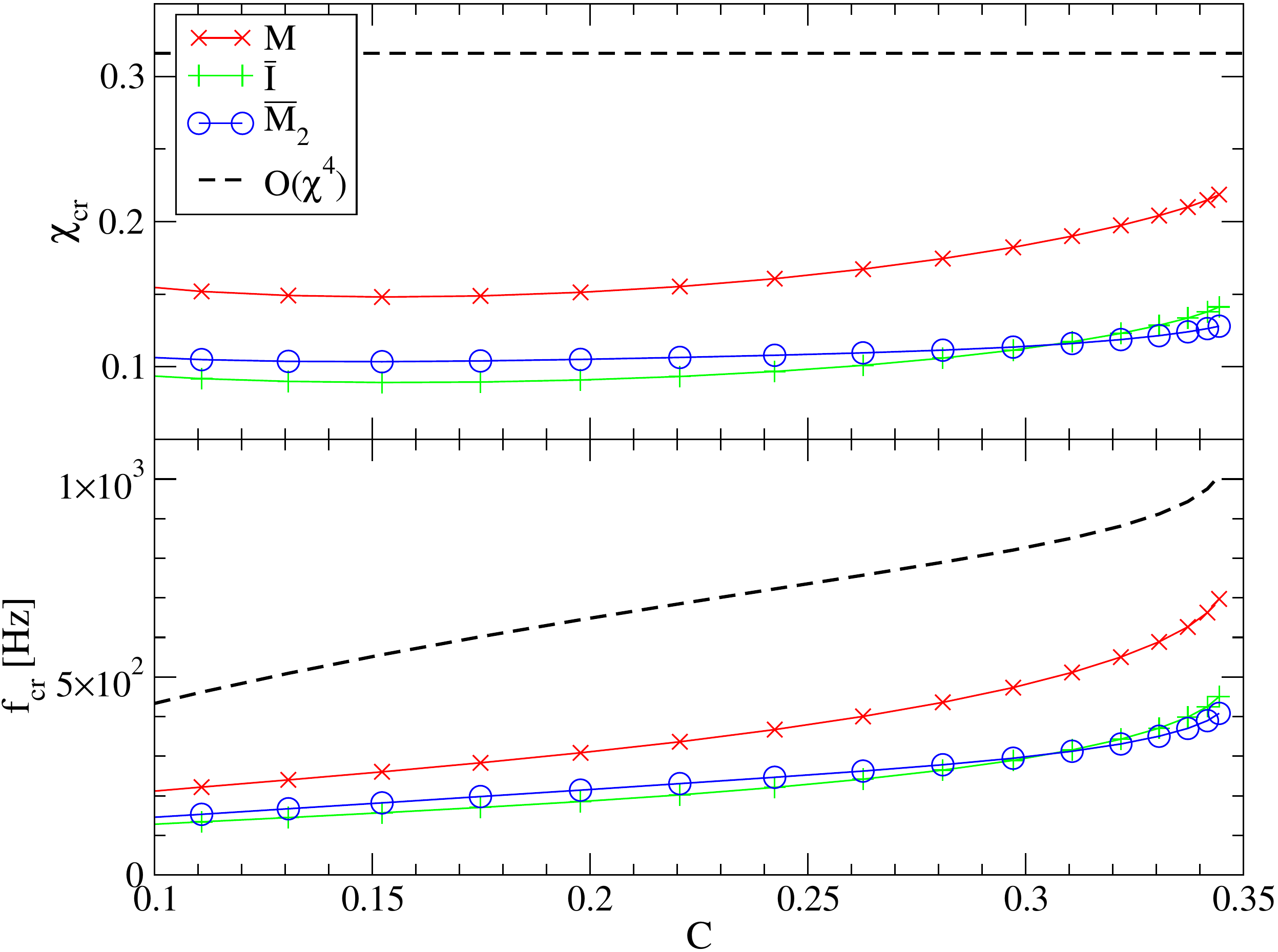}
\caption{
(Color online) (Top) Critical spin parameter $\chi_\mrm{cr}$ against stellar compactness, below which the difference between the $\mathcal{O}(\chi^0)$ and $\mathcal{O}(\chi^2)$ parts of the stellar mass (red cross), moment of inertia (green plus) and quadrupole moment (blue circle) with an APR EoS in the slow-rotation approximation is less than 1\%. The black dashed line roughly corresponds to $\chi_\mrm{cr}$ below which the difference between the $\mathcal{O}(\chi^0)$ and $\mathcal{O}(\chi^4)$ parts is less than 1\%, assuming that the latter is expressed as the leading order term times $\chi^4$. (Bottom) Same as the top panel, except for the critical spin frequency $f_\mrm{cr}$. 
\label{fig:leading-vs-spin-corr}
}
\end{figure}

\begin{figure}[htb]
\centering
\includegraphics[width=\columnwidth,clip=true]{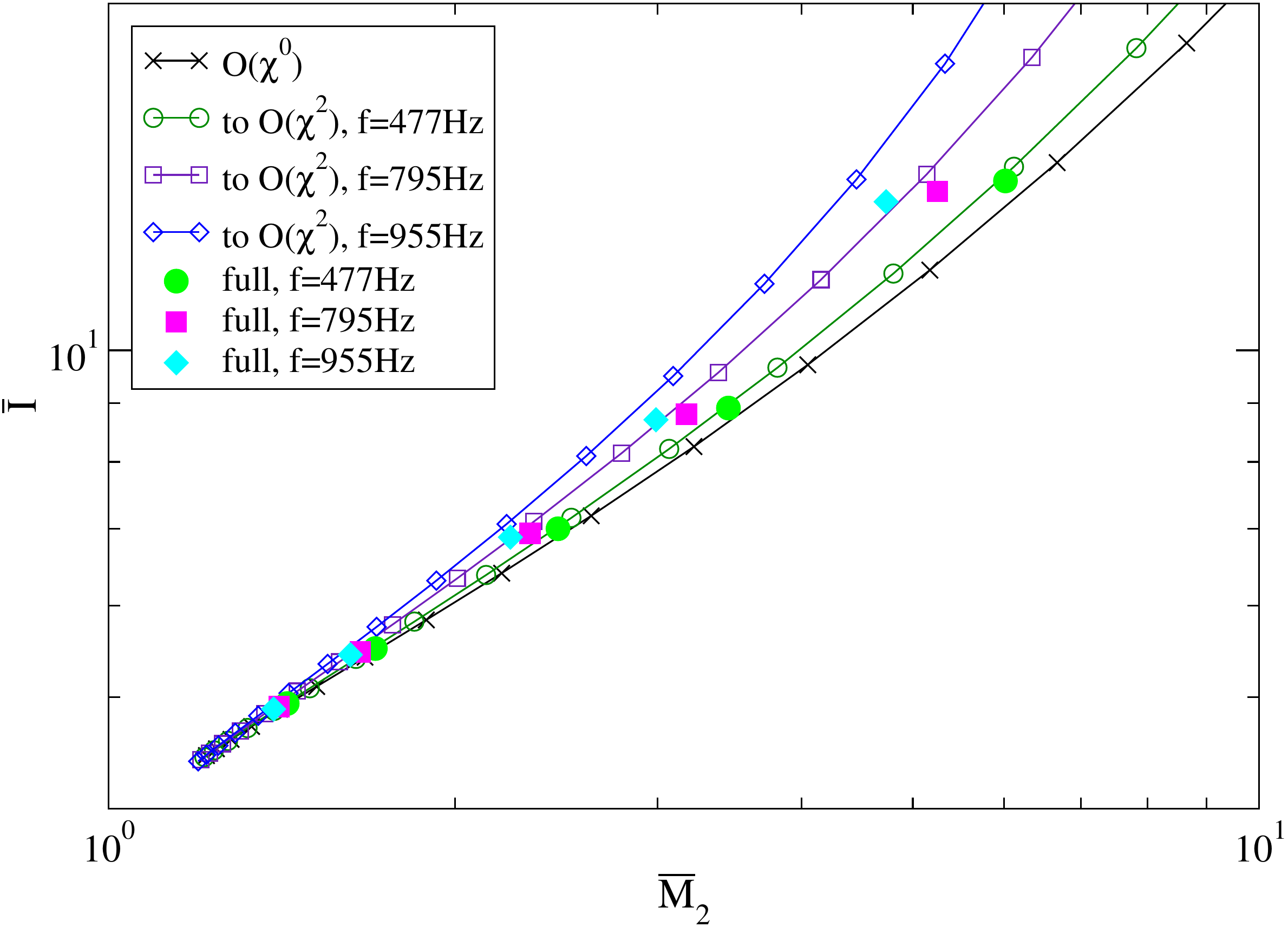}
\includegraphics[width=\columnwidth,clip=true]{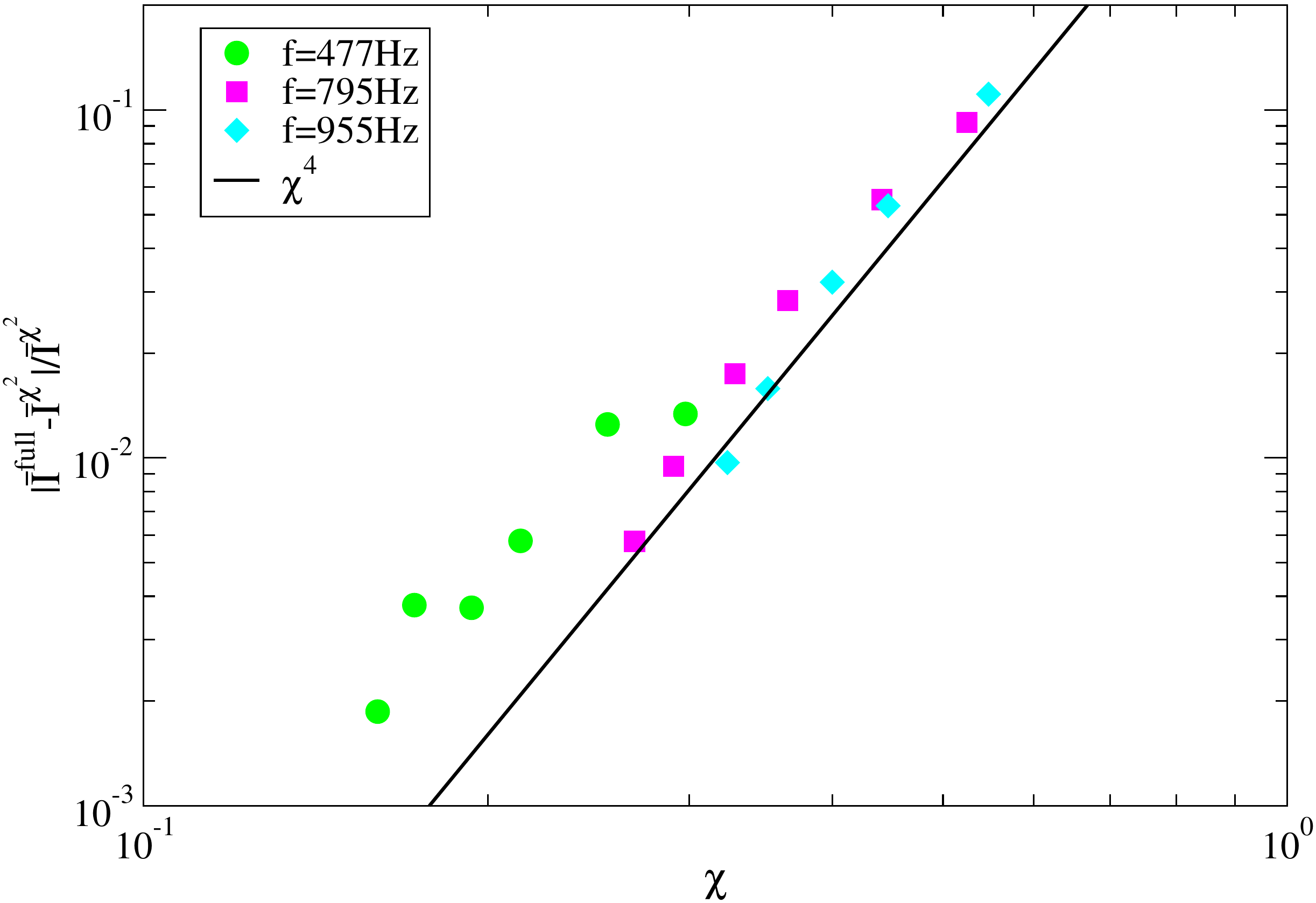}
\caption{
(Color online) (Top) $\bar{I}$--$\bar{M}_2$ relation for an APR EoS in the slow-rotation limit, including quadratic order spin corrections and full-order spin corrections using LORENE for various spin frequencies. (Bottom) Spin dependence of the fractional difference. Observe the scaling of $\chi^4$ as expected.  
\label{fig:IQ-Omega}
}
\end{figure}

\begin{figure}[htb]
\centering
\includegraphics[width=\columnwidth,clip=true]{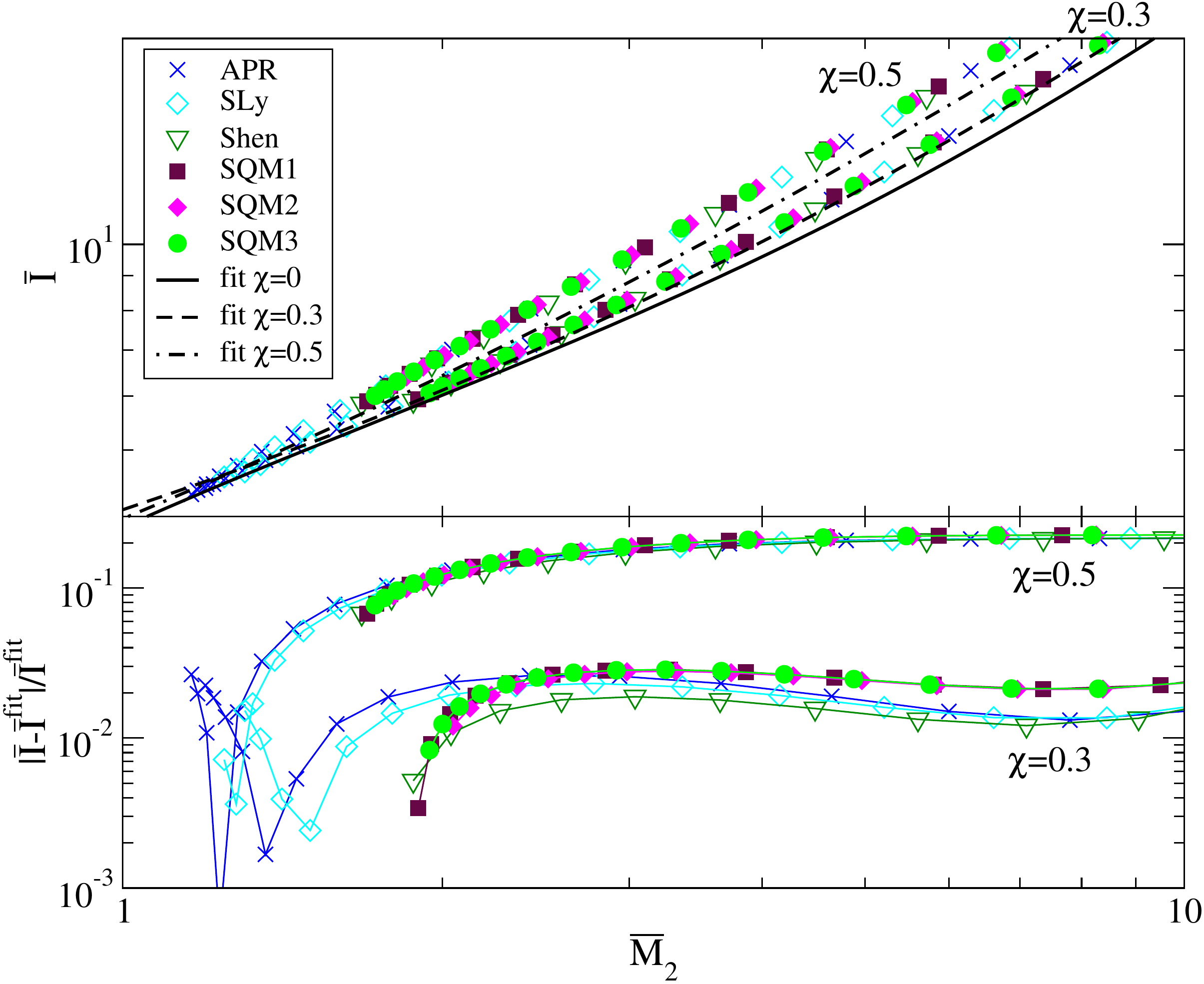}
\caption{
(Color online) (Top) $\bar{I}$--$\bar{M}_2$ relations computed in the slow-rotation approximation (including $\mathcal{O}(\chi^2)$ corrections) for various realistic NS and QS EoSs and with $\chi = 0.3$ and $0.5$. Observe that the QS relation is almost the same as the NS one. We also show the fit to the slow-rotation results (without ${\cal{O}}(\chi^{2})$ corrections) of~\cite{I-Love-Q-Science,I-Love-Q-PRD} and the fit to RNS data valid for all spins of~\cite{Pappas:2013naa}. (Bottom) Fractional difference between the data and the fits in~\cite{Pappas:2013naa}. 
\label{fig:IQ-chi}
}
\end{figure}

The slow-rotation expansion to quartic order in spin allows us to calculate the relative ${\cal{O}}(\chi^{2})$ spin corrections to the stellar mass, moment of inertia and quadrupole moment. Let us use these results to roughly estimate the value of $\chi$ or $f$ at which such corrections become important. Figure~\ref{fig:leading-vs-spin-corr} shows the critical spin parameter $\chi_\mrm{cr}$ and the critical spin frequency $f_\mrm{cr}$ at which the relative ${\cal{O}}(\chi^{2})$ corrections become equal to 1\% the leading-order in $\chi$ result, as a function of stellar compactness. We here choose an APR EoS and used the slow-rotation results only. Observe that the ${\cal{O}}(\chi^{2})$ corrections to the moment of inertia and the quadrupole moment are important only for spin frequencies larger than 100--450Hz. Such corrections become important when $\chi \sim 0.1 - 0.2$, as expected. This figure also shows a rough estimate of $\chi_\mrm{cr}$ and $f_\mrm{cr}$ at which the relative ${\cal{O}}(\chi^{4})$ corrections to the stellar mass, moment of inertia and quadrupole moment become equal to 1\% the leading order result. Of course, we do not formally know the ${\cal{O}}(\chi^{4})$ corrections, since this would require an ${\cal{O}}(\chi^{6})$ analysis of NSs in the slow-rotation approximation. We have thus assumed that such ${\cal{O}}(\chi^{4})$ corrections are simply $\chi^{4}$ times the leading order result. Observe that the ${\cal{O}}(\chi^{2})$ corrected results are valid up to spin frequencies of 1000Hz for highly-relativistic stars.

\begin{figure}[htb]
\centering
\includegraphics[width=8.5cm,clip=true]{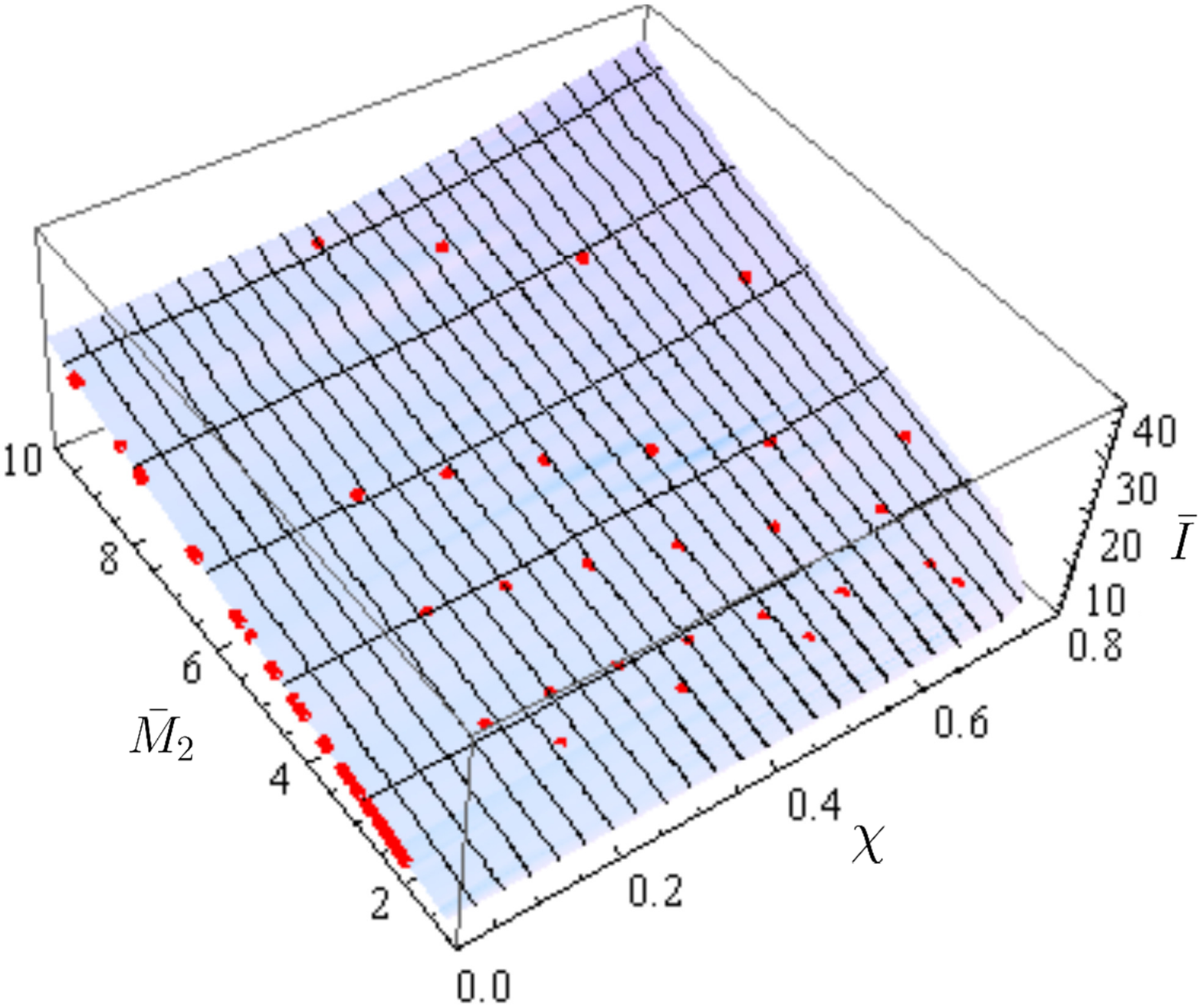}
\includegraphics[width=9.0cm,clip=true]{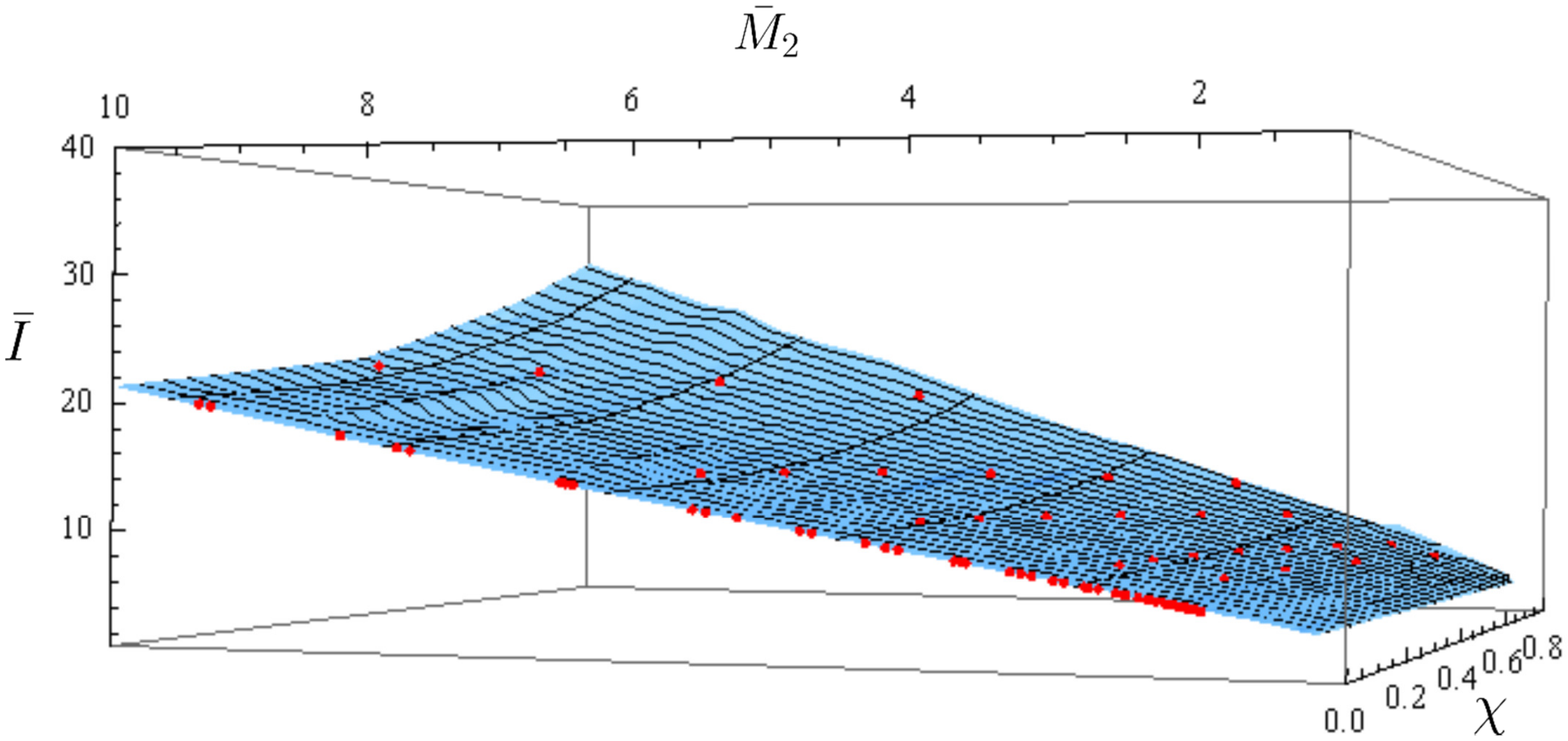}
\caption{
(Color online) (Top) $\bar{I}$--$\bar{M}_2$ relation as a function of $\chi$. The blue plane shows the relation for NSs, consistent with that found in~\cite{Pappas:2013naa}, while red points show the relation for QSs. Observe that the QS points lie on the NS plane, showing that the QS relation is almost identical to the NS one.  (Bottom) Same as the left panel but from a different viewpoint.
\label{fig:IQ-3D}
}
\end{figure}

\begin{figure}[htb]
\centering
\includegraphics[width=\columnwidth,clip=true]{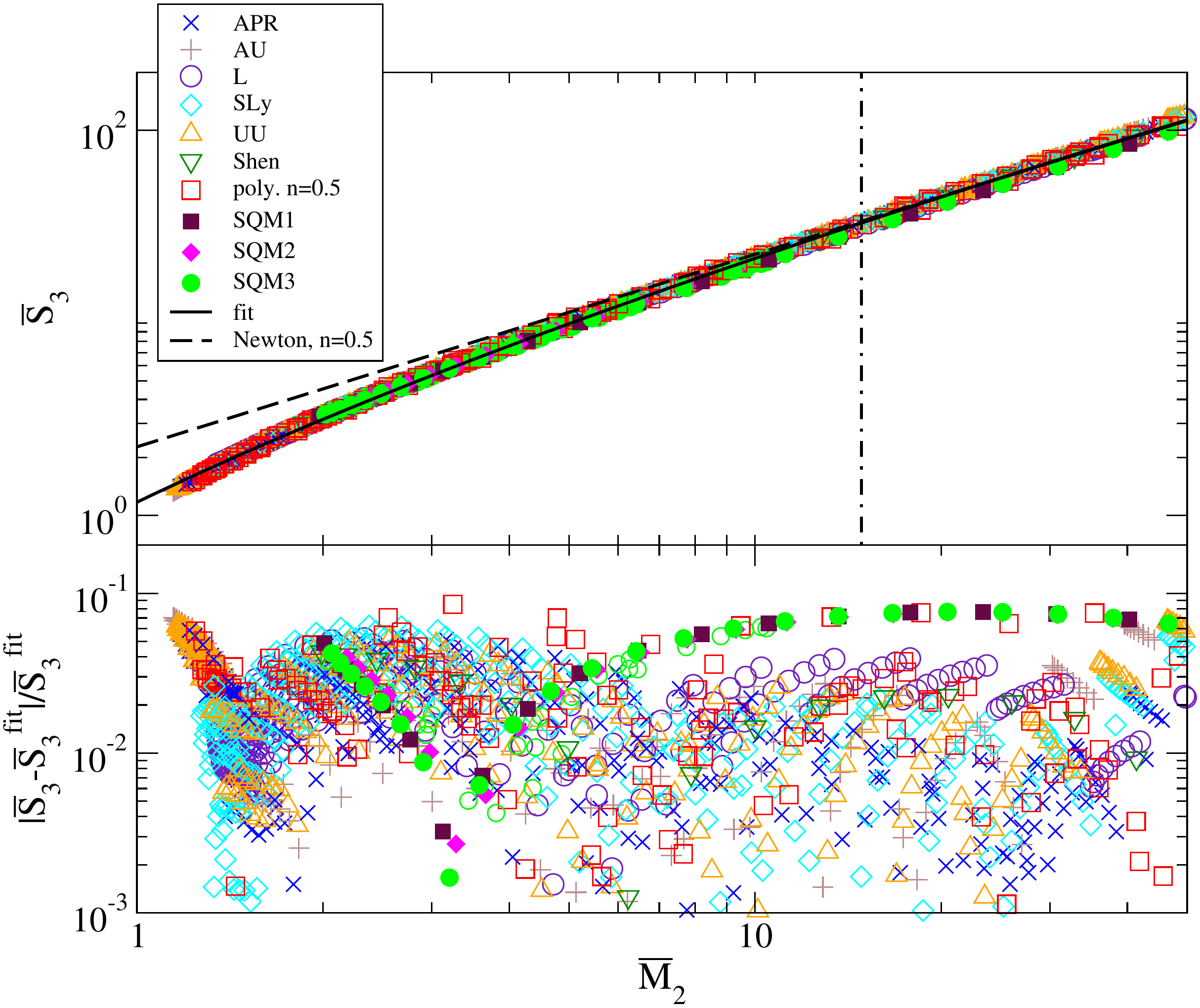}
\caption{
(Color online) (Top) $\bar{S}_3$--$\bar{M}_2$ relation with various realistic NS and QS EoSs and spins, together with the fit in Eq.~\eqref{eq:fit} and the Newtonian relation for the $n=0.5$ polytropes in~\cite{Stein:2013ofa}. The meaning of the dotted-dashed vertical line is the same as in Fig.~\ref{fig:M4-M2-intro}. Observe that the QS relation is almost the same as the NS one. (Bottom) Fractional difference between the data and the fit.
\label{fig:S3-M2}
}
\end{figure}

Let us now look at the spin dependence of the $\bar{I}$--$\bar{M}_2$ relation. The top panel of Fig.~\ref{fig:IQ-Omega} shows this relation computed both within the slow-rotation approximation (the black line corresponds to a leading-order-in-$\chi$ calculation, while the color lines correspond to a calculation valid to relative ${\cal{O}}(\chi^{2})$) and with the LORENE code (color circle, squares and diamonds), using an APR EoS and various spin frequencies. We have checked that the LORENE results are consistent with the fit of~\cite{doneva-rapid}. Observe that the ${\cal{O}}(\chi^{2})$ corrections to the $\bar{I}$--$\bar{M}_2$ relation in the slow-rotation approximation improve the agreement with the LORENE result. With these corrections, the slow-rotation results agree with the LORENE ones to better than $1\%$ for all stars with masses $M_{*} > 1 M_{\odot}$ and spin frequency smaller than $500$ Hz. One expects the slow-rotation approximation with $\chi^{2}$ corrections to disagree with the LORENE results at ${\cal{O}}(\chi^{4})$, which is borne out by the bottom panel of Fig.~\ref{fig:IQ-Omega}. This panel shows the fractional difference between the LORENE relation and the slow-rotation result to $\mathcal{O}(\chi^2)$, as a function of $\chi$. Again, this shows consistency between the slow-rotation and the LORENE calculations.

\subsection{$\bar{I}$--$\bar{M}_2$ and $\bar{S}_3$--$\bar{M}_2$ Relations}

Let us look at two previously-found approximately universal relations: the $\bar{I}$--$\bar{M}_2$ and $\bar{S}_3$--$\bar{M}_2$ relations. By approximate universality we here mean that these relations are approximately the same irrespective of the EoS used to calculate the moments. The $\bar{I}$--$\bar{M}_2$  relation was first found in~\cite{I-Love-Q-Science,I-Love-Q-PRD} in the slow-rotation and unmagnetized limit for both NSs and QSs. Reference~\cite{I-Love-Q-B} extended the calculation to magnetized NSs, while Refs.~\cite{doneva-rapid,Pappas:2013naa,Chakrabarti:2013tca} relaxed the slow-rotation approximation. The $\bar{S}_3$--$\bar{M}_2$ relation was first found by~\cite{Pappas:2013naa} for NSs.

An issue of whether universality (or EoS-independence to be precise) still holds for rapidly rotating stars arose in different investigations. Reference~\cite{doneva-rapid} constructed NS sequences by varying the dimensional spin frequency and the mass and found that universality then holds only for NSs with realistic modern EoSs that produce NSs of masses 1.4--2$M_\odot$ with radii 10.5--12.5km. From this finding,~\cite{doneva-rapid} concluded that the universal relation found in~\cite{I-Love-Q-Science,I-Love-Q-PRD} should not be used for rapidly rotating relativistic stars. On the other hand, Refs.~\cite{Pappas:2013naa,Chakrabarti:2013tca} constructed NS sequences by varying \textit{dimensionless} spin parameters, such as $\chi$, and found that not only is universality preserved, but the $\bar{S}_3$--$\bar{M}_2$ relation is also spin-insensitive. The calculations of~\cite{Pappas:2013naa,Chakrabarti:2013tca} then explained that whether universality holds for rapidly-rotating stars depends on the parameters chosen to describe the stellar sequence.

Recently, Ref.~\cite{Stein:2013ofa} further supported the results of~\cite{Pappas:2013naa,Chakrabarti:2013tca}. 
Reference~\cite{Stein:2013ofa} computed the $\bar{I}$--$\bar{M}_2$ and $\bar{S}_3$--$\bar{M}_2$ relations (as well as other relations) \emph{analytically}, albeit to leading-order in a weak-field (Newtonian) expansion with a certain isodensity approximation. Within their framework, EoS universality and spin-insensitivity was evident, when the stellar sequences were parameterized in terms of dimensionless spin quantities. However, whether this universality still holds in the relativistic regime remains an open question, which we address in this subsection.   

Let us first concentrate on the $\bar{I}$--$\bar{M}_2$ relation, and in particular, on the $\chi^2$ correction to the relation found in~\cite{I-Love-Q-Science,I-Love-Q-PRD}. The top panel of Fig.~\ref{fig:IQ-chi} presents the relation for slowly-rotating NSs and QSs valid to quadratic order in spin for various EoSs with $\chi=0.3$ and $\chi=0.5$. First, notice that the NS and QS sequences are almost identical. As in the NS case, when one fixes a dimensionless spin parameter, like $\chi$, then approximate EoS-universality is preserved also for QSs. Second, observe that the approximately universal $\bar{I}$--$\bar{M}_2$ relations are not spin-independent. The black solid line is the fit found in~\cite{I-Love-Q-Science,I-Love-Q-PRD} to the leading-order in $\chi$, slow-rotation result (ie.~without $\chi^{2}$ corrections). The black dashed and dotted-dashed curves are the fits found in~\cite{Pappas:2013naa} setting $\chi=0.3$ and $\chi=0.5$ respectively, but obtained by fitting to RNS data valid to all $\chi$. We have checked that the fit in~\cite{Pappas:2013naa} is consistent with the one in~\cite{Chakrabarti:2013tca}. The bottom panel shows the fractional difference between the numerical values and the $\chi$ fits. Observe that the slow-rotation relation to $\mathcal{O}(\chi^2)$ is valid to $\mathcal{O}(1\%)$ for $\chi<0.3$.

Let us now compare the NS and QS $\bar{I}$--$\bar{M}_2$ relation for different EoSs, but with results valid to all orders in spin (Fig.~\ref{fig:IQ-3D}). The blue plane shows the NS relation, which is consistent with that found in~\cite{Pappas:2013naa}. The red points show the QS relation at different points in $(\bar{I},\bar{M}_{2},\chi)$ space. Observe that the QS points lie on the NS plane. This proves that the QS relation is almost identical to the NS one.

{\renewcommand{\arraystretch}{1.2}
\begin{table*}
\begin{centering}
\begin{tabular}{cccccccc}
\hline
\hline
\noalign{\smallskip}
$y$ & $x$ &&  \multicolumn{1}{c}{$A_0$} &  \multicolumn{1}{c}{$B_1$}
&  \multicolumn{1}{c}{$B_2$} &  \multicolumn{1}{c}{$\nu_1$} &  \multicolumn{1}{c}{$\nu_2$}  \\
\hline
\noalign{\smallskip}
$(\bar{S}_3)^{1/3}$ & $\bar{M}_2$  && $-0.925$ & 1.98 & -- & 0.273 & --\\
$(\bar{M}_4)^{1/4}$ & $\bar{M}_2$  && $-0.413$ & 1.50 & -- & 0.466 & --\\
$\bar{M}_4/\bar{S}_3$ & $\bar{M}_2$  && $-0.335$ & $1.99 \times 10^{-3}$  & 1.47 & 2.15 & 0.891\\
$\delta e$  & $C$ && 0.61262 & $-7.2641 \times 10^{3}$  & $7.2185 \times 10^{3}$ & 3.7422 & 3.7354\\
\noalign{\smallskip}
\hline
\hline
\end{tabular}
\end{centering}
\caption{Estimated numerical coefficients for the fitting formula in Eq.~\eqref{eq:fit}. ``--'' indicates that inclusion of such parameter hardly affects the accuracy of the fit.}
\label{table:coeff}
\end{table*}
}

Let us now turn our attention to the $\bar{S}_3$--$\bar{M}_2$ relation. The top panel of Fig.~\ref{fig:S3-M2} shows this relation, not only for NSs but also for QSs, and various EoSs and spins. Observe that the QS relation is again almost identical to the NS one. Following~\cite{Pappas:2013naa}, we fit all these data to the polynomial
\be
\label{eq:fit}
y = A_0 + B_1 x^{\nu_1} + B_2 x^{\nu_2}\,,
\ee
with $y = (\bar{S}_3)^{1/3}$ and $x=\bar{M}_2$, with fitting parameters given in Table~\ref{table:coeff}. The new fit found here, which includes both NSs and QSs results, is very similar to the one found in~\cite{Pappas:2013naa} for NSs. In the bottom panel of Fig.~\ref{fig:S3-M2}, we present the fractional difference between the data and the fit. Observe that the relation is approximately universal, with variability of $\lesssim {\cal{O}}(10\%)$.

\subsection{$\bar{M}_4$--$\bar{M}_{2}$ and $\bar{M}_{4}/\bar{S}_{3}$--$\bar{M}_{2}$ Relations}

Let us now study whether higher multipoles satisfy approximately EoS independent relations for relativistic stars spinning at different frequencies. Reference~\cite{Stein:2013ofa} already found that there exists a universal $\bar{M}_4$--$\bar{M}_{2}$ relation to leading-order in a weak-field, Newtonian expansion, so let us investigate this relation first. The top panel of Fig.~\ref{fig:M4-M2-intro} shows the $\bar{M}_4$--$\bar{M}_2$ relation for various realistic NS and QS EoSs and various spins, computed with the LORENE and RNS codes, as well as in the slow-rotation approximation. The bottom panel shows the fractional difference between the data and the fit of Eq.~\eqref{eq:fit} with $y=(\bar{M}_4)^{1/4}$ and $x=\bar{M}_2$ and the coefficients given in Table~\ref{table:coeff}. Observe that the EoS-universality is slightly weaker than the $\bar{S}_3$--$\bar{M}_2$ relation, but still, it holds up to roughly 20\%. This larger variation is not an artifact of numerical error, since our calculations are valid to $\mathcal{O}(1\%)$. This indicates that the universality becomes worse as one considers multipole moment relations for higher $\ell$ modes, as first predicted in~\cite{Stein:2013ofa}. 

\begin{figure}[htb]
\centering
\includegraphics[width=\columnwidth,clip=true]{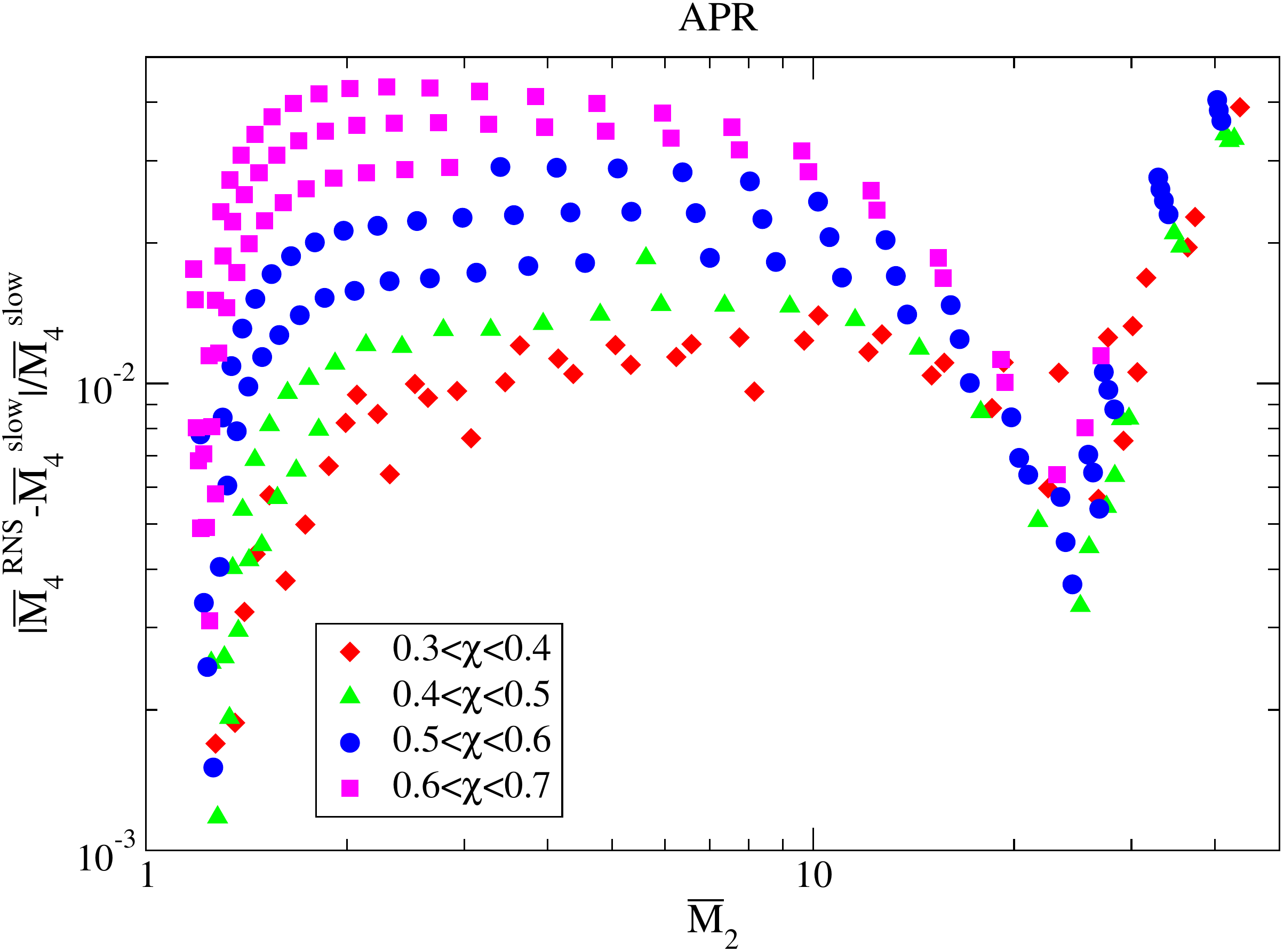}
\caption{
(Color online) Fractional difference between the $\bar{M}_4$--$\bar{M}_2$ relation for rapidly-rotating stars with RNS and the one in the slow-rotation limit for an APR EoS with various spin parameters.
\label{fig:M4-M2}
}
\end{figure}

The $20\%$ variability observed in Fig.~\ref{fig:M4-M2-intro} has two possible origins: EoS variability and spin-variability. In order to determine which of these dominates, Fig.~\ref{fig:M4-M2} attempts to assess the spin dependence of the $\bar{M}_{4}$--$\bar{M}_{2}$ relation. This figure shows the fractional difference between $\bar{M}_{4}$ computed with the RNS code and in the slow-rotation approximation, as a function of $\bar{M}_{2}$, clustered in groups of different $\chi$, using an APR EoS as a characteristic example. As expected, the difference becomes larger as one increases spins, reaching a maximum of 5\% accuracy for the largest $\chi$ models considered. Comparing this with the fractional difference in Fig.~\ref{fig:M4-M2-intro}, we conclude that the $20\%$ variability in the latter is dominated by EoS-variations and not spin effects. We recall that the multipole moments have a clear spin dependence if they are expressed in terms of the stellar compactness (see Fig.~\ref{fig:C-dep}). Our results indicate that such spin dependence seems to partially cancel if one expresses one multipole moment in terms of another. 

Not surprisingly, one can improve the accuracy of the fit, if one constructs NS and QS fits separately.
In particular, QSs have a relatively large variation from the fit, but this is because the amount of data for NSs that we used to make the fit is much larger than that for QSs, and hence, the fit was weighted more on the NS ones. Also, the fit does not capture quite well the behavior of the relation in the highly relativistic regime.
We have checked that by restricting ourselves to realistic EoSs (i.e. neglecting an $n=0.5$ polytropic EoS) and including $B_2$ and $\nu_2$ in the fit [Eq.~\eqref{eq:fit}], the variation in the relation becomes ~10\% for each NS and QS sequence.
 
\begin{figure}[htb]
\centering
\includegraphics[width=\columnwidth,clip=true]{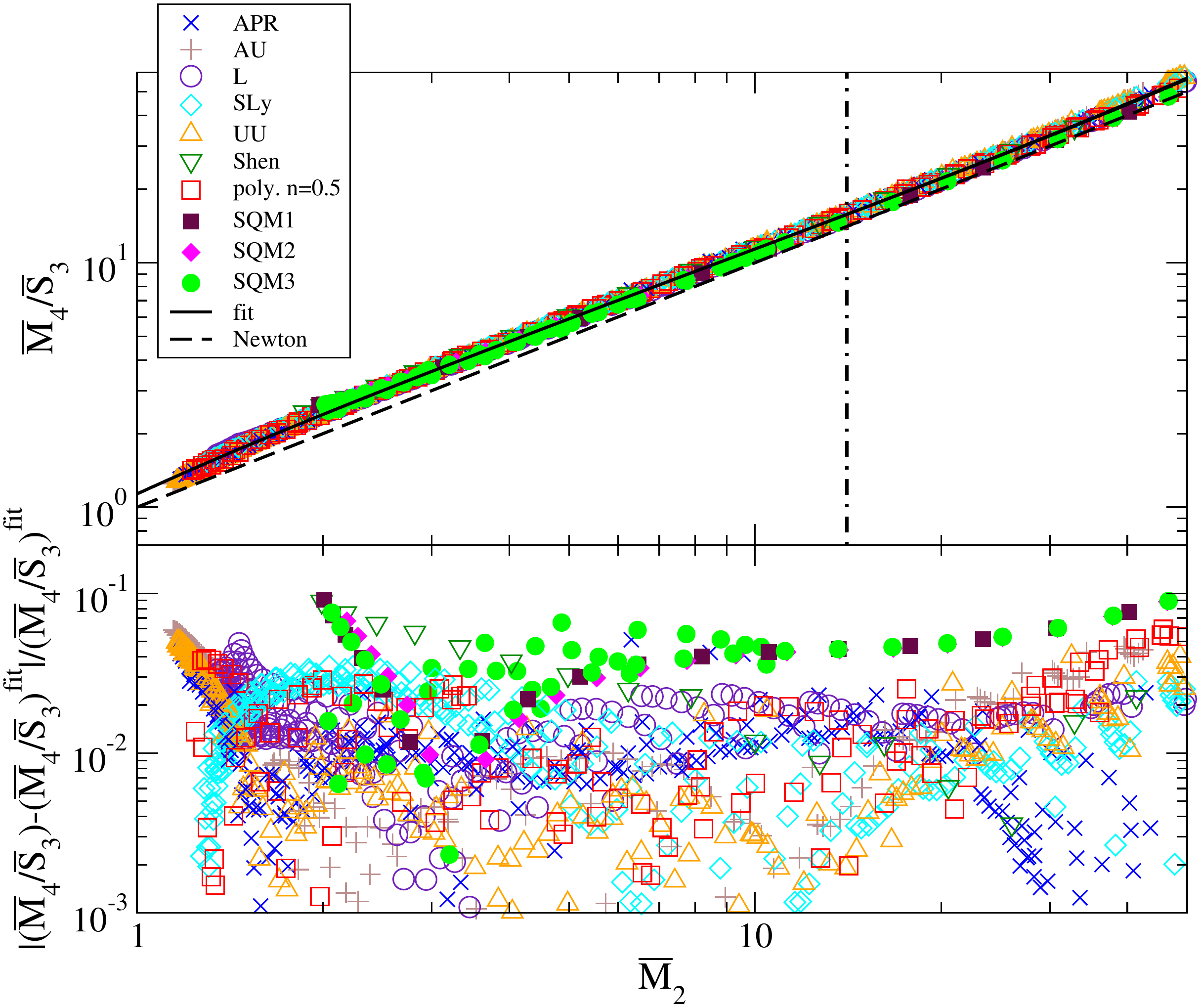}
\caption{
(Color online) (Top) $\bar{M}_4/\bar{S}_3$--$\bar{M}_2$ relation with various EoSs and spins, together with the fit in Eq.~\eqref{eq:fit} and the EoS-independent Newtonian relation. The meaning of the dotted-dashed line is the same as in Fig.~\ref{fig:M4-M2-intro}. (Bottom) Fractional difference from the fit. 
\label{fig:M4S3-M2}
}
\end{figure}

Reference~\cite{Stein:2013ofa} found that another relation between multipole moments is also EoS-universal in the Newtonian limit, namely $\bar{M}_4/\bar{S}_3$--$\bar{M}_2$. The top panel of Fig.~\ref{fig:M4S3-M2} shows this relation for various NS and QS EoSs and spins, computed with the LORENE and RNS codes, as well as in the slow-rotation approximation. This figure also shows a fit to all this data, given by Eq.~\eqref{eq:fit} with $y = \bar{M}_4/\bar{S}_3$ and the coefficients of Table~\ref{table:coeff}. The bottom panel shows the fractional difference between the numerical data and the fit. Observe that the EoS-universality for the $\bar{M}_4/\bar{S}_3$--$\bar{M}_2$ relation is stronger than that of the $\bar{M}_4$--$\bar{M}_2$ relation.

\subsection{Spectrum of Stellar Moments}
\label{sec:spectrum}

\begin{figure*}[htb]
\centering
\includegraphics[width=7.5cm,clip=true]{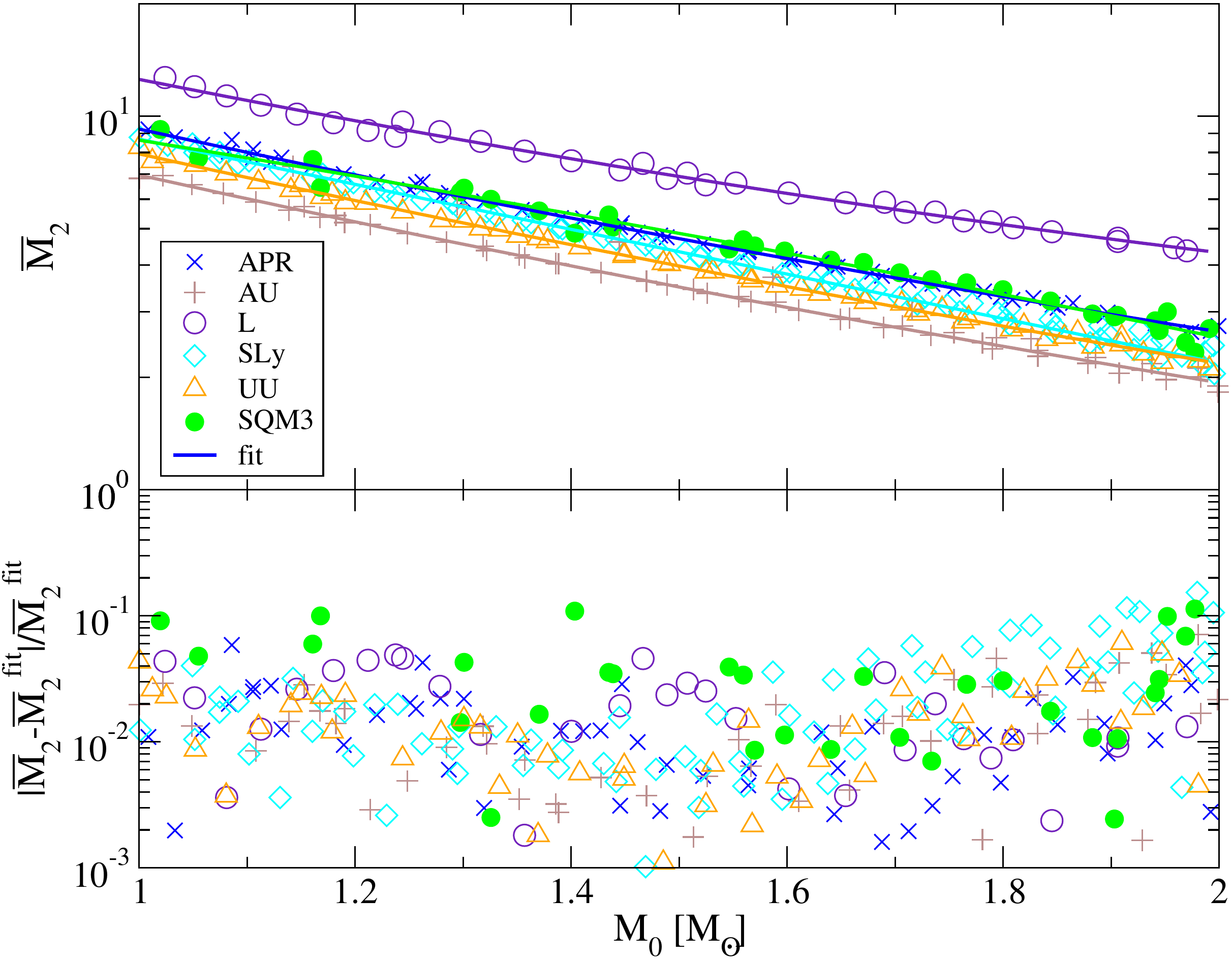}
\includegraphics[width=7.5cm,clip=true]{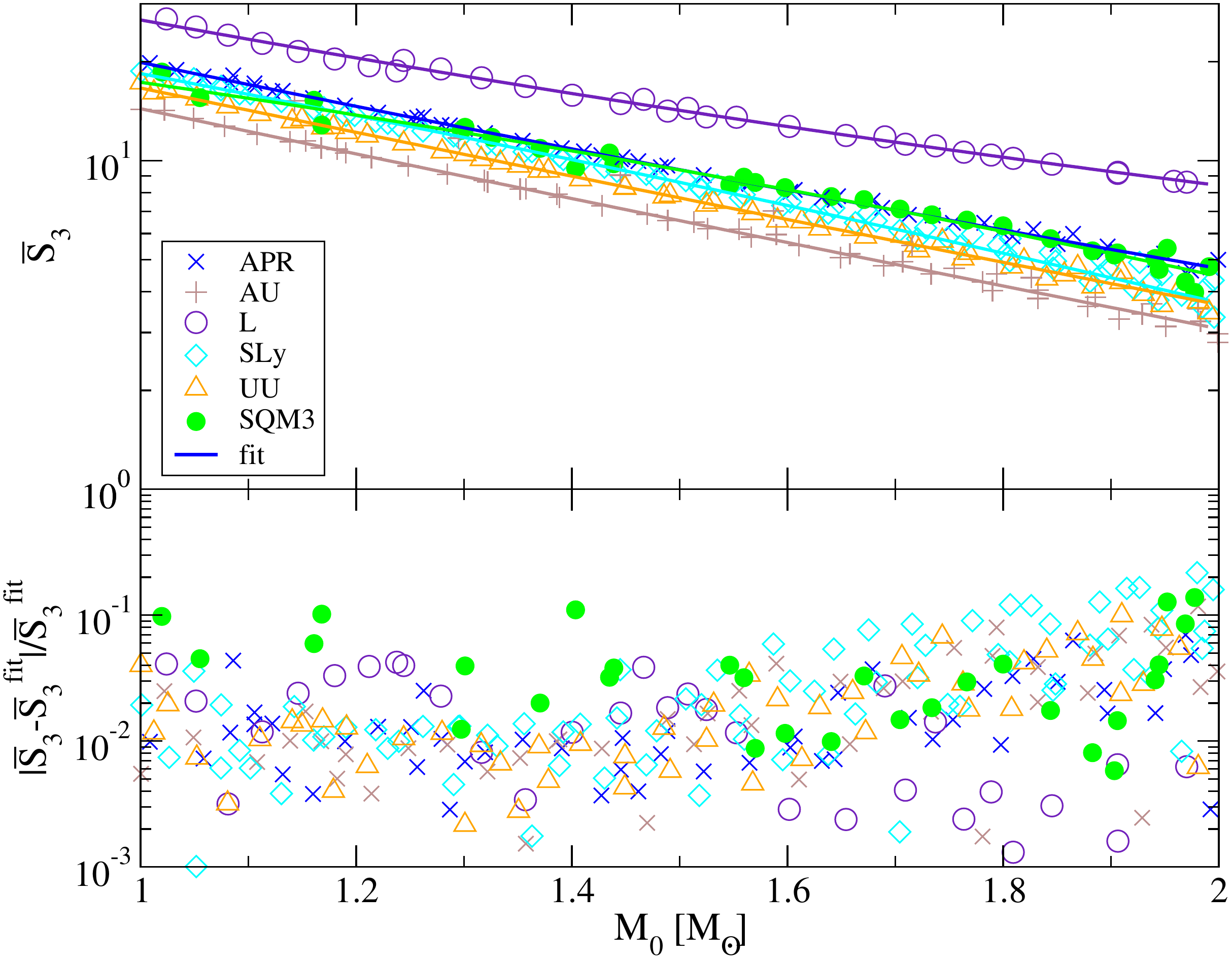}
\includegraphics[width=7.5cm,clip=true]{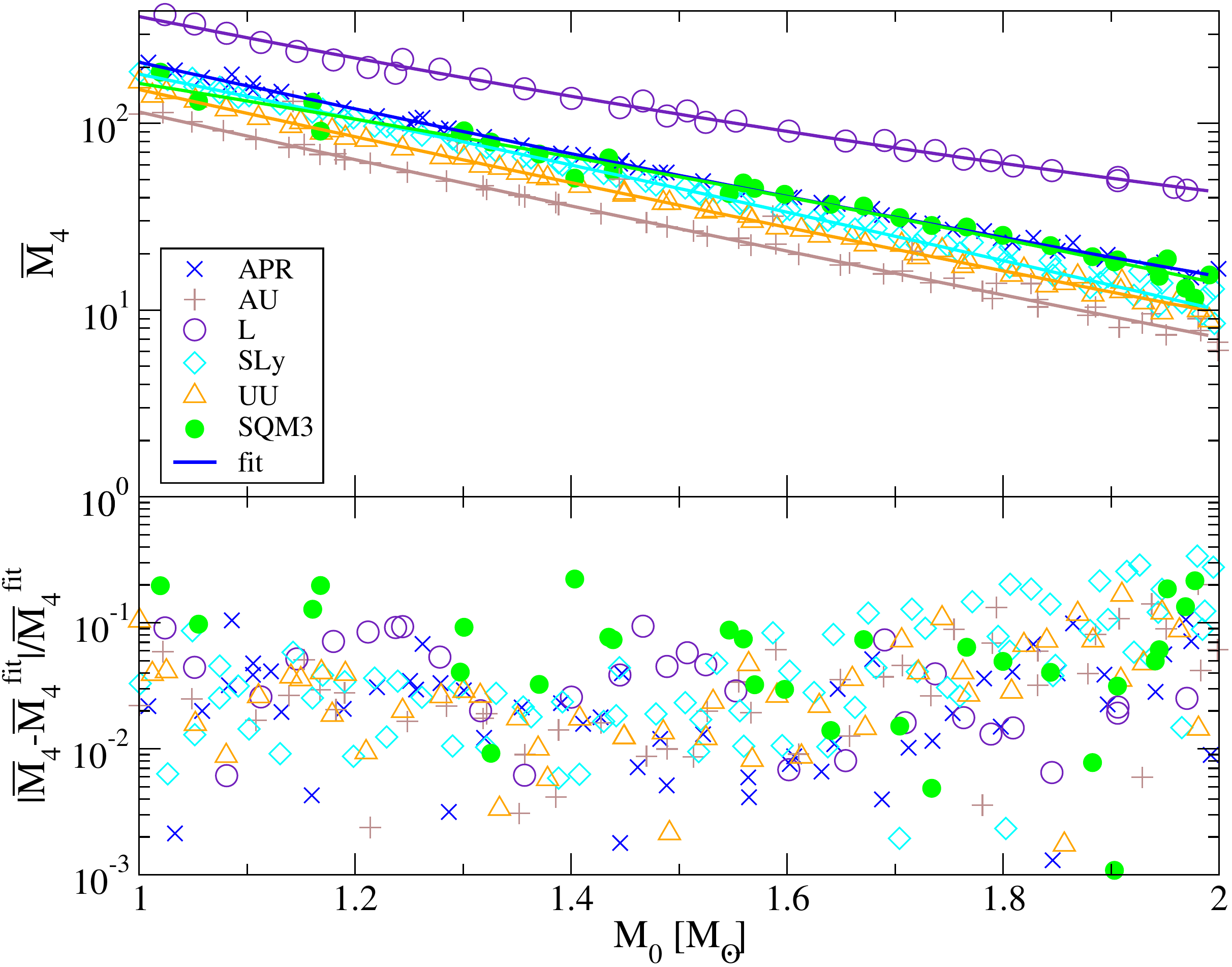}
\caption{
(Color online) (Top Left) $\bar{M}_2$ against $M_0$ for various EoSs and spins, together with the fractional difference from the fit created for each EoS. Observe that the relation for a given EoS depends very weakly on the spin. (Top Right) Same as the top panel, but for $\bar{S}_3$ against $M_0$. (Bottom) Same as the top panel, but for $\bar{M}_4$ against $M_0$.
\label{fig:M-dep}
}
\end{figure*}

{\renewcommand{\arraystretch}{1.2}
\begin{table*}
\begin{centering}
\begin{tabular}{c||ccc|ccc|ccc|ccc|ccc|ccc|ccc}
\hline
\hline
\noalign{\smallskip}
   &   & \text{APR} &   &   & \text{SLy} &   &   & \text{AU} &   &  & \text{L} &   &   & \text{UU} &  &   & \text{SQM3} &  \\ \hline
 $y_i$  & $a_i$ & $b_i$  & $c_i$  & $a_i$ & $b_i$  & $c_i$ & $a_i$ & $b_i$  & $c_i$ & $a_i$ & $b_i$  & $c_i$ & $a_i$ & $b_i$  & $c_i$ & $a_i$ & $b_i$  & $c_i$  \\ \hline
 $\bar{M}_2$  & 3.88 & -1.86  & 0.202  & 3.58  & -1.43  & 0.0174  & 3.62  & -1.88  & 0.201  & 4.12  & -1.84  & 0.259  & 3.72  & -1.83  & 0.181  & 3.10 & -0.810 & -0.136 \\
  $\bar{S}_3$ & 4.67  & -1.80  & 0.118  & 4.19  & -1.12  & -0.163  & 4.31  & -1.70  & 0.0530  & 4.88  & -1.80  & 0.215  & 4.41  & -1.65  & 0.0425  & 3.69 & -0.574 & -0.262 \\
  $\bar{M}_4$ & 8.59  & -3.53  & 0.295  & 7.76  & -2.37  & -0.180  & 7.91  & -3.36  & 0.192  & 9.05  & -3.62  & 0.483  & 8.16  & -3.33  & 0.191  & 6.91 & -1.48 & -0.328 \\
\noalign{\smallskip}
\hline
\hline
\end{tabular}
\end{centering}
\caption{Coefficients of the fit for the multipole moments as a function of $M_0$ given by Eq.~\eqref{eq:fit-M-dep}.}
\label{table-coeff-M-dep}
\end{table*}
}


In this subsection, we study the dependence of the NS and QS multipole moments on $M_0$ and $\chi$, and compare this dependence to that satisfied by a generic, stationary, and axisymmetric spacetime, including the Kerr BH limit. Reference~\cite{pappas-apostolatos} showed that for NSs, both $M_2$ and $S_3$ approximately scale with the spin parameter $\chi$ in the same way as the moments of a Kerr BH do, i.e.~the moments scale with the appropriate power of the spin parameter times a coefficient (larger than unity for NSs):
\ba
\label{mom1} 
M_2 &=& - \bar{M}_2(\mathrm{EoS},M_0,\chi) \chi^2 M_0^3 \nn \\ 
& \approx & - \bar{M}_2(\mathrm{EoS},M_0) \chi^2 M_0^3 \,, \\
\label{mom2}
S_3 &=& - \bar{S}_3 (\mathrm{EoS},M_0,\chi) \chi^3 M_0^4 \nn \\
 &\approx & - \bar{S}_3 (\mathrm{EoS},M_0) \chi^3 M_0^4 \,,
\ea
where in the second line of the above equations we have expanded in $\chi \ll 1$ and retained only the leading-order term. Such behavior is explicitly shown in Fig.~\ref{fig:M-dep}, where we show the mass dependence of $\bar{M}_2$ and $\bar{S}_3$ for various EoSs and spins. Observe that the relations depend only weakly on the spins for a given EoS sequence, but they are not universal with EoS. We here find that this behavior extends to $M_4$ as well (see also Eq.~\eqref{eq:M4}): 
\ba 
\label{mom3}
M_4 &=& \bar{M_4} (\mathrm{EoS},M_0, \chi) \chi^4 M_0^5 \nn \\
&\approx& \bar{M_4} (\mathrm{EoS},M_0) \chi^4 M_0^5\,,
\ea
where again in the second line we have expanded in $\chi \ll 1$ and retained only the leading-order term. This relation is also shown graphically in Fig.~\ref{fig:M-dep} as a function of $M_{0}$ for a variety of EoSs and spins.

These data can be fitted for any given EoS sequence with a function of the form
\be
\label{eq:fit-M-dep}
\ln y_i = a_i + b_i \left( \frac{M_0}{M_\odot} \right) + c_i \left( \frac{M_0}{M_\odot} \right)^2\,,
\ee
with $y_i = (\bar{M}_2,\bar{S}_3,\bar{M}_4)$ and the coefficients are given in Table~\ref{table-coeff-M-dep}. We show the fractional difference between the numerical data and the fit on the bottom panels of Fig.~\ref{fig:M-dep}. Observe that the fractional difference tends to increase somewhat as the mass is increased. 

The maximum spins considered in these figures is $\chi \sim 0.73$ for NSs and $\sim 0.9$ for QSs, which are realized for low-mass stars. If the spin corrections enter at $\mathcal{O}(\chi^2)$, one would expect the maximum fractional difference to be $\sim 50\%$ for NSs and $\sim 80\%$ for QSs. However, Fig.~\ref{fig:M-dep} shows that the fractional difference is always $\lesssim 10\%$ for NSs and $\lesssim 20\%$ for QSs for $M<1.6M_\odot$. Comparing this figure to Fig.~\ref{fig:C-dep}, where we showed the dependence of each multipole moment on the stellar compactness $C$, one can clearly see the spin-dependence is weaker in the former. Such spin-independence is realized not only for NSs but also for QSs.

The above findings lead us to the conclusion that the reduced multipole moments of NSs and QSs behave similarly to those of BHs. This behavior is intriguing and a bit puzzling because of how the moments are constructed in GR. Reference~\cite{fodor:2252} showed that for a stationary, asymptotically flat and axisymmetric spacetime, the multipole moments have the form
\ba   M_0&=&m_0,\nn \\
        S_1&=&m_1,\nn \\
        M_2&=&m_2,\nn \\
        S_3&=&m_3,\nn \\
        M_4&=&m_4 - \frac{1}{7} m_0^*\widetilde{M}_{20}, \\
        S_5&=&m_5 - \frac{1}{3}m_0^*\widetilde{M}_{30}-\frac{1}{21}m_1^*\widetilde{M}_{20},
\ea
where $*$ stands for complex conjugation, $\widetilde{M}_{ij}=m_im_j-m_{i-1}m_{j+1}$, and the $m_\ell$ quantities are complex coefficients that are related to the asymptotic expansion of the Ernst potential associated with the specific spacetime. One can see that as the $\ell$ number of the multipole moment increases, an extra contribution is added from the lower order moments in the form of the $\tilde{M}_{ij}$ correction terms. In the case of the Kerr spacetime, these complex coefficients become $m_\ell = M_0 (ia)^\ell$, where $a\equiv S_1/M_0$ is the Kerr angular momentum parameter. The fact that these coefficients have this particular form for the Kerr spacetime leads to a special form for the spectrum of the Kerr BH moments:
\be
M_{2\ell}= M_0 (ia)^{2\ell}, \quad i S_{2\ell+1}=M_0 (ia)^{2\ell+1}\,,
\ee
This special form for the spectrum of the multipole moments, another manifestation of the no-hair theorem (i.e.~the fact that the exterior spacetime of stationary and axisymmetric BHs can be completely described by the mass and angular momentum of the spacetime), is a result of all the correction terms $M_{ij}$ vanishing due to the form of $m_\ell$ for Kerr BHs. NSs and QSs seem to exhibit such behavior, except that the non-vanishing contribution of $M_{ij}$ seems to be almost perfectly canceled by the lower-order-in-spin contribution of $m_\ell$, rather than by $M_{ij}$ being zero. 

This result, apart from its connection to the universal behavior of the moments, is also relevant to attempts to construct analytic, stationary and axisymmetric spacetimes that represent the exterior of NSs and QSs based on the Ernst formalism. This is because it provides specific constraints on the form of the Ernst potential for such spacetimes.   

\subsection{Comparison with the Newtonian Limit}
\label{sec:Newton}

Reference~\cite{Stein:2013ofa} proved analytically that certain relations, like the $\bar{M}_{4}$--$\bar{M}_{2}$ relation and the $\bar{M}_{4}/\bar{S}_{3}$--$\bar{M}_{2}$ relation, are EoS-universal and spin-insensitive in the ``Newtonian limit.'' By the latter, we here mean that Ref.~\cite{Stein:2013ofa} worked to leading (Newtonian) order in a weak-field expansion of GR. This reference also employed a certain isodensity approximation~\cite{Lai:1993ve}, which assumes isodensity surfaces can be modeled as self-similar ellipsoids of a given eccentricity. In this subsection, we compare our fully relativistic results without the isodensity approximation to the Newtonian results of~\cite{Stein:2013ofa}.  

Before carrying out such comparisons, it is instructive to derive a rough estimate of how the multipole moments scale in the slow-rotation limit (see also~\cite{Pappas:2013naa}).  For a rotating star with angular velocity (or angular spin frequency) $\Omega$, dimensional arguments suggest that the
(physical) $\ell$-th multipole moments scales as
\begin{align}
 M_{2\ell} & \propto M_0 R_\mrm{eq}^{2\ell} \left( \frac{\Omega}{\Omega_\mathrm{K}}
 \right)^{2\ell} , \\
 S_{2\ell + 1} & \propto M_0 R_\mrm{eq}^2 \Omega \times R_\mrm{eq}^{2\ell} \left(
 \frac{\Omega}{\Omega_\mathrm{K}} \right)^{2\ell}\,,
\end{align}
to leading order in $\Omega$ and up to a constant of order unity
dependent on the EoS and angular velocity, where $\Omega_\mathrm{K} \equiv \sqrt{M_0/R_\mrm{eq}^3}$ is the Kepler or mass-shedding angular velocity and where we recall that $R_{\mrm{eq}}$ is the equator radius. The angular momentum scales as $S_1 \propto M_0 R_\mrm{eq}^2
\Omega$, and thus, one expects
\begin{equation}
\label{eq:C-scaling}
 \bar{M}_{2\ell}, \ \bar{S}_{2\ell + 1} \propto C^{-\ell} \propto \bar{M}_2^{\ell}\,.
\ee
As stated above, the proportionality constant depends on the angular velocity and the EoS via the density profile and susceptibility to rotational deformations. EoS-universality and spin-insensitivity exist if this constant is approximately independent of the EoS and the angular velocity. 

Figure~\ref{fig:C-dep} showed that the dimensionless moment of inertia $\bar{I}$ scales with a certain power of compactness and is almost completely spin-insensitive; let us try to analytically understand this better. To do so, let us follow~\cite{Stein:2013ofa}, consider polytropic EoSs and use the elliptical isodensity approximation~\cite{Lai:1993ve}\footnote{Reference~\cite{Stein:2013ofa} defined the compactness as the stellar mass over the geometric mean radius instead of the mass over the equatorial radius.} to find the moment of inertia to Newtonian order:
\be
\bar{I} = \frac{2}{3} \frac{\mathcal{R}_{n,2}}{|\vartheta'| \xi_1^5} \frac{1}{C^2}\,,
\ee
where $n$ is the polytropic index, $\vartheta (\xi)$ is the Lane-Emden function, $\xi_1$ corresponds to the stellar surface and ${\cal{R}}_{n,\ell}$ is an integral of $\vartheta^{n} \xi^{\ell + 2}$~\cite{Stein:2013ofa}. This relation is completely spin-independent and it scales as $C^{-2}$, as found numerically in Fig.~\ref{fig:C-dep}. 

Figure~\ref{fig:C-dep} also showed that the dimensionless multipoles $\bar{M}_{2}$, $\bar{S}_{3}$ and $\bar{M}_{4}$ decrease with increasing $\chi$ for fixed stellar compactness; let us try to understand this better. Again following~\cite{Stein:2013ofa}, the leading-order weak-field expansion of $\bar{M}_{2\ell + 2}$ and $\bar{S}_{2\ell + 1}$ can be written in terms of $C$ and $\chi$ as
\ba
\bar{M}_{2\ell + 2}(C,\chi) &=& \bar{M}_{2\ell + 2}^{(0)} (C) \left[1 - (\ell +1) A_n(C) \chi^2 + \mathcal{O}(\chi^4)\right]\,, \nn \\
\\
\bar{S}_{2\ell + 1}(C,\chi) &=& \bar{S}_{2\ell + 1}^{(0)} (C) \left[1 - \ell A_n (C) \chi^2 + \mathcal{O}(\chi^4) \right]\,, 
\ea
where $\bar{M}_{2\ell + 2}^{(0)}$ and $\bar{S}_{2\ell + 1}^{(0)}$ denote $\bar{M}_{2\ell + 2}$ and $\bar{S}_{2\ell + 1}$ in the non-spinning limit and $A_{n}>0$ (for $n < 5$) is a constant that depends on $\vartheta'(\xi_{1})$, ${\cal{R}}_{n,2}$ and is linear in $C$. From this expression, we see that the spin corrections to $\bar{M}_{2}$, $\bar{S}_{3}$ and $\bar{M}_{4}$ are indeed of quadratic order and they generically reduce the dimensionless multipole moments, as we found numerically in Fig.~\ref{fig:C-dep}.

The dashed lines in Figs.~\ref{fig:M4-M2-intro},~\ref{fig:S3-M2} and~\ref{fig:M4S3-M2} show the Newtonian relation for an $n=0.5$ polytrope, found in~\cite{Stein:2013ofa}. Observe that the relativistic results approach nicely the Newtonian ones as one decreases the stellar compactness (or equivalently, increases $\bar{M}_2$). The slight difference between the fit and the Newtonian relation in the Newtonian regime is because the fit is constructed using various realistic EoSs and an $n=0.5$ polytropic EoS, while the Newtonian relations shown in Figs.~\ref{fig:M4-M2-intro} and~\ref{fig:S3-M2} are only valid for an $n=0.5$ polytrope. If one compares the latter with a fully relativistic calculation with the same $n=0.5$ polytropic EoS, the agreement is much better.

\begin{figure}[htb]
\centering
\includegraphics[width=\columnwidth,clip=true]{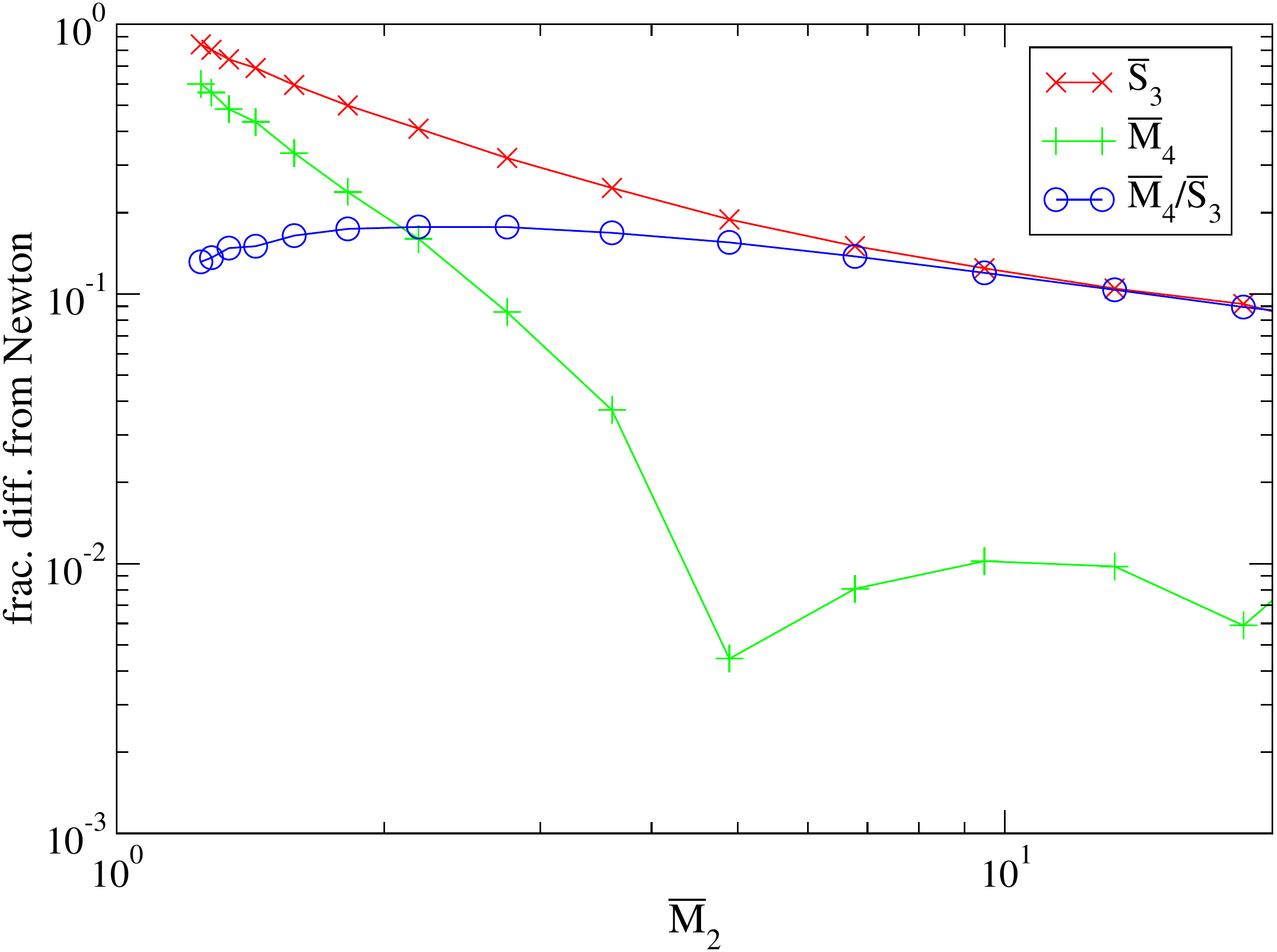}
\caption{
(Color online) Fractional difference between the numerical values and the Newtonian ones for the $\bar{S}_3$--$\bar{M}_2$, $\bar{M}_4$--$\bar{M}_2$ and $\bar{M}_4/\bar{S}_3$--$\bar{M}_2$ relations with the $n=0.5$ polytropes. 
\label{fig:diff-n05-Newton}
}
\end{figure}

Figure~\ref{fig:diff-n05-Newton} shows the fractional difference between the relativistic and Newtonian results for an $n=0.5$ polytrope. Observe that the $\bar{M}_4$--$\bar{M}_2$ relation approaches the Newtonian relation faster than the $\bar{S}_3$--$\bar{M}_2$ relation. This suggests that the relativistic relations may be closer to the Newtonian ones as one considers higher multipole moments. Observe also that the fractional difference for the $\bar{M}_4/\bar{S}_3$--$\bar{M}_2$ relation is always less than 20\% and is dominated by the difference for the $\bar{S}_3$--$\bar{M}_2$ relation in the Newtonian limit.

\section{Implications for X-ray Observations}
\label{sec:X}

Recently, it was suggested~\cite{Baubock:2013gna,Psaltis:2013zja} that the universal relations among multipole moments may help in determining the NS (or QS) mass and radius independently with X-ray pulse and atomic line profiles from millisecond pulsars~\cite{Morsink:2007tv,Baubock:2012bj,Psaltis:2013zja,Baubock:2013gna,Psaltis:2013fha}. Such independent measurements are important because they would constrain the EoS in the nuclear and supranuclear density regime. At first, this may seem counter-intuitive; how can relations that are EoS-universal be used to measure the NS radius, and thus, constrain the EoS? 

The answer is that the universal relations break certain degeneracies between $(M_{0},R_\mrm{eq},I,M_{2})$ in millisecond pulsar observations to quadratic order in spin. Let us explain how this degeneracy-breaking occurs in more detail. The X-ray profiles depend not only on $M_{0}$ and $R_\mrm{eq}$, but also on the stellar eccentricity $e$, moment of inertia $I$, the quadrupole moment $M_{2}$, the spin frequency $f$ and the observer inclination angle $\theta_0$. Since the eccentricity can in turn be written entirely in terms of the mass, the equatorial radius, the moment of inertia, the spin frequency and the quadrupole moment, the profile depends on the parameter vector $(M_{0},R_\mrm{eq},I,M_{2}, f, \theta_0)$. A given observation cannot be used to measure all 5 parameters independently, but by using the universal relations, one can eliminate $M_{2}$ in favor of $I$. Moreover, for stars with compactness in $(0.1,0.4)$, there exists an approximately universal relation between $I$ and $C$, which then means the profiles depend only on $(M_{0},R_\mrm{eq},f,\theta_0)$. One can then fit for these four parameters independently. 

Such a procedure was first proposed in~\cite{Baubock:2013gna} for X-ray atomic line observations and used in~\cite{Psaltis:2013fha} for X-ray pulse observations. Let us first discuss the former. Using the observations reported in~\cite{2002Natur.420...51C}, Ref.~\cite{2006Natur.441.1115O} placed a constraint on the NS EoS. However, this constraint did not take the effect of spin into account. Reference~\cite{Baubock:2012bj} used a metric valid to $\mathcal{O}(\chi^2)$ and showed that the stellar quadrupole moment can affect the qualitative features of atomic line profiles significantly, while the effect of stellar eccentricity is less significant. Although it is difficult to quantify such effects, the hexadecapole moment and the spin corrections to the quadrupole moment might not be negligible. The effect of the octupole moment and the spin corrections to the moment of inertia through frame-dragging are likely to be negligible, since Ref.~\cite{Morsink:2007tv} showed that the effect of frame-dragging to linear order in spin on X-ray line profiles is small.

Let us now discuss X-ray pulse observations. Reference~\cite{Psaltis:2013zja} used a slow-rotation approximation valid to ${\cal{O}}(\chi^{2})$~\cite{Hartle:1968ht} and found that the stellar eccentricity $e$ and the quadrupole moment $M_2$ modify the pulse profile by 10--30\% and 1--5\% respectively for NSs with a mass of $1.8M_\odot$ and a spin frequency of 600Hz. Since the goal of future X-ray missions, like NICER~\cite{2012SPIE.8443E..13G} and LOFT~\cite{2012AAS...21924906R,2012SPIE.8443E..2DF}, is to measure the mass and radius to 5\% accuracy, both of these quantities had to be included in the analysis. The quartic-order in spin expansion developed here allows us to estimate the systematic errors in~\cite{Psaltis:2013zja} due to the assumption that the pulsar is a NS and their ${\cal{O}}(\chi^{2})$ truncation. 

Before doing so, let us first calculate the $e$--$C$ relation for QSs and compare them to those for NSs, which were found to be EoS-universal to $\mathcal{O}(1\%)$~\cite{Baubock:2013gna}. Following~\cite{Hartle:1968ht}, we define the stellar eccentricity as
\be
\label{eq:ecc}
e \equiv \sqrt{\frac{R_\mrm{eq}^2}{R_\mrm{pol}^2}-1}\,,
\ee
where $R_\mrm{eq}$ and $R_\mrm{pol}$ are the gauge-invariant circumferential equatorial and polar radius respectively. One obtains a gauge-invariant parametrization of the stellar surface by embedding it in flat spacetime~\cite{Hartle:1968ht}, which amounts to finding a circumferential radius $R_\mrm{circ}$ in terms of the original coordinates $(R,\theta)$. Extending~\cite{Hartle:1968ht}, one obtains such circumferential radius at the stellar surface valid to quartic order in spin via
\ba
R_\mrm{circ}^* (\theta) &=& R_* + [\xi_{20}^* + (R_* k_{22}^*+\xi_{22}^*) P_2 (\cos \theta)] \epsilon^2 \nn \\
& & + \frac{1}{2} \left[2 \xi_{40}^* + 2 ( R_* k_{42}^* + k_{22}^* \xi_{20}^* +  \xi_{42}^*) P_2 (\cos \theta)  \right. \nn \\
& & \left. + ( 2 k_{22}^* \xi_{22}^*-R_* k_{22}^*{}^2) P_2 (\cos \theta)^2 \right. \nn \\
& & \left. + 2(R_* k_{44}^* + \xi_{44}^*) P_4 (\cos \theta)  \right] \varepsilon^4 + \mathcal{O}(\epsilon^6)\,,
\ea
where $A^*$ means $A(R_{*})$ for any $A$. Next, we decompose a quantity $A$ as 
\ba
A = A^{(0)} [1 + \delta A \; \chi^2 + \mathcal{O}(\chi^4)]\,.
\ea
Inserting this expansion with $A=e$ into Eq.~\eqref{eq:ecc}, one finds
\ba
e^{(0)} &=& \sqrt{-\frac{3(R_* k_{22}^* + \xi_{22}^*)}{R_*}}\,, \\
\delta e &=& \frac{1}{24 R_* (R_* k_{22}^* + \xi_{22}^*)} [(12 k_{42}^*+5 k_{44}^* - 24 k_{22}^*{}^2) R_*^2 \nn \\
& & +(5 \xi_{44}^* - 36 k_{22}^* \xi_{22}^*+ 12  \xi_{42}^*) R_*-12 \xi_{20}^* \xi_{22}^* - 21 \xi_{22}^*{}^2]\,, \nn \\
\ea
which is the $e$--$C$ relation. 

\begin{figure}[htb]
\centering
\includegraphics[width=\columnwidth,clip=true]{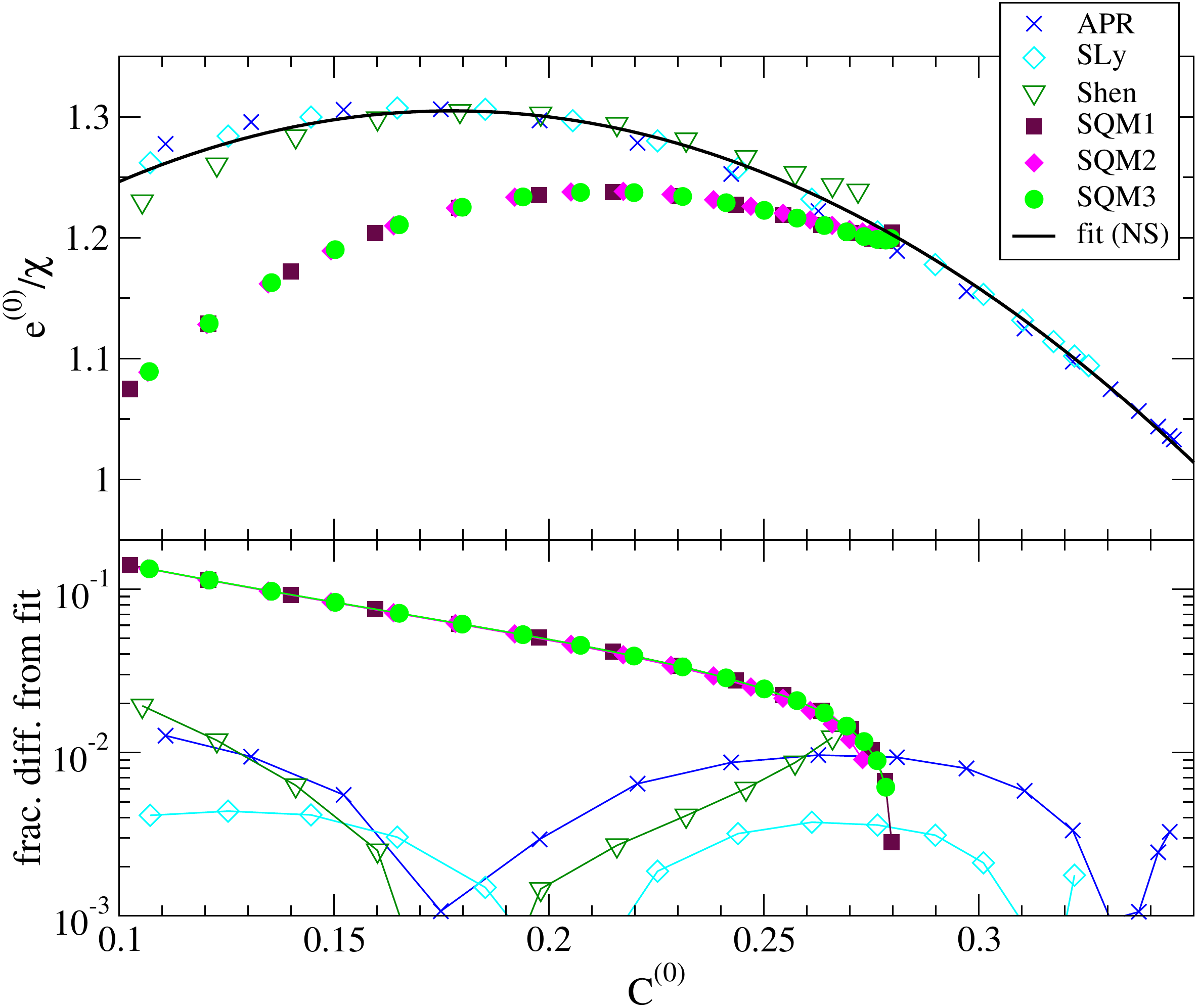}
\caption{
(Color online) (Top) Stellar eccentricity-compactness relation in the slow-rotation limit for various realistic EoSs and the fit for the NS relation given by Eq.~\eqref{eq:fit2}. (Bottom) Fractional difference between the numerical values and the NS fit. Observe that the QS relation is different from the NS one by $\mathcal{O}(10\%)$.
\label{fig:ecc-C}
}
\end{figure}

The top panel of Fig.~\ref{fig:ecc-C} shows the relation for both NSs and QSs in the slow-rotation limit. The NS data can be fitted to
\be
\label{eq:fit2}
\frac{e^{(0)}}{\chi} = A_0 + A_1 C^{(0)} + A_2 \left( C^{(0)} \right)^2\,,
\ee
with coefficients $A_0=0.997$, $A_1=3.47$ and $A_2=-9.97$. This fit is also shown in the top panel of Fig.~\ref{fig:ecc-C}. The bottom panel shows the fractional difference between the data and the fit. For NSs, this difference is $\mathcal{O}(1\%)$, in agreement with~\cite{Baubock:2013gna}. However, the fractional difference of QSs is $\mathcal{O}(10\%)$. Such a large variation originates from a similar loss of universality in the $\bar{I}$--$C$ relation for QSs, which is presented in Fig.~6 of~\cite{I-Love-Q-PRD}. The QS relation is close to that for $n=0$ polytropes, but this is different than the NS relation by $\mathcal{O}(10\%)$. Therefore, a 5\% accuracy requirement for the extraction of the mass and radius, requires that one assumes the pulsar is a NS. Otherwise, the non-universality of the $e$--$C$ or $\bar{I}$--$C$ relation spoils the degeneracy-breaking, leading to a systematic error in the $M_{*}$ and $R_{*}$ measurement that dominates the statistical error.

Let us now estimate the error introduced in the extraction of $M_{0}$ and $R_\mrm{eq}$ due to truncating the analysis at ${\cal{O}}(\chi^{2})$. Naively, one expects the spin corrections to be $\mathcal{O}(\chi^2)$ in $M_{2}$ and $e$. The  fastest-spinning pulsar currently-observed, J1748-2446ad, has a spin period of 1.4ms (716Hz)~\cite{2006Sci...311.1901H}. Assuming that the mass is 1.4$M_\odot$, the dimensionless spin parameter is then $\chi = 0.39$ (APR), 0.35 (SLy), 0.53 (Shen) and 0.40 (SQM3). Therefore, the neglected spin terms would lead to a $\sim \mathcal{O}(30\%)$ correction. 

Depending on how a given quantity affects the pulse profile, such a $30\%$ correction may or may not be relevant. When applied to the quadrupole moment, such ${\cal{O}}(\chi^{2})$ corrections modify the pulse profile by $\sim 0.3\%$ -- $1.5\%$. Here, we assumed that the quadrupole moment contribution to the pulse profile is similar to that of a $1.8M_\odot$ NS with a 600Hz spin frequency, which should give a conservative estimate since the contribution should be larger for NSs with a smaller mass (or a larger radius) and a larger spin frequency. Such effects can thus be neglected, if the accuracy goal for the measurement of $(M_{0},R_\mrm{eq})$ with NICER and LOFT observations is $5\%$. A similar analysis suggests similar  modifications to the pulse profile due to the hexadecapole moment. When applied to the stellar eccentricity, however, the ${\cal{O}}(\chi^{2})$ corrections could be $\sim 3\%$ -- $9\%$, which would have to be included. 

Quartic-in-spin corrections increase the number of intrinsic parameter needed in X-ray observations to $(M_{0},R_\mrm{eq},I,M_{2},e,S_3,M_4,f)$ (the first four have spin corrections), but this set can be reduced as follows. First, from the $\bar{S}_3$--$\bar{M}_2$ and $\bar{M}_4$--$\bar{M}_2$ relations, we can express $S_3$ and $M_4$ in terms of $M_0$, $I$, $M_2$ and $f$. Depending on the accuracy goal of the observation, additional assumptions may be needed. For example, if one wishes to determine the stellar mass and radius to 5\% accuracy, as expected with NICER and LOFT, one needs to assume that the star is a NS with $5 < \bar{M}_2 < 10$, and check the consistency of such assumptions after parameter extraction. Otherwise, the EoS-variation in the $\bar{M}_4$--$\bar{M}_2$ relation given in the fit presented in this paper is too large, and systematic errors in the relation would dominate the error budget. If this is the case, however, one can create a more accurate fit by restricting attention to NSs only (i.e.~removing QSs from the data to be fitted) and perhaps also restricting the range of $\bar{M}_2$ considered. Once  $S_3$ and $M_4$ have been expressed in terms of $M_0$, $I$ and $M_2$, one can use the $\bar{I}$--$\bar{M}_2$ relation valid to all orders in spin to rewrite $M_2$ in terms of $M_0$ and $I$. This relation is given by Eq.~(8) of~\cite{Pappas:2013naa} or Eq.~(6) of~\cite{Chakrabarti:2013tca}. Finally, one can then use the $\bar{I}$--$C$ relation in~\cite{Baubock:2013gna} to rewrite $I$ in terms of $M_0$, $R_\mrm{eq}$ and $f$. After doing all of this, one is left with only 4 parameters: $(M_{0},R_\mrm{eq},e,f)$. 

\begin{figure}[htb]
\centering
\includegraphics[width=\columnwidth,clip=true]{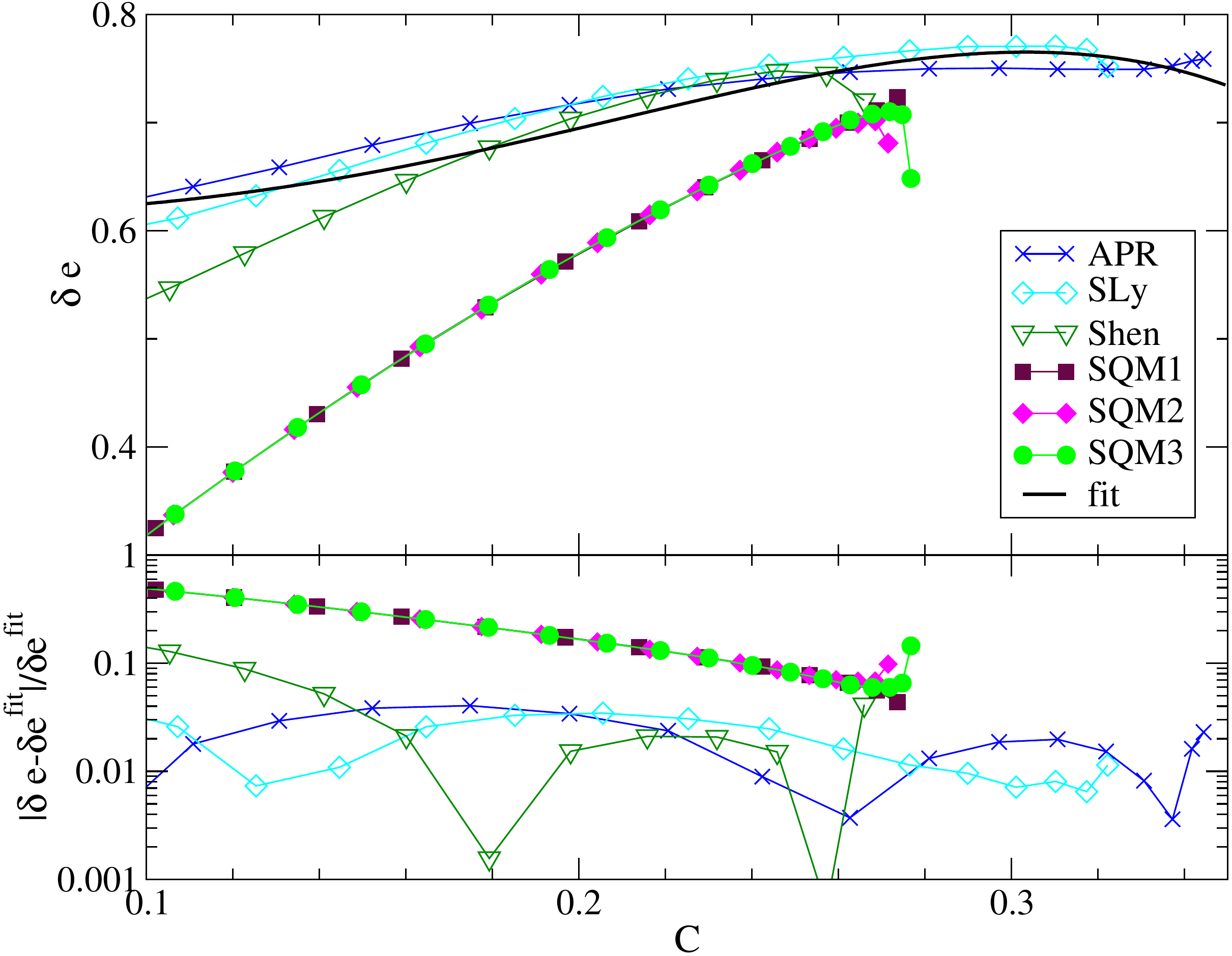}
\caption{
(Color online) (Top) Compactness dependence on the spin correction to the stellar eccentricity for various realistic EoSs, together with the fit given in Eq.~\eqref{eq:fit} for the NS sequence. (Bottom) Fractional difference from the fit.
\label{fig:deltaecc-C}
}
\end{figure}

We now need to express $e$ in terms of the other three parameters. $e^{(0)}$ is given in terms of $C^{(0)}$ through Eq.~\eqref{eq:fit2}, and hence we need to derive the spin correction $\delta e (C)$ to the $e$--$C$ relation:
\be
e (C) = e^{(0)} (C) \left[ 1 + \delta e (C) \chi^2 + \mathcal{O}(\chi^4) \right]\,,
\ee
where recall that $C = C^{(0)} [1 + \delta C \chi^2 + \mathcal{O}(\chi^4)]$. The top panel of Fig.~\ref{fig:deltaecc-C} presents the $\delta e$--$C$ relation. Observe that the relation is not EoS-independent if one includes QSs, just like the $e^{(0)}$--$C^{(0)}$ relation in Fig.~\ref{fig:ecc-C}. The figure shows that $\delta e < 0.8$, and hence such corrections modify the pulse profile by $0.8 \times 0.53^2 \times 30\% \sim 7\%$ at most. This then confirms that such ${\cal{O}}(\chi^{2})$ effects on the eccentricity will be important for rapidly rotating pulsars, if one aims at 5\% accuracy with NICER and LOFT. We create a fit given by Eq.~\eqref{eq:fit} with $y=\delta e$ and $x=\bar{M}_2$ just for the NS sequence with the  coefficients given in Table~\ref{table:coeff}. 
 We show the fractional difference from the fit in the bottom panel. Observe that the NS sequence is universal to 10\% (5\%) for $C>0.1$ ($C>0.15$). The $e^{(0)}$--$C^{(0)}$ (or the $e^{(0)}$--$C$ relation, replacing $C^{(0)}$ by $C$) and $\delta e$--$C$ relations allow us to finally reduce the intrinsic parameter space to just $(M_{0},R_\mrm{eq},f)$.

\section{Conclusions and Discussions} 
\label{sec:conclusions}

We investigated the existence of approximate EoS-universal, no-hair like relations for NSs and QSs, i.e.~relations between high and low-order multipole moments that are approximately independent of the EoS. We calculated NS and QS solutions and their multipole moments to $\ell=4$ order in three different ways: using two fully relativistic numerical codes valid, in principle, for arbitrarily fast spinning stars, and using a slow-rotation approximation to quartic order in spin. Such a calculation allowed us to confirm that the LORENE and RNS codes are indeed consistent with each other, in terms of the calculation of multipole moments. We were also able to establish that these codes become highly inaccurate for computations of higher multiple moments as the spin decreases below $\chi = 0.1$--$0.2$ due to numerical noise.

We found that the $\bar{M}_4$--$\bar{M}_2$ relation only depends very weakly on the EoSs and spins, just like the $\bar{S}_3$--$\bar{M}_2$ relation found in~\cite{Pappas:2013naa}. Therefore, the NS and QS current octupole and mass hexadecapole moments can be expressed in terms of the first three moments to $\mathcal{O}(10\%)$ accuracy. However, the universality is weaker for the former compared to the latter. This suggests that universal relations may not exist for higher order $\ell$ modes, or at the very least, that they will deteriorate with increasing $\ell$ number. This is consistent with leading-order Newtonian results in a weak-field expansion~\cite{Stein:2013ofa}. We also found that the $\bar{M}_4$--$\bar{M}_2$  relation is very close to these Newtonian results, even for highly relativistic NSs with large compactness. 

Our results have both theoretical and practical applications. On the theoretical side, it may not be possible to find mathematical theorems that prove the existence of truly EoS-independent no-hair like theorems for NSs or QSs. Of course, the relations found have always been of an approximate nature, but we have here found evidence that the universality seems to break as one considers higher $\ell$ multipoles. 

On a practical side, the relations found here may be important in the measurement of the mass and radius of NSs with future X-ray observations, including NICER and LOFT. In particular, the hexadecapole moment and the quadratic spin corrections to the quadrupole moment and the stellar eccentricity may dominate the error budget of such measurements over statistical error. The universal relations found here should allow us to reduce the number of parameters and break degeneracies among them. It would be important to investigate if one could come up with relations with less EoS- and spin-variation by changing the normalization of the multipole moments. For example, one could define new dimensionless multipole moments via $\bar{M}^\mrm{new}_\ell \equiv (-1)^{\ell/2} M_\ell/(M_0^{\ell +1} \chi^\ell C^k)$ and $\bar{S}^\mrm{new}_\ell \equiv (-1)^{(\ell-1)/2} S_\ell/(M_0^{\ell +1} \chi^\ell C^k)$, and find the $k$ that minimizes the variation. Since some of the calculations performed in this paper are order of magnitude estimates, detailed calculations would be desirable, where one solves the null geodesic equations for the NSs and QSs (valid to quartic order in spin) constructed here, using a ray-tracing algorithm~\cite{Baubock:2011ke} and evaluate the impact of multipole moments, stellar eccentricity and spin corrections on the X-ray pulse profile.  

Reference~\cite{I-Love-Q-Science,I-Love-Q-PRD} showed that the universality holds not only between the stellar moment of inertia and quadrupole moment, but also among the quadrupolar tidal Love number~\cite{love,hinderer-love,damour-nagar,binnington-poisson}. Reference~\cite{Yagi:2013sva} found that the universality holds among the higher-$\ell$ tidal Love numbers. These results, together with the ones shown in this paper, suggest that universal relations also exist among higher-$\ell$ rotation-induced multipole moments and higher-$\ell$ tidal Love numbers.

A straightforward extension of this work would be to investigate the relations among multipole moments at even higher $\ell$ order, such as $\ell=5$ and 6. It would be interesting to see if the EoS-universality indeed deteriorates, as we found here for lower $\ell$ modes. It would also be interesting to see if the higher-$\ell$ relations approach the Newtonian limit faster with increasing $\ell$. In order to achieve this goal, one would have to extend the slow rotation expansion to higher order in spin, or alternatively, attempt to extract higher-order multipole moments with LORENE or RNS. The latter may be feasible for rapidly-rotating stars with spin frequencies near the mass-shedding limit, but it would be extremely difficult for slowly-spinning stars due to numerical noise. 

Perhaps, what would be even more interesting is to understand physically why the EoS-universality is realized for the low-order multipoles in the first place. One way to understand this would be to investigate which part of the EoS is most responsible for the universality. Another approach would be to consider the Newtonian limit again, where one can tackle the problem (semi-)analytically, and break some of the approximations used in~\cite{Stein:2013ofa}. In particular, it would be interesting to investigate how breaking the elliptical isodensity approximation of~\cite{Lai:1993ve} impacts the EoS universality. Work along these lines is currently in progress.

An analytic model to describe the NS and QS exterior spacetimes may be useful in astrophysical observations. References~\cite{Stute:2003aj,berti-stergioulas} calculated a three-parameter solution, based on~\cite{Manko:2000ud}, by using the formalism developed by Ernst~\cite{Ernst:1967wx,Ernst:1967by}; they found that such a solution describes nicely the exterior spacetime of a rapidly-rotating NS. This solution cannot accurately capture the features of the exterior spacetime of a slowly-rotating NS because its quadrupole moment does not vanish in the non-rotating limit. Reference~\cite{Pappas:2012nv} extended these studies by considering a four-parameter solution (the two-soliton solution), found in~\cite{1995JMP....36.3063M}, which includes the three-parameter solution of~\cite{Manko:2000ud} as a special case. Since this solution depends on four free parameters, it can describe the stellar exterior spacetime to octupole order~\cite{Pappas:2012nt}. However, its hexadecapole moment in general has a term that depends on the spin-squared, which is absent in NS and QS spacetimes, as we proved in this paper. The slowly-rotating stellar solution valid to quartic order in spin found in this paper allows one to analytically express the stellar exterior spacetime with the correct hexadecapole moment. It would be interesting to extend these studies to find an analytic exterior spacetime for NSs and QSs valid to hexadecapole order and higher order without using the slow-rotation expansion.

When calculating the multipole moments, we assumed that the stars are uniformly-rotating and unmagnetized. Newly-born NSs and hypermassive NSs formed after NS binary mergers are expected to be differentially rotating, and magnetars can have magnetic fields as large as $10^{17}$G. Therefore, it may be interesting to relax the uniformly-rotating and unmagnetized assumptions. To relax the former, one could use a given rotation law, such as that in~\cite{komatsu_eh1989}. To relax the latter, one could consider magnetic field configurations similar to those in~\cite{I-Love-Q-B}. Probably, the magnetic field will become progressively more important, as one considers higher-$\ell$ multipole moments. This is because the rotationally-induced multipole moments should become smaller in magnitude as one increases $\ell$. One could also study how differential rotation and magnetic field affect the relations among multipole moments in the Newtonian limit, namely extending the work in~\cite{Stein:2013ofa}.

\acknowledgments
The authors thank Dimitrios Psaltis, Leo Stein and John L. Friedman for useful discussions.  KK is supported by JSPS Postdoctoral Fellowships for Research Abroad. GP would like to thank Kostas Kokkotas, and acknowledges financial support from the European Research Council under the European Union's Seventh Framework Programme (FP7/2007-2013) / ERC grant agreement n. 306425 ``Challenging General Relativity''. NY acknowledges support from NSF grant PHY-1114374 and NSF CAREER Grant PHY-1250636, as well as support provided by the National Aeronautics and Space Administration from grant NNX11AI49G, under sub-award 00001944.  The authors thank the YITP long-term workshop ``Gravitational Waves and Numerical Relativity'' and the YKIS2013 conference ``Gravitational Waves'', where the collaboration started. Some calculations used the computer algebra-systems MAPLE, in combination with the GRTENSORII package~\cite{grtensor}. Other calculations were carried out with the XTENSOR package for MATHEMATICA~\cite{2008CoPhC.179..586M,2009GReGr..41.2415B}.

\appendix

\section{Source Functions for the Equations of Stellar Structure at Quartic Order in Spin}
\label{app:exterior}

In this appendix, we present lengthy expressions for the source terms of the perturbed Einstein equations at quartic order in spin. 
First, the ones for the $\ell=2$ mode are given by
\bw
\allowdisplaybreaks
\ba
S_{\xi 42} &=&  \frac{1}{21 \nu'}\, \left\{ 28\,\omega_{11}   \left[ \frac{1}{2}\,\omega_{11}  R \left( -\frac{5}{7}\,\xi_{22}  +\xi_{20}   \right) \nu'  -R \left( -\frac{5}{7}\,\xi_{22}  +\xi_{20}   \right) \omega'_{11}  -{\frac {18}{7}}\, \omega_{33}  R+ \frac{5}{7}\,\omega_{11}  R k_{22} \right. \right. \nn \\
& & \left. \left.  + \left( \frac{5}{7}\,\xi_{22}  +h_{20}  R-\frac{5}{7}\,h_{22}  R-\xi_{20}   \right) \omega_{11}  +R\omega_{31}
   \right] R{{\rm e}^{-\nu  }}+
 \left( -3\, \left( \xi_{22}   \right) ^{2}-21\,\xi_{22}  \xi_{20}   \right) \nu''  -8\,{R}^{4} \left( \omega_{11}
   \right) ^{4}{{\rm e}^{-2\,\nu  }} \right. \nn \\
& & \left. +
 \left( -12\,\xi_{22}  -42\,\xi_{20}   \right) h'_{22}  +84\,h_{22}  h_{20}  -42\,\xi_{22}  h'_{20}  +12\, \left( h_{22}   \right) ^{2} \right\}  \,, \\
S_{m 42} &=& \frac{1}{252\,M  +1008\,\pi \,p  {R}^{3} } \left\{ -768\, \left[ -\frac{3}{16}\, \left( \pi \,p {R}^{3}+\frac{1}{4}\,M   \right)  \left( R-2\,M   \right) ^{2} \omega'_{11} \omega'_{33} \right. \right. \nn \\
& & \left. \left.  +\frac{1}{16}\, \left( R-2\,M   \right) ^{2} \left[  \left( \pi \,p {R}^{3}+\frac{1}{4}\,M  
 \right) k_{22}  + \left( -\frac{1}{4}\,h_{22}  +\frac{7}{4}\,h_{20}   \right) M \right. \right. \right. \nn \\
& & \left. \left. \left. -
7\,\pi \, \left[  \left( \frac{1}{7}\,h_{22}  - h_{20}   \right) p +\rho 
 \left( -h_{20}  +h_{20,c}-\frac{2}{7}\,h_{22}   \right)  \right] {R}^{3} \right]  \left(\omega'_{11} \right) ^{2}+{\frac {7}{16}}\, \left( \pi \,p {R}^{3}+\frac{1}{4}\,M   \right) 
 \left( R-2\,M   \right) ^{2} \omega'_{31} \omega'_{11} \right. \right. \nn \\
& & \left. \left.
  +\omega_{11}  \pi \, \left\{ {\frac{7}{8}}\,\omega_{11}   \left( -h_{20}  +h_{20,c}+\frac{2}{7}\,h_{22}   \right) R
 \left( R-2\,M   \right) ^{2} \rho'  +\omega_{11}   \left( \pi \,p {R}^{3}+\frac{1}{4}\,M   \right)  \left( \rho  +p  \right)  \left( R-2\,M  
 \right) k_{22} \right. \right. \right. \nn \\ 
& & \left. \left. \left. -3\, \left( \pi \,P  {R}^{3}+\frac{1}{4}\,M   \right)  \left( \rho  +p  \right)  \left( R-2\,M  
 \right) \omega_{33}  +\frac{7}{4}\, \left( \rho  +p  \right)  \left[  \left( h_{20,c}-3\,h_{20}  +{\frac {8}{7}}\,h_{22}  
 \right) \omega_{11}  -2\,\omega_{31}   \right]   M^{2} \right. \right. \right. \nn \\
& & \left. \left. \left. + \left\{ 
 \left[ 7\,\pi \, \left( h_{20,c}-3\,h_{20}  +{\frac {8}{7}}\,h_{22}   \right) {R}^{3} p^{2}+ \left[ 21\, \left( -\frac{5}{3}\,h_{20}
  +h_{20,c}+{\frac {4}{21}}\,h_{22}   \right) \pi \,{R}^{3}\rho -{\frac {7}{8}}\,h_{20,c}\,R+{\frac {21}{8}}\,h_{20}  R \right. \right. \right. \right. \right. \right. \nn \\
& & \left. \left. \left. \left. \left. \left.
 -h_{22}
  R-\frac{7}{4}\,m_{20}   \right] p  +14\,\rho  \left[ \pi \,{R}^{3} \left( -h_{20}  +h_{20,c}-\frac{2}{7}\,h_{22}  
 \right) \rho -\frac{1}{8}\,m_{20}  -\frac{1}{16}\,h_{20,c}\,R-{\frac {9}{56}}\,h_{22}  R+\frac{3}{16}\, h_{20}  R \right]  \right] \omega_{11}  \right. \right. \right. \right. \nn \\
& & \left. \left. \left. \left.
-14\, \left( \rho +p  \right) \omega_{31}  R \left( -\frac{1}{8}+\pi \,p {R}^{
2} \right)  \right\} M  -7\, \left\{  \left[ \frac{1}{2}\,\pi \,
 \left( h_{20,c}\,R-3\,h_{20}  R+{\frac {8}{7}}\,
h_{22}  R+2\,m_{20}   \right) {R}^
{2} p^{2}  \right. \right.  \right. \right.  \right. \nn \\
& &  \left. \left. \left. \left. \left.   +\frac{3}{2}\,\pi \, \left( -\frac{5}{3}\,h_{20}  R+\frac{2}{3}\,m_{20}  +{\frac {4}{21}}\,h_{22}  R+h_{20,c}\,R \right) \rho  {R}^{2}p  \right. \right.  \right. \right.  \right. \nn \\
& &  \left. \left. \left. \left. \left.  + \left[ \pi \,{R}^{3} \left( -h_{20}  +h_{20,c}-\frac{2}{7}\,h_{22}  
 \right) \rho -{\frac {5}{56}}\,h_{22}  R-\frac{1}{8}\,m_{20}   \right] \rho 
 \right] \omega_{11}  -\omega_{31}  \pi \,{R}^{3}p  \left( \rho +P
   \right)  \right\} R \right\} R \right] {R}^{3}{
{\rm e}^{-\nu  }}  \right. \nn \\
& & \left. -1024\, \left\{ -{\frac {3}{256}}\,
 \left( \pi \,p {R}^{3}+\frac{1}{4}\,M  
 \right)  \left( R-2\,M   \right) ^{2} \left( \omega'_{11} \right)^{4} \right. \right. \nn \\
& & \left. \left. -\frac{1}{8}\, \left( \omega_{11}   \right) ^{2}\pi \, \left[  \left( \frac{1}{4}\,\rho
  +\frac{1}{4}\,p  \right) M  + \left( \pi \,p {R}^{2}\rho +
\pi \,{R}^{2}  p^{2}+\frac{5}{8}\,\rho   \right) R \right] R \left( R-2\,M   \right) 
 \left( \omega'_{11}   \right) ^{2} \right. \right. \nn \\
& & \left. \left. +
 \left( \omega_{11}   \right) ^{4}\pi \, \left\{ {
\frac {9}{32}}\,R \left( R-2\,M   \right) \rho' + \left( \rho +p 
 \right)  \left[  \left( \frac{1}{4}\,\pi \,\rho {R}^{2}-{\frac 
{27}{32}}+\frac{1}{4}\,\pi \,p {R}^{2} \right) M \right. \right. \right. \right. \nn \\
& & \left. \left. \left. \left. +\pi \, \left[ \pi \,{R}^{2} p^{2}+ \left( \pi \,\rho {R}^{2}-{\frac {27}{8}}
 \right) p -\frac{5}{4}\,\rho  \right] {R}^{3}
 \right]  \right\}  \right\} {R}^{6} \left( R-2\,M  
 \right) {{\rm e}^{-2\,\nu  }}  \right. \nn \\
& & \left. + \left( 144\, \left( h_{22}   \right) ^{2}-1008\,h_{22}  h_{20}   \right)  M^{2}+ \left[ 2016\,\pi \, \left( \frac{2}{7}\, \left( h_{22}   \right) ^{2}-2\,h_{22}  h_{20}   \right) {R}^{3}p \right. \right. \nn \\
& & \left. \left. +4032\,\pi \,{R}^{3} \left( -h_{20}  +h_{20,c}-\frac{2}{7}\,h_{22}
   \right) h_{22}  \rho  -72\, \left( h_{22}   \right) ^{2}R+
 \left( 504\,h_{20}  R-504\,m_{20}   \right) h_{22}   \right] M \right. \nn \\
& & \left. -1008\,\pi \, \left[  \left( \frac{2}{7}\, \left( h_{22}   \right) ^{2}R+ \left( -2\,h_{20}  R+2\,m_{20}   \right) h_{22}   \right) p
  +2\, \left( -\frac{2}{7}\,h_{22}  R+h_{20,c}\,R+m_{20}  -h_{20}  R
 \right) \rho h_{22}   \right] {R}^{3} \right\}\,, \\
S_{h 42} &=& \frac{1}{42  {R}^{5} \left( R-2\,M   \right)}\, \left\{ 6\,{R}^{3}\xi_{22}   \left( R-2\,M   \right) ^{2} \left( \xi_{22}  +7\,
\xi_{20}   \right)  \nu'''
  -5\, \omega'_{11}  \left\{ -2\, \left( \xi_{22}  -\frac{7}{5}
\,\xi_{20}   \right)  \left( R-2\,M  
 \right) R \omega''_{11} \right. \right. \nn \\
& & \left. \left.   + \omega'_{11}  \left( 
\xi_{22}  -\frac{7}{5}\,\xi_{20}   \right) 
 \left( R-2\,M   \right) R \nu'  + \left[ 2\, \left( \xi_{22}  -\frac{7}{5}\, \xi_{20}   \right) R M' \right. \right. \right. \nn \\
& & \left. \left. \left.  +
 \left( 4\,Rk_{22}  -4\,h_{22}  R+
{\frac {28}{5}}\,h_{20}  R-{\frac {14}{5}}\,\xi_{20}  +2\,\xi_{22}   \right) M \right. \right. \right. \nn \\
& & \left. \left. \left. -{\frac {14}{5}}\,R \left( m_{20}  +h_{20}  R-\frac{5}{7}\,h_{22}  R+\frac{5}{7}\,Rk_{22}
  -\xi_{20}  +\frac{5}{7}\,\xi_{22}  -\frac{5}{7}\,m_{22}   \right)  \right] \omega'_{11}  \right. \right. \nn \\
& & \left. \left.  +{\frac {36}{5}}\, \left( R-2\,M   \right) R \left( \omega'_{33}
  -{\frac {7}{18}}\, \omega'_{31}
   \right)  \right\}  \left( R-2\,M  
 \right) {R}^{5}{{\rm e}^{-\nu  }} \right. \nn \\
& & \left. +12\, \left[ \left( \frac{7}{2}\,\xi_{20}  +\xi_{22}  
 \right)  \left( R-2\,M   \right) {R}^{2} k'_{22}  -2\,\xi_{22}  R \left( \xi_{22}  +7\,\xi_{20}   \right) M'  +4\,\xi_{22}  
 \left( \xi_{22}  +7\,\xi_{20}  
 \right) M \right. \right. \nn \\
& & \left. \left. 
 - \left(  \left( \xi_{22}   \right) ^{2}+ \left( 7\,m_{20}  +2\, m_{22}  +7\,\xi_{20}   \right) \xi_{22}
  +7\,\xi_{20}  m_{22}   \right) R \right]  \left( R-2\,M   \right) {
R}^{2} \nu'' \right. \nn \\
& & \left. +12\, \left( \frac{7}{2}\,
\xi_{20}  +\xi_{22}   \right) 
 \left( R-2\,M   \right) ^{2}{R}^{3} \left( 2+R \nu'   \right) k'_{22}  -12\,{R}^{2}\xi_{22}   \left( R \nu'  +1 \right)  \left( R-2\,M   \right)  \left( \xi_{22}  +7\,\xi_{20}
   \right)  M'' \right. \nn \\
& & \left. +24\, \left( \frac{7}{2}\,\xi_{20}  +\xi_{22}
   \right)  \left( R-2\,M   \right) ^{2}{R}^{3} h''_{22}  +84\, \xi_{22}  {R}^{3} \left( R-2\,M   \right) 
^{2} h''_{20} \right. \nn \\
& & \left. -24\, \left\{ \left( R-2\,M   \right) {R}^{2} \left[  \left( \frac{7}{2}\, \xi_{20}  +\xi_{22}   \right) R M'  + \left( -\frac{7}{2}\,\xi_{20}  -2\,Rk_{22}  -\xi_{22}  
 \right) M  + \left( Rk_{22}  +\frac{7}{2}\,
m_{20}  +m_{22}   \right) R
 \right] k'_{22}  \right. \right. \nn \\
& & \left. \left.  -2\, \left\{ -2\,\xi_{22}   \left( \xi_{22}  +7\, \xi_{20}  \right) M  + \left[  \left( \xi_{22}   \right) ^{2}+ \left( 7\,\xi_{20}  +\frac{7}{2}\,m_{20}  +m_{22}  
 \right) \xi_{22}  +\frac{7}{2}\,\xi_{20}  m_{22}   \right] R \right\} R M' \right. \right. \nn \\
& & \left. \left. + \left( \frac{7}{2}\,\xi_{20}  +\xi_{22}   \right)  \left( R-2\,M   \right) {R}^{2} m'_{22}  +\frac{7}{2}\,{R}^{2}\xi_{22}   \left( R-2\,M   \right) m'_{20}  -6\,\xi_{22}   \left( \xi_{22}  +7\,\xi_{20}   \right) M^{2} \right. \right. \nn \\
& & \left. \left. +\frac{3}{2}\, \left[ \frac{8}{3}\, \left( \xi_{22}
   \right) ^{2}+ \left( h_{22}  R+\frac{8}{3}\,m_{22}  +{\frac {56}{3}}\,\xi_{20}  +{\frac {28}{3}}\,m_{20}   \right) \xi_{22}  +{\frac {28}{3}}\,\xi_{20} m_{22}   \right] RM \right. \right. \nn \\
& & \left. \left. -\frac{3}{4}\, \left[ \frac{2}{3}\,
 \left( \xi_{22}   \right) ^{2}+ \left( \frac{14}{3}\,\xi_{20}  +h_{22}  R+{\frac {28}{3}}\, m_{20}  +\frac{5}{3}\,m_{22}   \right) \xi_{22}  +{\frac {56}{3}}\, \left( \frac{1}{7}\,m_{22}
  +m_{20}  +\frac{1}{2}\,\xi_{20}   \right) m_{22}   \right] {R}^{2} \right\} R
\nu' \right. \nn \\
& & \left. +12\, \left( R-2\,M   \right) ^{2}{R}^{4} \left( k'_{22}   \right) ^{2}+84\, \left[ -\frac{4}{7}\, \left( \frac{7}{2}\,\xi_{20}
  +\xi_{22}   \right) R M'  +\frac{2}{7}\, \left( R-2\,M   \right) {R}^{2} h'_{22}  + \left( R-2\,M   \right) {R}^{2} h'_{20} \right. \right. \nn \\
& & \left. \left. +
 \left( 4\,\xi_{20}  +{\frac {8}{7}}\,Rk_{22}
  +{\frac {8}{7}}\,\xi_{22}  
 \right) M  -\frac{4}{7}\, \left( \frac{3}{4}\,\xi_{22}  +{\frac {21}{4}}\,\xi_{20}  +Rk_{22}
  +\frac{7}{2}\,m_{20}  +m_{22}   \right) R \right]  \left( R-2\,M   \right) {
R}^{2} k'_{22} \right. \nn \\
& & \left. -168\,R \left\{ \frac{2}{7}\,
 \left( \frac{7}{2}\,\xi_{20}  +\xi_{22}  
 \right)  \left( R-2\,M   \right) {R}^{2} h'_{22}  +{R}^{2}\xi_{22}  
 \left( R-2\,M   \right)  h'_{20}
  +\frac{6}{7}\,\xi_{22}   \left( \xi_{22}
  +7\,\xi_{20}   \right) M \right. \right. \nn \\
& & \left. \left.  -\frac{3}{7}\, \left[  \left( \xi_{22}   \right) ^{2
}+ \left( 7\,\xi_{20}  +\frac{2}{3}\,m_{22}  +\frac{7}{3}\,m_{20}   \right) \xi_{22}  +\frac{7}{3}\,\xi_{20}  m_{22}  
 \right] R \right\}  M' \right. \nn \\
& & \left. -168\, \left( R-2\,M   \right) {R}^{2} \left[  \left( -\frac{4}{7}\,h_{22}
  R-2\,h_{20}  R-\frac{4}{7}\,\xi_{22}
  -2\,\xi_{20}   \right) M \right. \right. \nn \\
& & \left. \left. +R \left( h_{20}  R+\frac{2}{7}\,h_{22}  R+{\frac {5}{14}}\,\xi_{22}  +m_{20}
  +2\,\xi_{20}  +\frac{2}{7}\,m_{22}
   \right)  \right]  h'_{22}  \right. \nn \\
& & \left.  -168\, \left( R-2\,M   \right) {R}^{2}
 \left[  \left( -2\,h_{22}  R-2\,\xi_{22}   \right) M  + \left( h_{22}  R+\frac{1}{2}\,\xi_{22}  +m_{22}  
 \right) R \right]  h'_{20}  \right. \nn \\
& & \left.  -24\,
 \left( \frac{7}{2}\,\xi_{20}  +\xi_{22}  
 \right)  \left( R-2\,M   \right) {R}^{2} m'_{22}  -84\,{R}^{2}\xi_{22}  
 \left( R-2\,M   \right)  m'_{20}
  +288\,\xi_{22}   \left( \xi_{22}
  +7\,\xi_{20}   \right)  M ^{2}  \right. \nn \\
& & \left.  -1008\, \left\{ \frac{1}{7}\, \left( \xi_{22}
   \right) ^{2}+ \left( \xi_{20}  +\frac{1}{7}\,m_{22}  +\frac{3}{14}\,h_{22}  R+{\frac {4}{21}}\,Rk_{22}  +\frac{1}{2}\,m_{20}   \right) \xi_{22} \right. \right. \nn \\
& & \left. \left. + \left( \frac{2}{3}\,Rk_{22}
  +\frac{1}{2}\,m_{22}  +h_{22}  R \right) \xi_{20}  + \left[ \frac{3}{14}\, \left( h_{22}   \right) ^{2}+ \left( h_{20}  +\frac{2}{7}\,k_{22}   \right) h_{22}  +\frac{1}{3}\, \left( k_{22}   \right) ^{2}
 \right] {R}^{2} \right\} RM \right. \nn \\
& & \left. +504\, \left\{  \left( \frac{3}{14}\,h_{22}  R+\frac{1}{14}\,m_{22}  +{
\frac {4}{21}}\,Rk_{22}  +\frac{1}{2}\,m_{20}   \right) \xi_{22}  + \left( \frac{2}{3}\,Rk_{22}
  +\frac{1}{2}\,m_{22}  +h_{22}  R \right) \xi_{20}  +\frac{2}{21}\, \left( m_{22}
   \right) ^{2} \right. \right. \nn \\
& & \left. \left. +\frac{2}{3}\,m_{20}  m_{22}  + \left[ \frac{3}{14}\, \left( h_{22}  
 \right) ^{2}+ \left( h_{20}  +\frac{2}{7}\,k_{22}   \right) h_{22}  +\frac{1}{3}\, \left( k_{22}
   \right) ^{2} \right] {R}^{2} \right\} {R}^{2}
 \right\}\,, \\
S_{v 42} &=& {\frac {1}{504 \left( 4 \pi \,p  {R}^{3} + M   \right) {R} \left( R-2\,M   \right)^{2}}}\, \left\{ 21504\, \left\{ \frac{1}{16}\, \left\{  \left( -{\frac {3}{56}}\,h_{22}  -{\frac {3}{56}}\,k_{22}
  +\frac{1}{4}\,h_{20}   \right)  M^{2}  \right. \right. \right. \nn \\
& & \left. \left. \left.+ \left[ \frac{1}{4}\,\pi \, \left( h_{20,c}
-{\frac {8}{7}}\,k_{22}  +3\,h_{20}   \right) {R}^{3} p  +\frac{1}{4}\,{R}^{3}\pi \,
 \left( -h_{20}  +h_{20,c} \right) \rho  -{\frac {1}{56}}\,Rk_{22}  +\frac{1}{14}\,h_{22}
  R-\frac{1}{16}\,m_{20}   \right] M  \right. \right. \right. \nn \\
& & \left. \left. \left. +{R}^{2} \left[ {\pi }^{2}{R}^{4} \left( -\frac{2}{7}\,k_{22}
  +\frac{6}{7}\,h_{22}  +h_{20,c}-h_{20}   \right)  p^{2}  \right. \right. \right. \right. \nn \\
& & \left. \left. \left. \left. +
 \left[ {R}^{3}\pi \, \left( -h_{20}  +h_{20,c}
 \right) \rho  +\frac{3}{14}\,h_{22}  R-\frac{1}{2}\,h_{20}  R+\frac{1}{14}\,Rk_{22}  -\frac{1}{4}\, m_{20}   \right] \pi \,R p \right. \right. \right. \right. \nn \\
& & \left. \left. \left. \left. -{\frac {1}{112}}\,h_{22}  +{\frac {1}{56}}\,k_{22}  -\frac{1}{16}\,h_{20}   \right]  \right\}  \left( R-2\,M   \right) ^{2} \left( \omega'_{11}
   \right) ^{2}+{\frac {3}{56}}\, \left[  \left( -\frac{7}{3}\,\pi \,p  {R}^{3}+{\frac {7}{12}}\,M  -{\frac {7}{12}}\,R \right) \omega'_{31}
 \right. \right. \right. \nn \\
& & \left. \left. \left. + \left( \pi \,p  {R}^{3}+\frac{1}{4}\,R-\frac{1}{4}
\,M   \right) \omega'_{33}  -\frac{5}{2}\,\omega_{33}   \right]  \left( \pi \,p
  {R}^{3}+\frac{1}{4}\,M   \right)  \left( R-2\,M   \right) ^{2}\omega'_{11} \right. \right. \nn \\
& & \left. \left. +\pi \,\omega_{11}  R \left\{ -\frac{1}{4}\, \left( R-2\,M   \right) ^{2} \left( \pi \,p  {
R}^{3}+\frac{1}{4}\,R-\frac{1}{4}\,M   \right)  \left( h_{20,c}-h_{20}  +\frac{2}{7}\,h_{22}   \right) \omega_{11}  R\rho' \right. \right. \right. \nn \\
& & \left. \left. \left.  +
 \left( \rho  +p   \right)  \left\{  \left[  \left( \frac{1}{8}\,h_{20,c}+\frac{3}{14}\,h_{22}  -\frac{5}{8}
\,h_{20}  +{\frac {3}{28}}\,k_{22}   \right) \omega_{11}  -\frac{1}{4}\,\omega_{31}
  +{\frac {3}{28}}\,\omega_{33}  
 \right] M^{3} \right. \right. \right. \right. \nn \\
& & \left. \left. \left. \left. -\frac{1}{2}\, \left[ \pi \,\omega_{11}   \left( h_{20,c}-{\frac {8}{7}}\, k_{22}  +3\,h_{20}   \right) {R}^{2}p
  + \left( \pi \,{R}^{2} \left( -h_{20}  +h_{20,c} \right) \rho  +{\frac {9}{28}}\,k_{22}  -{\frac {15}{8}}\,h_{20}  +
\frac{3}{8}\,h_{20,c}+\frac{4}{7}\,h_{22}   \right) \omega_{11}  \right. \right. \right. \right. \right. \nn \\
& & \left. \left. \left. \left. \left.
  -\frac{3}{4}\,\omega_{31}  +{\frac {9}{28}
}\,\omega_{33}   \right] R M^{2}-2\, \left[ 2\,{\pi }^{2}{R}^{5} \left[  \left( -\frac{1}{7}\,k_{22}  +\frac{6}{7}\,h_{22}  -h_{20}
  +h_{20,c} \right) \omega_{11} +
\frac{3}{7}\,\omega_{33}  -\omega_{31}  
 \right]  p^{2}  \right. \right. \right. \right. \right. \nn \\
& & \left. \left. \left. \left. \left. + \left[  \left( {R}^{3}\pi \, \left( -h_{20}  +h_{20,c} \right) \rho
  +\frac{3}{14}\,Rk_{22}  +\frac{3}{7}\,h_{22}
  R-\frac{1}{4}\,m_{20}  +\frac{1}{8}\,h_{20,c}\,
R-{\frac {9}{8}}\,h_{20}  R \right) \omega_{11} \right. \right. \right. \right. \right. \right. \nn \\
& & \left. \left. \left. \left. \left. \left.
  +\frac{3}{14}\,R \left( -\frac{7}{3}\,\omega_{31}  +\omega_{33}   \right)  \right] \pi \,{R}^{2
}p  + \left( -\frac{1}{8}\,{R}^{3}\pi \, \left( -h_{20}
  +h_{20,c} \right) \rho  -\frac{1}{32}\,m_{20}  -{\frac {5}{112}}\,Rk_{22}  +{
\frac {7}{32}}\,h_{20}  R  \right. \right. \right. \right. \right. \right. \nn \\
& & \left. \left. \left. \left. \left. \left. -\frac{1}{32}\,h_{20,c}\,R-\frac{1}{28}
\,h_{22}  R \right) \omega_{11}  -{\frac {3}{112}}\,R \left( -\frac{7}{3}\,\omega_{31}  + \omega_{33}   \right)  \right] RM  \right. \right. \right. \right. \nn \\
& & \left. \left. \left. \left. +
 \left[ 2\, \left[  \left( m_{20}  +h_{20,c}\,R-\frac{1}{7}\,Rk_{22}  +\frac{6}{7}\,h_{22}  R- h_{20}  R \right) \omega_{11}  +\frac{3}{7}\,R
 \left( -\frac{7}{3}\,\omega_{31}  +\omega_{33}   \right)  \right] {\pi }^{2}{R}^{3} p^{2} \right. \right. \right. \right. \right. \nn \\
& & \left. \left. \left. \left. \left. +\pi \,R \left[  \left( {R}^{3}\pi \, \left( -h_{20}  +h_{20,c} \right) \rho  -\frac{3}{4}
\,h_{20}  R+\frac{1}{4}\,h_{20,c}\,R+\frac{1}{4}\,m_{20}
  +\frac{3}{7}\,h_{22}  R+ \frac{1}{14}\,Rk_{22}
   \right) \omega_{11}  \right. \right. \right. \right. \right. \right. \nn \\
& & \left. \left. \left. \left. \left. \left. +\frac{3}{14}\,R
 \left( -\frac{7}{3}\,\omega_{31}  +\omega_{33}   \right)  \right] p  +\frac{1}{16}\, \left( h_{20}
  +\frac{1}{7}\,h_{22}  -\frac{2}{7}\,k_{22}
   \right) \omega_{11}   \right] {R}^{3} \right\}  \right\}  \right\} {R}^{3}{{\rm e}^{-\nu  }
} \right. \nn \\
& & \left. -16384\, \left\{  \left\{ -{\frac {1}{256}}\, \left( R-2\,M   \right) ^{2} \left( -\frac{1}{4}\,M  -\frac{3}{16}\,R+\pi \,
p  {R}^{3} \right)  \left( \pi \,p  {R}^{3}+ \frac{1}{4}\,M   \right)  \left( \omega'_{11}   \right) ^{4} \right. \right. \right. \nn \\
& & \left. \left. \left.  +\pi \, \left( \omega_{11}
   \right) ^{4} \left\{ {\frac {9}{64}}\, \left( R-2\,M
   \right)  \left( \pi \,p  {R}^{3}+ \frac{1}{4}\,R- \frac{1}{4}\,M   \right) R\rho'  \right. \right. \right. \right. \nn \\
& & \left. \left. \left. \left.  + \left( \pi \,p  {R}^{3}+\frac{1}{4}\,M   \right)  \left[  \left( -\frac{1}{4}\,\pi \,p  {R}^{
2}-\frac{1}{4}\,\pi \,\rho  {R}^{2}+{\frac {27}{64}} \right) M \right. \right. \right. \right. \right. \nn \\
& & \left. \left. \left. \left. \left. 
  + \left[ {R}^{4}{\pi }^{2} p^{2}+ \left( \pi \,\rho  {R}^{2}-{
\frac {15}{8}} \right) \pi \,{R}^{2}p  -{\frac {27}{64}}-\frac{3}{16}\,\pi \,\rho  {R}^{2} \right] R \right]  \left( \rho
  +p   \right)  \right\}  \right\} {R}^{6}{{\rm e}^{-2\,\nu  }}  \right. \right. \nn \\
& & \left. \left.-{\frac {63}{256}}\,h_{22}
   \left( \frac{1}{4}\,h_{22}  -\frac{2}{7}\,k_{22}  +h_{20}   \right) M^{2}+{\frac {63}{128}}\,R \left[ h_{22}
  \pi \, \left( h_{20}  +\frac{1}{7}\,h_{22}  -\frac{2}{7}\,k_{22}   \right) {R}^{2}p \right. \right. \right. \nn \\
& & \left. \left. \left.
  +h_{22}  \pi \, \left( h_{20}
  +\frac{1}{7}\,h_{22}  -\frac{2}{7}\,k_{22}
   \right) {R}^{2}\rho  +{\frac {1}{56}}
\, \left( h_{22}   \right) ^{2}-\frac{1}{4}\,h_{22}
  h_{20}  -\frac{1}{2}\, \left( h_{20}
  -\frac{2}{7}\,k_{22}   \right) k_{22}
   \right] M \right. \right. \nn \\
& & \left. \left. +{\frac {27}{64}}\,
 \left[  \left( h_{22}   \right) ^{2}{\pi }^{2}{R}^{4} p^{2} -{\frac {7}{12}}\,h_{22}
  \pi \,{R}^{2} \left( -\frac{2}{7}\,h_{22}  -\frac{2}{7}\,k_{22}  +h_{20}  
 \right) p \right. \right. \right. \nn \\
& & \left. \left. \left. -{\frac {7}{12}}\, \left( h_{20}  +\frac{1}{7}\,h_{22}  -\frac{2}{7}\,k_{22}  \right)  \left( -\frac{1}{2}\,k_{22}  +{R}^{2}h_{22}  \pi \,\rho  -\frac{1}{2}\,h_{22}
   \right)  \right] {R}^{2} \right\}  \left( R-2\,M   \right)  \right\}\,. \\
\ea
%
Similarly, the ones for the $\ell=4$ mode are given by
%
\allowdisplaybreaks
\ba
S_{\xi 44} &=&  \frac{1}{35  \left( \nu'   \right) {{\rm e}^{\nu  }}} \, \left\{ 4\,{R}^{4} \left( \omega_{11} 
 \right) ^{4}{{\rm e}^{-\nu  }}-9\,{{\rm e}^{\nu  }} \left( \xi_{22}   \right) ^{2}\nu''  +12\, \nu' \xi_{22}  {R}^{2}
 \left( \omega_{11}  \right) ^{2}  -24\,{R}^{2}\omega_{11} \xi_{22}  \omega'_{11} -36\,{{\rm e}^{\nu  }}\xi_{22}  h'_{22}  +36
\,{{\rm e}^{\nu  }} \left( h_{22}  
 \right) ^{2} \right. \nn \\
& & \left.-24\,R \left( \omega_{11}  \right) ^{2}
\xi_{22}  +24\,{R}^{2} \left( \omega_{11}  \right) ^{2}h_{22}  -24\,{R}^{2} \left( \omega_{11}  \right) ^{2}k_{22} +120\,{R}^{2}\omega_{11} \omega_{33}
  \right\}\,, \\
S_{m 44} &=& \frac{2\left( R-2\,M   \right)}{105\,M  +420\,\pi \,p
  {R}^{3}}\, \left\{ 24\, \left\{ -\frac{5}{4}\, \left( \omega'_{11}
   \right)  \left( R-2\,M   \right) 
 \left( \pi \,p  {R}^{3}+\frac{1}{4}\,M  
 \right) \omega'_{33} \right. \right. \nn \\
& & \left. \left.  + \left( R-2\,M   \right)  \left[  \left( 
\pi \,p  {R}^{3}+\frac{1}{4}\,M   \right) k_{22}  -\frac{3}{2}\,h_{22}  
 \left[ \frac{1}{6}\,M  + \left( \frac{2}{3}\,p  +\rho
   \right) \pi \,{R}^{3} \right]  \right]  \left( \omega'_{11}   \right) ^{2} \right. \right. \nn \\
& & \left. \left.  +16\, \left[ 
\frac{1}{32}\,R\omega_{11}  h_{22}  
 \left( R-2\,M   \right) \rho'  +\omega_{11}   \left( p  +\rho
   \right)  \left( \pi \,p  {R}^{3}+\frac{1}{4}\,M   \right) k_{22}  -\frac{5}{4}\,
 \left( p  +\rho   \right)  \left( \pi \,
p  {R}^{3}+\frac{1}{4}\,M   \right) \omega_{33} \right. \right. \right. \nn \\
& & \left. \left. \left. -\frac{3}{2}\,\omega_{11}  h_{22}   \left[  \left( {\frac {11}{48}}\,p  +{\frac {11}{48}}\,\rho   \right) M  + \left( {\frac {11}{12}}\,{R}^{2}\pi \, p^{2}+{R}^{2}\pi \, \rho^{2}+{\frac {23}{12}}\,{R}^{2}\pi \,\rho  p  -\frac{1}{8}\,\rho   \right) R \right] 
 \right] \omega_{11}  \pi \,R \right\} {R}^{3}{{\rm e}
^{-\nu  }}  \right. \nn \\
& & \left. +  \left\{ \left( R-2\,M  
 \right) ^{2} 
\left( \pi \,p  {R}^{3}+\frac{1}{4}\,M \right)  
\left( \omega'_{11}   \right) ^{4}  \right. \right. \nn \\
& & \left. \left. +48\, \left[  \left( \frac{1}{4}\,p  +\frac{1}{4}\,\rho   \right) M  + \left( {R}^{2}\pi 
\, p^{2}+{R}^{2}\pi \,\rho  p  -\frac{1}{4}\,\rho   \right) R
 \right]  \left( \omega_{11}   \right) ^{2}\pi \,
 \left( R-2\,M   \right) R \left( \omega'_{11}   \right) ^{2} \right. \right. \nn \\
& & \left. \left. +512\, \left\{ {\frac {1}{128}}\,
R \left( R-2\,M   \right) \rho'  + \left[  \left( \frac{1}{4}\,\pi \,\rho  {R}^{2}-{
\frac {3}{128}}+\frac{1}{4}\,\pi \,{R}^{2}p   \right) M \right. \right. \right. \right. \nn \\
& & \left. \left. \left. \left.
  +\pi \,{R}^{3} \left[ {R}^{2}\pi \, p^{2}+ \left( \pi \,\rho  {R}^{2}-{
\frac {3}{32}} \right) p  -\frac{3}{8}\,\rho  
 \right]  \right]  \left( p  +\rho  
 \right)  \right\}  \left( \omega_{11}   \right) ^{4}
\pi  \right\} {R}^{6}{{\rm e}^{-2\,\nu  }} \right. \nn \\
& & \left. +216\,
 \left( h_{22}   \right) ^{2} \left[ \frac{1}{6}\,M  + \left( \frac{2}{3}\,p  +\rho  
 \right) \pi \,{R}^{3} \right]  \right\}\,, \\
S_{h 44} &=& {\frac {1}{1680 {R} \left( R-2\,M   \right) \left( 4 \pi \,p  {R}^{3}+ M   \right) ^{3}}}\, \left\{ 49152\,{R}^{3} \left\{ -\frac{5}{4}\, \left( R-2\,
M   \right)  \left( \pi \,p  {R}^{3}+\frac{1}{4}\,M   \right) ^{2} \left[ \frac{M^2}{16}\, + \left( \frac{1}{2}\,\pi \,p  {R}^{3}+\frac{3}{8}\,R \right) M \right. \right. \right. \nn \\
& & \left. \left. \left. +{\pi }^{2} p^{2}{R}^{6}-\frac{3}{16}\,{R}^{2} \right]  \omega'_{11}  \omega'_{33}  + \left( R-2\,M   \right)  \left\{ 
 \left[ {\frac {1}{256}}\,  M^{4}+ \left( {\frac {9}{512}}\,R+\frac{1}{16}\,\pi \,p  {R}^{3}
 \right) M^{3} \right. \right. \right. \right. \nn \\
& & \left. \left. \left. \left. +\frac{3}{8}\, \left( {R}^{4}
{\pi }^{2} p^{2}+\frac{3}{8}\,\pi \,p
  {R}^{2}-{\frac {7}{128}} \right) {R}^{2} M^{2}+{R}^{3} \left( {R}^{6}{\pi }^{3}
 p^{3}+{\frac {9}{32}}\,{R}^{4}{\pi 
}^{2} p^{2}+{\frac {3}{256}}-{\frac
{9}{128}}\,\pi \,p  {R}^{2} \right) M  \right. \right. \right. \right. \nn \\
& & \left. \left. \left. \left.  +{R}^{4} \left( -{\frac {9}{64}}\,{R}^{4}{\pi }^{2} p^{2}+{R}^{8}{\pi }^{4} p^4-{\frac {3}{1024}} \right)  \right] k_{22}
  +\frac{1}{2}\, \left\{ {\frac {7}{256}}\, M^{4}+ \left( {\frac {13}{64}}\,\pi \,p  {R}^{3}+{\frac {3}{512}}\,R-{\frac {3}{64}}\,{R}^{3}\pi \,
\rho   \right)  M^{3} \right. \right. \right. \right. \nn \\
& & \left. \left. \left. \left.  -\frac{3}{8}\,{R}^{2} \left[ -{R}^{4}{\pi }^{2} p^{2}+\pi \,{R}^{2} \left( \pi \,\rho  {R}^{2}
-{\frac {11}{16}} \right) p  -\frac{3}{16}\,\pi \,\rho  {R}^{2}+{\frac {9}{128}} \right]  M^{2} \right. \right. \right. \right. \nn \\
& & \left. \left. \left. \left.  -\frac{3}{4}\, \left[ -\frac{1}{3}\,{R}^{6}{\pi }^{3} p^{3}+{\pi }^{2}{R}^{4} \left( \pi \,\rho  {R}^{2}-\frac{5}{8} \right) p^{2}+
 \left( -\frac{1}{4}\,{R}^{4}{\pi }^{2}\rho  +{\frac {9}{32}}
\,\pi \,{R}^{2} \right) p  -{\frac {7}{256}}+\frac{1}{16}\,
\pi \,\rho  {R}^{2} \right] {R}^{3}M \right. \right. \right. \right. \nn \\
& & \left. \left. \left. \left.   
+{R}^{4} \left[ {R}^{8}{\pi }^{4} p^{4}+\frac{3}{8}\,{R}^{6}{\pi }^{3} p^{3}+
 \left( -{\frac {15}{64}}\,{R}^{4}{\pi }^{2}+\frac{3}{8}\,{R}^{6}{\pi }^{3}
\rho   \right)   p^{2}+{\frac {3}{128}}\,\pi \,p  {R}^{2}-{\frac {3}{512}}+
{\frac {3}{256}}\,\pi \,\rho  {R}^{2} \right] 
 \right\} h_{22}   \right\}  \left( \omega'_{11}   \right) ^{2} \right. \right. \nn \\
& & \left. \left. +{\frac {15}{16}}\, \left( R
-2\,M   \right)  \left( \pi \,p  {R}^{3}+\frac{1}{4}\,R-\frac{1}{4}\,M   \right)  \left( \pi \,p  {R}^{3}+\frac{1}{4}\,M   \right) ^{2}\omega_{33}
  \omega'_{11} \right. \right. \nn \\
& & \left. \left. +16\,\omega_{11}  \pi \, \left\{ \frac{1}{32}\,\omega_{11}   \left[ \frac{1}{16}\, M^{2}+
 \left( \frac{1}{2}\,\pi \,p  {R}^{3}-\frac{3}{8}\,R \right) M  +{\pi }^{2} p^{2}{R}^{6}+\frac{3}{16}\,{R}^{2} \right] h_{22}   \left( R-2\,M   \right) R \left( \pi \,p  {R}^{3}+\frac{1}{4}\,M
   \right) \rho' \right. \right. \right. \nn \\
& & \left. \left.\left. +
 \left\{ \omega_{11}   \left[ {\frac {1}{256}}\,
 M^4+ \left( \frac{1}{16}\,\pi \,p  {R}^{3}+{\frac {3}{512}}\,R \right) M^{3}+\frac{3}{8}\,{R}^{2} \left( {R}^{4}{\pi }^{2} p^{2}+\frac{1}{8}\,\pi \,p  {R}^{2}+
{\frac {3}{128}} \right)  M^2  \right. \right. \right. \right. \right. \nn \\
& & \left. \left.\left. \left.\left. +{R}^
{3} \left( -{\frac {3}{256}}+{\frac {3}{32}}\,{R}^{4}{\pi }^{2}
 p^2+{R}^{6}{\pi }^{3} p^3-{\frac {3}{128}}\,\pi \,p {R}^{2} \right) M  +{R}^{4} \left( {R}^{8}{
\pi }^{4} p^4+{\frac {3}{1024}}-{
\frac {3}{64}}\,{R}^{4}{\pi }^{2} p^2 \right)  \right] k_{22} \right. \right. \right. \right. \nn \\
& & \left. \left.\left. \left. -\frac{5}{4}\, \left[ \frac{1}{16}\,
 M^2+ \left( \frac{1}{2}\,\pi \,p  {R}^{3}-\frac{3}{8}\,R \right) M  +{\pi }^{2} \left( 
p   \right) ^{2}{R}^{6}+\frac{3}{16}\,{R}^{2} \right]  \left( \pi \,p  {R}^{3}+\frac{1}{4}\,M   \right) ^{2}
\omega_{33} \right. \right. \right. \right. \nn \\
& & \left. \left.\left. \left. +\frac{1}{8}\,\omega_{11}  h_{22}   \left[ {\frac {25}{256}}\, M^{4} + \left( \frac{5}{8}\,\pi \,p  {R}^{3}-{3}{16}\,{R}^{3}\pi \,\rho  +{\frac {9}{64}}\,R \right) 
 M^3 \right. \right. \right. \right. \right. \nn \\
& & \left. \left.\left. \left. \left. -\frac{3}{2}\, \left[ -\frac{1}{4}\,{R}^{4}{\pi }^{2} p^2+\pi \,{R}^{2}
 \left( -{\frac {5}{16}}+\pi \,\rho  {R}^{2} \right) p
  +\frac{1}{16}\,\pi \,\rho  {R}^{2}+{\frac {1
}{64}} \right] {R}^{2} M^2  \right. \right. \right. \right. \right. \nn \\
& & \left. \left.\left. \left. \left.  -3\,
 \left[ \frac{2}{3}\,{R}^{6}{\pi }^{3} p^3
+{\pi }^{2}{R}^{4} \left( \pi \,\rho  {R}^{2}-\frac{3}{4}
 \right)  p^2+ \left( \frac{1}{16}\,\pi \,
{R}^{2}-\frac{1}{4}\,{R}^{4}{\pi }^{2}\rho   \right) p  -\frac{1}{16}\,\pi \,\rho  {R}^{2}+{\frac {7}{256}}
 \right] {R}^{3}M \right. \right. \right. \right. \right. \nn \\
& & \left. \left.\left. \left. \left. + \left[ {R}^{8}{\pi }^{4} p^4+\frac{3}{2}\,{R}^{6}{\pi }^{3} p^{3}+ \left( \frac{3}{2}\,{R}^{6}{\pi }^{3}\rho
  -{\frac {9}{8}}\,{R}^{4}{\pi }^{2} \right)  p^2-{\frac {3}{32}}\,\pi \,p  {R}^{2}-{\frac {3}{64}}\,\pi \,\rho  {R}^{2}+
{\frac {3}{128}} \right] {R}^{4} \right]  \right\}  \left( \rho  +p   \right)  \right\} R \right\} {{\rm e}^{-
\nu  }} \right. \nn \\
& & \left. -1024\,{R}^{6} \left\{  \left( R-2\,M   \right) ^{2} \left[ {\frac {1}{256}}\, M^4+ \left( \frac{1}{16}\,\pi \,p  {R}^{3}+
{\frac {3}{128}}\,R \right)  M^3 \right. \right. \right. \nn \\
& & \left. \left. \left. + 
 \left( \frac{3}{16}\,\pi \,{R}^{4}p  +\frac{3}{8}\,{\pi }^{2} p^2{R}^{6} \right) M^{2}+{R}^{3} \left( {R}^{6}{\pi }^{3} p^3-{\frac {3}{256}}-{\frac {3}{32}}\,\pi 
\,p  {R}^{2}+\frac{3}{8}\,{R}^{4}{\pi }^{2} p^2 \right) M \right. \right. \right. \nn \\
& & \left. \left. \left. +{R}^{4} \left( {R}^
{8}{\pi }^{4} p^4-\frac{3}{16}\,{R}^{4}{\pi }^{2} p^2+{\frac {3}{1024}}
 \right)  \right]  \left( \omega'_{11}   \right) ^{4} \right. \right. \nn \\
& & \left. \left. -6\, \left( \omega_{11}  
 \right) ^{2} \left( R-2\,M   \right) ^{2}\pi \,
 \left( \rho  +p   \right) {R}^{2}
 \left[ \frac{1}{16}\, M^2+ \left( \frac{1}{2}\,
\pi \,p  {R}^{3}-\frac{1}{32}\,R \right) M  +{\pi }^{2} p^2{R}^{6}+{\frac {1}{64
}}\,{R}^{2} \right]  \left( \omega'_{11}   \right) ^{2} \right. \right. \nn \\
& & \left. \left. -256\, \left\{ \frac{1}{32}\, \left( R-2\,M   \right) R \left( \pi \,p  {R}^{3}+\frac{1}{4}\,M
   \right)  \left[ \frac{1}{16}\, M^{2}+ \left( \frac{1}{2}\,\pi \,p  {R}^{3}-\frac{3}{16}\,R
 \right) M  +{\pi }^{2} p^2{R}^{6}+{\frac {3}{32}}\,{R}^{2} \right] \rho' \right. \right. \right. \nn \\
& & \left. \left. \left. + \left[  \left( -{\frac {3}{2048}}+{\frac {1}{
256}}\,\pi \,\rho  {R}^{2}+{\frac {1}{256}}\,\pi \,p
  {R}^{2} \right)  M^{4}+\frac{1}{16}\,R \left[ {R}^{4}{\pi }^{2} p^2+{R}^{2}\pi \, \left( \pi \,\rho  {R}^{2}
-\frac{3}{4} \right) p  -\frac{3}{8}\,\pi \,\rho  {R}^
{2}+{\frac {9}{128}} \right]  M^3 \right. \right. \right. \right. \nn \\
& & \left. \left. \left. \left. +
\frac{3}{8}\,{R}^{2} \left[ -{\frac {3}{512}}+{R}^{6}{\pi }^{3} p^3+{R}^{4} \left( \pi \,\rho  {R}^{2}-{\frac {7}{8}} \right) {\pi }^{2} p^{2}+ \left( -\frac{1}{2}\,{R}^{4}{\pi }^{2}\rho  +{\frac {3}{32}}\,\pi \,{R}^{2} \right) p  
 \right]  M^2 \right. \right. \right. \right. \nn \\
& & \left. \left. \left. \left. +\pi \,{R}^{5}
 \left[ {\pi }^{3}{R}^{6} p^4+{
\pi }^{2}{R}^{4} \left( \pi \,\rho  {R}^{2}-\frac{3}{4}
 \right)  p^3+ \left( {\frac {21}{
128}}\,\pi \,{R}^{2}-\frac{3}{8}\,{R}^{4}{\pi }^{2}\rho  
 \right)  p^2+ \left( {\frac {3}{
32}}\,\pi \,\rho  {R}^{2}-{\frac {3}{512}} \right) p
  +{\frac {3}{256}}\,\rho   \right] M \right. \right. \right. \right. \nn \\
& & \left. \left. \left. \left.
  + \left[ {R}^{8}{\pi }^{4} p^{5}+ \left( \pi \,\rho  {R}^{2}-\frac{3}{8}
 \right) {\pi }^{3}{R}^{6} p^4+\frac{3}{16}\,{R}^{4}{\pi }^{2} p^3+ \left( 
-{\frac {9}{256}}\,\pi \,{R}^{2}+\frac{3}{16}\,{R}^{4}{\pi }^{2}\rho   \right)  p^2-{\frac {3}{
1024}}\,p \right. \right. \right. \right. \right. \nn \\
& & \left. \left. \left. \left. \left. -{\frac {3}{1024}}\,\rho  
 \right] \pi \,{R}^{6} \right]  \left( \rho  +p
   \right)  \right\}  \left( \omega_{11}   \right) ^{4}\pi  \right\} {{\rm e}^{-2\,\nu  
}}+6912\,{R}^{4} \left( \pi \,p  {R}^{3}+\frac{1}{4}\,M
   \right)  \left( h_{22}   \right) 
^{2}\pi \, \left( R-2\,M   \right) ^{2} \rho' \right. \nn \\
& & \left. +193536\, \left( R-2\,M  
 \right) R \left[ \frac{1}{16}\, M^2+
 \left( \frac{1}{2}\,\pi \,p  {R}^{3}+{\frac {1}{56}}\,R
 \right) M  +{\pi }^{2} p^2{R}^{6}-{\frac {1}{112}}\,{R}^{2} \right]  \left( k_{22}   \right) ^{2} \right. \nn \\
& & \left. +55296\, \left( R-2\,M  \right) h_{22}  R \left[ \frac{1}{8}\, M^{2}+ \left( {\frac {3}{32}}\,R-\frac{1}{4}\,{R}^{3
}\pi \,\rho   \right) M  +{R}^{2}
 \left( {R}^{4}{\pi }^{2} p^2+\frac{1}{8}
\,\pi \,\rho  {R}^{2}+\frac{1}{4}\,\pi \,p  {R
}^{2}-\frac{1}{16} \right)  \right] k_{22} \right. \nn \\
& & \left. -147456\,
 \left( h_{22}   \right) ^{2} \left[ -{\frac {5}{256
}}\, M^4+ \left( -{\frac {55}{128}
}\,\pi \,p  {R}^{3}-{\frac {15}{128}}\,{R}^{3}\pi \,
\rho  +{\frac {33}{256}}\,R \right)  M^3 \right. \right. \nn \\
& & \left. \left. +\frac{3}{16}\,{R}^{2} \left[ 2\,{R}^{4}{\pi }^{2}
 p^2+ \left( 3\,{R}^{4}{\pi }^{2}
\rho  +{\frac {49}{16}}\,\pi \,{R}^{2} \right) p
  -\frac{3}{16}-{\frac {7}{16}}\,\pi \,\rho  {R}^{2}+{R}^{4}{\pi }^{2} \rho^2
 \right]  M^2 \right. \right. \nn \\
& & \left. \left. -\frac{3}{8}\,{R}^{3} \left[ -\frac{5}{3}\,{R}^{6}{\pi }^{3} p^3+{R}^{4
}{\pi }^{2} \left( \pi \,\rho  {R}^{2}-\frac{9}{4} \right) 
 p^2+ \left( \frac{7}{4}\,{R}^{4}{\pi }^{2
}\rho  -\frac{1}{16}\,\pi \,{R}^{2} \right) p  +\frac{1}{2}\,{R}^{4}{\pi }^{2} \rho^2-{\frac {7}{16}}\,\pi \,\rho  {R}^{2}+{\frac {3}{32}} \right] M \right. \right. \nn \\
& & \left. \left. +{R}^{4} \left[ {R}^{8}{\pi 
}^{4} p^4+\frac{3}{16}\,{R}^{6}{\pi }^{3}
 p^3+ \left( -{\frac {27}{64}}\,{R
}^{4}{\pi }^{2}+\frac{3}{16}\,{R}^{6}{\pi }^{3}\rho   \right) 
 p^2+ \left( -{\frac {3}{32}}\,
\pi \,{R}^{2}+\frac{3}{16}\,{R}^{4}{\pi }^{2}\rho   \right) p
  +{\frac {3}{64}}\, \left( \pi \,\rho  {R}^{2}-\frac{1}{2} \right) ^{2} \right]  \right]  \right\}\,, \nn \\
\\
S_{v 44} &=& {\frac {1}{420 {R} \left( R-2\,M   \right) \left( 4 \pi \,p  {R}^{3}+M  
 \right)}}\, \left\{ 768\, \left\{ -\frac{5}{4}\, \left( R-2\,M   \right)  \left( \omega'_{11}   \right)  \left( \pi \,p  {R}^{3}+\frac{1}{4}\,R-\frac{1}{4}
\,M   \right)  \left( \pi \,p  {R}^{3}
+\frac{1}{4}\,M   \right) \omega'_{33} \right. \right. \nn \\
& & \left. \left. + \left[  \left[ -\frac{1}{4}\, M^2+ \left( \frac{1}{16}\,R-\frac{3}{4}\,\pi \,p  {R}^{3} \right) M
  +{R}^{2} \left( {R}^{4}{\pi }^{2} p^2+{\frac {3}{64}}+\frac{5}{8}\,\pi \,p  {
R}^{2} \right)  \right] k_{22} \right. \right. \right. \nn \\
& & \left. \left. \left. +\frac{1}{2}\, \left( {\pi }
^{2} p^2{R}^{6}+\frac{1}{4}\,\pi \,{R}^{4}
p  -{\frac {3}{64}}\,{R}^{2}+{\frac {5}{32}}\,RM
  -\frac{1}{16}\, M^2
 \right) h_{22}   \right]  \left( R-2\,M   \right)  \left( \omega'_{11}   \right) ^{2} \right. \right. \nn \\
& & \left. \left. +{\frac {15}{16}}\, \left( R-2\,M   \right) \omega_{33}   \left( \pi \,p
  {R}^{3}+\frac{1}{4}\,M   \right) \omega'_{11}  +16\,R\pi \, \left\{ \frac{1}{32}\, \left( R-
2\,M   \right) R \left( \pi \,p  {R}^{
3}+\frac{1}{4}\,R-\frac{1}{4}\,M   \right) h_{22}  \omega_{11}  \rho' \right. \right. \right. \nn \\
& & \left. \left. \left. +
 \left[  \left[ -\frac{1}{4}\, M^2+
 \left( -\frac{3}{4}\,\pi \,p  {R}^{3}+\frac{1}{4}\,R \right) M
  +{R}^{2} \left( {R}^{4}{\pi }^{2} p^2-{\frac {3}{64}}+\frac{5}{8}\,\pi \,p  {
R}^{2} \right)  \right] \omega_{11}  k_{22} \right. \right. \right. \right. \nn \\
& & \left. \left. \left. \left.  -\frac{5}{4}\, \left( \pi \,p  {R}^{3}+\frac{1}{4}\,R-\frac{1}{4}\,M
   \right)  \left( \pi \,p  {R}^{3}+\frac{1}{4}\,M   \right) \omega_{33} \right. \right. \right. \right. \nn \\
& & \left. \left. \left. \left. +\frac{1}{8}\,
 \left[ -\frac{1}{16}\, M^2-{\frac {5}{16}
}\,RM  +{R}^{2} \left( \frac{3}{16}+\frac{1}{4}\,\pi \,p  {R}^{2}+{R}^{4}{\pi }^{2} p^{2} \right)  \right] h_{22}  \omega_{11}   \right]  \left( \rho  +p  
 \right)  \right\} \omega_{11}   \right\} {R}^{3}{
{\rm e}^{-\nu  }} \right. \nn \\
& & \left. -16\,{R}^{6} \left\{  \left( R-2\,M
   \right) ^{2} \left( \pi \,p  {R}^{3
}+\frac{1}{4}\,M   \right)  \left( \omega'_{11}
   \right) ^{4}  -256\, \left[ \frac{1}{32}\,R \left( R-2\,M
   \right) \rho' \right. \right. \right. \nn \\
& & \left. \left. \left. +
 \left( \pi \,p  {R}^{3}+\frac{1}{4}\,M  
 \right)  \left( \pi \,\rho  {R}^{2}+\pi \,p  {R}^{2}-\frac{3}{8} \right)  \left( \rho  +p   \right)  \right] \pi \, \left( \omega_{11}   \right) ^{4} \right\}  \left( \pi \,p  {R}^{3
}+\frac{1}{4}\,R-\frac{1}{4}\,M   \right) {{\rm e}^{-2\,\nu  }} \right. \nn \\
& & \left. +864\,R \left( R-2\,M   \right)  \left( k_{22}   \right) ^{2}-1728\, \left( R-2\,M   \right) h_{22}   \left( \pi \,p  {R}^{3}+{R}^{3}\pi \,\rho  -\frac{1}{4}\,R-\frac{1}{2}\,M
   \right) k_{22} \right. \nn \\
& & \left. -2304\, \left( h_{22}   \right) ^{2} \left[ -{\frac {7}{16}}\,
 M^2+ \left( -\frac{1}{8}\,R+\frac{3}{4}\,{R}^{3}
\pi \,\rho  +\frac{3}{4}\,\pi \,p  {R}^{3}
 \right) M  +{R}^{2} \left( \frac{3}{16}-\frac{1}{8}\,\pi \,p  {R}^{2}+{R}^{4}{\pi }^{2} p^2-\frac{3}{8}\,\pi \,\rho  {R}^{2} \right)  \right] 
 \right\}\,.
\ea
\ew
%
%

\section{Exterior Solutions to Cubic Order in Spin}
\label{app:ext-sol-O3}
 The exterior solutions for $\nu$, $M$, $\omega_{11}$, $h_{20}$, $h_{22}$, $k_{22}$, $\omega_{31}$ and $\omega_{33}$ are given in~\cite{hartle1967,benhar}. We present them again here for completeness
\ba
\nu^\ext &=& \ln \left( 1 - \frac{2 M_*}{R}  \right)\,, 
\qquad
M^\ext(R) = M_*\,, \\
\omega_{11}^\ext (R) &=& \Omega - \frac{2J}{R^3}\,, \\
h_{20}^\ext(R) &=& -\frac{C_{20}^\ext}{R-2 M_*} + \frac{J^2}{R^3 (R-2 M_*)}\,, \\ 
h_{22}^\ext(R) &=& \left( 1+ \frac{M_*}{R} \right) \frac{J^2}{M_* R^3}+C_{22}^\ext Q_2^2\left( \frac{R}{M_*}-1 \right)\,, \nn \\
\\
k_{22}^\ext(R) &=& - \frac{J^2}{R^4} + \frac{2 C_{22}^\ext M_*}{R (R-2 M_*)}Q_2^1\left( \frac{R}{M_*}-1 \right) - h_{22}^\ext(R)\,, \nn \\
\\
\omega_{31}^\ext (R) &=& \omega_{31}^{\ext, \Part} + C_{31}^\ext \omega_{31}^{\ext, \Hom}\,,  \\ 
\omega_{33}^\ext (R) &=& \omega_{33}^{\ext, \Part} + C_{33}^\ext \omega_{33}^{\ext, \Hom}\,,   
\ea
where $Q_\ell^m$ is the associated Legendre function of the second kind. The functions $\omega_{31}^{\ext, \Hom}$ and $\omega_{31}^{\ext, \Part}$ are the homogeneous and particular solutions to the equations of structure for $\omega_{31}$ in the exterior:
\ba
\omega_{31}^{\ext, \Hom} &=& \frac{2 M_*^2}{R^3}\,, \\
\omega_{31}^{\ext, \Part} &=& - \frac{3 J C_{22}^\ext (8 M_*^3 -12 M_* R^2 +5 R^3)}{5 (R^3 M_*^3)} \ln \left( 1- \frac{2 M_*}{R} \right) \nn \\
& & - \frac{2 J}{5 M_*^2 R^7} ( 6 J^2 M_*^2 + 2 J^2 M_* R - 21 C_{22}^\ext M_* R^5 \nn \\
& &  - 16 C_{22}^\ext M_*^2 R^4 + 15 C_{22}^\ext R^6 + 6 C_{22}^\ext M_*^3 R^3 )\,. 
\ea
Similarly, $\omega_{33}^{\ext, \Hom}$ and $\omega_{33}^{\ext, \Part}$ are the homogeneous and particular solutions for $\omega_{33}$ in the exterior: 
\allowdisplaybreaks
\bw
\ba
\omega_{33}^{\ext, \Hom} &=& \frac{7}{96 M_*^3 R^3} \left[ (120 M_*^2 -150 M_* R^4 +45 R^5) R^3 \ln \left(1-\frac{2M_*}{R} \right) + 8 M_*^5 + 20 M_*^4 R+60 M_*^3 R^2 - 210 M_*^2 R^3 + 90 M_* R^4 \right]\,, \nn \\
\\	
\omega_{33}^{\ext, \Part} &=& \frac{J}{480 R^7 M_*^9} \left\{ 2 M_* J^2 (576 M_*^8+32 M_*^7 R -160 M_*^6 R^2 +140 M_*^4  R^4 + 350 M_*^3 R^5 + 1050 M_*^2  R^6 -3675 M_* R^7 +1575 R^8) \right. \nn \\
& & \left. + 32 C_{22}^\ext M_*^5  R^3 (36 M_*^5 +52 M_*^4 R +34 M_*^3  R^2 +210 M_*^2 R^3 -735 M_*  R^4 + 315 R^5) \right. \nn \\
& & \left. + \left[ 525 R^7 J^2 (8 M_*^2-10 M_*  R +3 R^2) + 48 C_{22}^\ext M_*^4 R^4 (8 M_*^5 + 20 M_*^4 R \right. \right. \nn \\
& & \left. \left.  - 12 M_*^3 R^2 + 280 M_*^2 R^3 - 350 M_* R^4 + 105 R^5) \right] \ln \left(1 - \frac{2M_*}{R} \right) \right\}\,.
\ea
\ew
The quantities $M_*$, $J$, $C_{20}^\ext$, $C_{22}^\ext$, $C_{31}^\ext$ and $C_{33}^\ext$ are integration constants that are to be determined by matching the exterior to the interior solution at the stellar surface. In particular, $M_*$ and $J$ correspond to the stellar mass and (the magnitude of) the spin angular momentum in the slow-rotation limit, respectively.

\section{Exterior Solutions at Quartic Order in Spin}
\label{app:ext-sol-O4}

The homogeneous and particular solutions for $h_{42}$, $v_{42}$, $h_{44}$ and $v_{44}$ are given by
\bw
\allowdisplaybreaks
\ba
h_{42}^{\ext, \Hom} &=& \frac{1050 M_*^{11} R^8 + 700 M_*^{12} R^7 - 525 M_*^8 R^{11} + 2625 M_*^9 R^{10} - 3850 M_*^{10} R^9}{280 M_*^9 (R-2 M_*)^2 R^8} \nn \\
& &  -\frac{420 M_*^{10} R^5 - 420 M_*^9 R^6 + 105 M_*^8 R^7}{112 R^4 (R-2 M_*) M_*^{10}} \ln\left( 1-\frac{2 M_*}{R} \right)\,, \\
h_{42}^{\ext, \Part} &=& -{\frac {105}{4 {{\it M_*}}^{9} \left( 
R-2\,{\it M_*} \right) ^{2}{R}^{8}}}\, \left\{ {J}^{4}{R}^{11}-{\frac {44}{7}}\,{J}^{4}{
\it M_*}\,{R}^{10}+{\frac {575}{42}}\,{J}^{4}{{\it M_*}}^{2}{R}^{9}-{
\frac {157}{14}}\,{J}^{4}{{\it M_*}}^{3}{R}^{8} \right. \nn \\
& & \left. +{\frac {176}{105}}\,{J}
^{2} \left[  \left( {\frac {15}{352}}\,C_{20}^\ext+{\frac {183}{88}}\,C_{22}^\ext \right) {R}^{4}+{J}^{2} \right] {{\it M_*}}^{4}{R}^{7}+{\frac {
47}{105}}\,{J}^{2} \left[  \left( -{\frac {75}{94}}\,C_{20}^\ext-{\frac {
2262}{47}}\,C_{22}^\ext \right) {R}^{4}+{J}^{2} \right] {{\it M_*}}^{5}{R}
^{6} \right. \nn \\
& & \left. +{\frac {4}{21}}\,J \left[  \left( -\frac{3}{4}\,C_{31}^\ext+{\frac {21}{4}}
\,C_{33}^\ext \right) {R}^{6}+\frac{11}{4}\, \left( C_{20}^\ext+{\frac {964}{11}}\,C_{22}^\ext \right) J{R}^{4}+{J}^{3} \right] {{\it M_*}}^{6}{R}^{5} \right. \nn \\
& & \left. +{\frac 
{4}{105}}\, \left[ -\frac{4}{7}\,(C_{22}^\ext)^2{R}^{2}+J \left( C_{33}^\ext-8\,C_{31}^\ext \right)  \right] {{\it M_*}}^{13}{R}^{4} \right. \nn \\
& & \left. +{\frac {4}{49}}\,
 \left[ -\frac{6}{5}\,(C_{22}^\ext)^2{R}^{8}-77\,J \left( -{\frac {5}{44}}\,C_{31}^\ext +C_{33}^\ext \right) {R}^{6}-\frac{7}{4}\, \left( C_{20}^\ext+{\frac {8916}{
35}}\,C_{22}^\ext \right) {J}^{2}{R}^{4}+{J}^{4} \right] {{\it M_*}}^{7}{R
}^{4} \right. \nn \\
& & \left. +{\frac {32}{147}}\, \left[ {\frac {7}{10}}\,C_{22}^\ext\, \left( -{
\frac {93}{14}}\,C_{22}^\ext+C_{20}^\ext \right) {R}^{4}+{\frac {7}{8}}\,J C_{33}^\ext \,{R}^{2}+{J}^{2}C_{22}^\ext \right] {{\it M_*}}^{12}{R}^{3} \right. \nn \\
& & \left. -{
\frac {20}{147}}\, \left[ {\frac {21}{5}}\,C_{22}^\ext\, \left( -{\frac {
117}{70}}\,C_{22}^\ext+C_{20}^\ext \right) {R}^{8}-{\frac {805}{8}}\,
 \left( -{\frac {44}{575}}\,C_{31}^\ext+C_{33}^\ext \right) J{R}^{6}+{
\frac {21}{50}}\, \left( -{\frac {9076}{105}}\,C_{22}^\ext+C_{20}^\ext
 \right) {J}^{2}{R}^{4} \right. \right. \nn \\
& & \left. \left. +{J}^{4} \right] {{\it M_*}}^{8}{R}^{3}-{\frac {
36}{245}}\, \left[ -{\frac {49}{3}}\,C_{22}^\ext\, \left( -{\frac {365}{
294}}\,C_{22}^\ext+C_{20}^\ext \right) {R}^{8}+{\frac {5495}{72}}\, \left( C_{33}^\ext -{\frac {4}{157}}\,C_{31}^\ext \right) J{R}^{6} \right. \right. \nn \\
& & \left. \left. +{\frac {14}{27}}
\,{J}^{2} \left( C_{20}^\ext-{\frac {688}{35}}\,C_{22}^\ext \right) {R}^{4}+
{J}^{4} \right] {{\it M_*}}^{9}{R}^{2}+{\frac {272}{735}}\, \left[ -{
\frac {259}{34}}\,C_{22}^\ext\, \left( -{\frac {585}{518}}\,C_{22}^\ext+C_{20}^\ext \right) {R}^{8} \right. \right. \nn \\
& & \left. \left. +{\frac {77}{17}}\, \left( {\frac {3}{44}}\,C_{31}^\ext +C_{33}^\ext \right) J{R}^{6}+{\frac {7}{34}}\,{J}^{2} \left( 6\,
C_{22}^\ext+C_{20}^\ext \right) {R}^{4}+{J}^{4} \right] {{\it M_*}}^{10}R \right. \nn \\ 
& & \left. +
 \left[ {\frac {8}{15}}\,C_{22}^\ext\, \left( -{\frac {5}{98}}\,C_{22}^\ext+
C_{20}^\ext \right) {R}^{8}+{\frac {47}{105}}\,J \left( {\frac {24}{47}}
\,C_{31}^\ext+C_{33}^\ext \right) {R}^{6}-{\frac {16}{105}}\, \left( {
\frac {21}{5}}\,C_{22}^\ext+C_{20}^\ext \right) {J}^{2}{R}^{4}+{\frac {176}{
735}}\,{J}^{4} \right] {{\it M_*}}^{11} \right\}  \nn \\
& & -{\frac {105}{8 {R}^{4} \left( R-2\,{\it M_*} \right){{\it M_*}}^
{10}}}\, \left\{ \frac{1}{14}\,{J}^{2} \left( C_{20}^\ext+{\frac {244}{
5}}\,C_{22}^\ext \right) {{\it M_*}}^{4}{R}^{7}+{J}^{4}{R}^{7}-\frac{2}{7}\,
 \left( C_{20}^\ext+{\frac {316}{5}}\,C_{22}^\ext \right) {J}^{2}{{\it M_*}}^
{5}{R}^{6} \right. \nn \\
& & \left. -{\frac {37}{7}}\,{\it M_*}\,{J}^{4}{R}^{6}+\frac{2}{7}\,J \left[ 
 \left( \frac{7}{2}\,C_{33}^\ext-\frac{1}{2}\,C_{31}^\ext \right) {R}^{2}+ \left( {\frac {
528}{5}}\,C_{22}^\ext+C_{20}^\ext \right) J \right] {{\it M_*}}^{6}{R}^{5}+{
\frac {127}{14}}\,{J}^{4}{{\it M_*}}^{2}{R}^{5} \right. \nn \\
& & \left. -{\frac {3904}{245}}\,
 \left[ {\frac {3}{244}}\,(C_{22}^\ext)^2{R}^{4}+{\frac {1295}{3904}}\,
J \left( -{\frac {4}{37}}\,C_{31}^\ext+C_{33}^\ext \right) {R}^{2}+{J}^{2}C_{22}^\ext \right] {{\it M_*}}^{7}{R}^{4}-5\,{J}^{4}{{\it M_*}}^{3}{R}^{4}-
{\frac {64}{245}}\,(C_{22}^\ext)^2{{\it M_*}}^{12}{R}^{3} \right. \nn \\
& & \left. -{\frac {32}{
245}}\, \left[ {\frac {35}{8}}\,C_{22}^\ext\, \left( -\frac{3}{2}\,C_{22}^\ext+C_{20}^\ext \right) {R}^{4}-{\frac {4445}{64}}\, \left( C_{33}^\ext-{\frac {8}{
127}}\,C_{31}^\ext \right) J{R}^{2}+{J}^{2}C_{22}^\ext \right] {{\it M_*}}^{
8}{R}^{3} \right. \nn \\
& & \left. +{\frac {32}{245}}\, \left[ 14\,C_{22}^\ext\, \left( -{\frac {29
}{56}}\,C_{22}^\ext+C_{20}^\ext \right) {R}^{4}-{\frac {1225}{32}}\,JC_{33}^\ext\,{R}^{2}+{J}^{2}C_{22}^\ext \right] {{\it M_*}}^{9}{R}^{2}\right. \nn \\
& & \left. +{\frac {16}{
245}}\,C_{22}^\ext\, \left[  \left( -\frac{13}{2}\,C_{22}^\ext-21\,C_{20}^\ext \right) 
{R}^{4}+{J}^{2} \right] {{\it M_*}}^{10}R+ \left[ {\frac {32}{35}}\,(C_{22}^\ext)^{2}{R}^{4}-{\frac {32}{49}}\,{J}^{2}C_{22}^\ext \right] {{\it M_*
}}^{11} \right\} \ln \left( 1- \frac{2M_*}{R} \right) \nn \\
& & + \frac{9 (C_{22}^\ext)^2 R (R^2+3 M_* R - 5 M_*^2) (R-2 M_*)}{14 M_*^4} \left[ \ln \left( 1- \frac{2M_*}{R} \right) \right]^2\,, \\
v_{42}^{\ext, \Hom} &=& \frac{1050 M_*^8 R^{10} - 4200 M_*^9 R^9 + 4550 M_*^{10} R^8 - 700 M_*^{11} R^7}{280 M_*^8 (R-2 M_*)^2 R^8}  +\frac{420 M_*^{10} R^4 - 630 M_*^9 R^5 + 210 R^6 M_*^8}{112 R^4 (R-2 M_*) M_*^9} \ln \left( 1- \frac{2M_*}{R} \right)\,, \\
v_{42}^{\ext, \Part} &=& \frac{15}{2 {{\it M_*}}^{8} \left( R-2\,{\it M_*} \right) ^{2}{R}^{8}}\, \left\{ {J}^{4}{R}^{10}-{\frac {17}{4}}\,{\it M_*}\,{J}^{4}{R}^{9
}+{\frac {61}{12}}\,{{\it M_*}}^{2}{J}^{4}{R}^{8}-{{\it M_*}}^{3}{J}^{4}
{R}^{7} \right. \nn \\
& & \left. -\frac{3}{10}\, \left[  \left( -\frac{5}{3}\,C_{20}^\ext-{\frac {52}{3}}\,C_{22}^\ext
 \right) {R}^{4}+{J}^{2} \right] {J}^{2}{{\it M_*}}^{4}{R}^{6}-\frac{2}{15}\,{J
}^{2} \left[  \left( 15\,C_{20}^\ext+162\,C_{22}^\ext \right) {R}^{4}+{J}^{2
} \right] {{\it M_*}}^{5}{R}^{5} \right. \nn \\
& & \left. +{\frac {2}{105}}\, \left[ {\frac {27}{
2}}\,(C_{22}^\ext)^2{R}^{8}+{\frac {105}{2}}\,J \left( C_{33}^\ext- C_{31}^\ext \right) {R}^{6}+{\frac {455}{4}}\, \left( C_{20}^\ext+{\frac {1004}{
91}}\,C_{22}^\ext \right) {J}^{2}{R}^{4}+{J}^{4} \right] {{\it M_*}}^{6}{R
}^{4} \right. \nn \\
& & \left. +\frac{2}{15}\, \left[ {\frac {44}{7}}\,(C_{22}^\ext)^2{R}^{2}+J \left( -8
\,C_{31}^\ext+C_{33}^\ext \right)  \right] {{\it M_*}}^{12}{R}^{4} \right. \nn \\
& & \left. +{\frac {
272}{105}}\, \left[ {\frac {21}{34}}\,C_{22}^\ext\, \left( -{\frac {131}{
42}}\,C_{22}^\ext+C_{20}^\ext \right) {R}^{4}-{\frac {7}{136}}\,J \left( -8
\,C_{31}^\ext+C_{33}^\ext \right) {R}^{2}+{J}^{2}C_{22}^\ext \right] {{\it M_*}
}^{11}{R}^{3} \right. \nn \\
& & \left. +\frac{4}{7}\, \left[ -\frac{3}{10}\,(C_{22}^\ext)^2{R}^{8}-{\frac {119}{
16}}\, \left( -{\frac {16}{17}}\,C_{31}^\ext+C_{33}^\ext \right) J{R}^{6}-{
\frac {7}{12}}\, \left( {\frac {64}{7}}\,C_{22}^\ext+C_{20}^\ext \right) {J}
^{2}{R}^{4}+{J}^{4} \right] {{\it M_*}}^{7}{R}^{3} \right. \nn \\
& & \left. +{\frac {34}{105}}\,
 \left[ -{\frac {126}{17}}\, \left( -\frac{3}{4}\,C_{22}^\ext+C_{20}^\ext \right) C_{22}^\ext \,{R}^{8}+{\frac {2135}{136}}\, \left( -{\frac {52}{61}}\,C_{31}^\ext +C_{33}^\ext \right) J{R}^{6}-{\frac {14}{17}}\, \left( -{\frac {37}{
5}}\,C_{22}^\ext+C_{20}^\ext \right) {J}^{2}{R}^{4} \right. \right. \nn \\
& & \left. \left. +{J}^{4} \right] {{\it M_*
}}^{8}{R}^{2}-{\frac {304}{105}}\, \left[ -{\frac {63}{19}}\, \left( -
{\frac {121}{84}}\,C_{22}^\ext+C_{20}^\ext \right) C_{22}^\ext\,{R}^{8}+{\frac 
{105}{304}}\, \left( C_{33}^\ext-\frac{2}{3}\,C_{31}^\ext \right) J{R}^{6} \right. \right. \nn \\
& & \left. \left. -{\frac {
7}{76}}\, \left( -{\frac {66}{35}}\,C_{22}^\ext+C_{20}^\ext \right) {J}^{2}{
R}^{4}+{J}^{4} \right] {{\it M_*}}^{9}R+ \left[ -{\frac {52}{5}}\,
 \left( -{\frac {27}{14}}\,C_{22}^\ext+C_{20}^\ext \right) C_{22}^\ext\,{R}^{8}
-\frac{3}{10}\, \left( {\frac {8}{9}}\,C_{31}^\ext+C_{33}^\ext \right) J{R}^{6} \right. \right. \nn \\
& & \left. \left. +{
\frac {8}{15}}\, \left( C_{20}^\ext-{\frac {517}{35}}\,C_{22}^\ext \right) {
J}^{2}{R}^{4}+{\frac {184}{105}}\,{J}^{4} \right] {{\it M_*}}^{10}  \right\} \nn \\
& & + {\frac {15}{4 {R}^{4} \left( R-2\,{\it M_*} \right) {{\it M_*}}^{9} }}\, \left\{ {J}^{4}{R}^{6}+\frac{1}{2}\, \left( {\frac {52}{5}}\,
C_{22}^\ext+C_{20}^\ext \right) {J}^{2}{{\it M_*}}^{4}{R}^{6}-{\frac {13}{4}}
\,{J}^{4}{\it M_*}\,{R}^{5}  -\frac{3}{2}\, \left( C_{20}^\ext+{\frac {164}{15}}\, C_{22}^\ext \right) {J}^{2}{{\it M_*}}^{5}{R}^{5} \right. \nn \\
&  &\left. +\frac{5}{2}\,{J}^{4}{{\it M_*}}^{2
}{R}^{4}+ \left[ {\frac {18}{35}}\,(C_{22}^\ext)^2{R}^{4}+J \left( C_{33}^\ext-C_{31}^\ext \right) {R}^{2}+ \left( {\frac {384}{35}}\,C_{22}^\ext+
C_{20}^\ext \right) {J}^{2} \right] {{\it M_*}}^{6}{R}^{4}-{\frac {96}{35}
}\,{{\it M_*}}^{11}(C_{22}^\ext)^2{R}^{3} \right. \nn \\
& & \left. +{\frac {16}{35}}\, \left[ \frac{3}{8}
\,(C_{22}^\ext)^2{R}^{4}-{\frac {455}{64}}\, \left( C_{33}^\ext-{\frac {12
}{13}}\,C_{31}^\ext \right) J{R}^{2}+{J}^{2}C_{22}^\ext \right] {{\it M_*}}^
{7}{R}^{3} \right. \nn \\
& & \left. +{\frac {104}{35}}\, \left[ -{\frac {21}{26}}\, \left( {
\frac {11}{14}}\,C_{22}^\ext+C_{20}^\ext \right) C_{22}^\ext\,{R}^{4}+{\frac {
175}{208}}\, \left( -\frac{4}{5}\,C_{31}^\ext+C_{33}^\ext \right) J{R}^{2}+{J}^{2} C_{22}^\ext \right] {{\it M_*}}^{8}{R}^{2} \right. \nn \\
& & \left. +{\frac {88}{35}}\, \left[ 
 \left( -{\frac {89}{44}}\,C_{22}^\ext+{\frac {63}{22}}\,C_{20}^\ext
 \right) {R}^{4}+{J}^{2} \right] C_{22}^\ext\,{{\it M_*}}^{9}R-{\frac {144
}{35}}\, \left[  \left( -{\frac {29}{12}}\,C_{22}^\ext+\frac{7}{6}\,C_{20}^\ext
 \right) {R}^{4}+{J}^{2} \right] C_{22}^\ext\,{{\it M_*}}^{10} \right\} \nn \\
& & + \frac{9 (C_{22}^\ext)^2 (3 R^4 + 4 M_* R^3 - 40 M_*^2 R^2 + 40 M_*^3 R + 16 M_*^4)}{56 M_*^4} \left[ \ln \left( 1- \frac{2M_*}{R} \right) \right]^2\,, \\
h_{44}^{\ext, \Hom} &=& \frac{3}{64 M_*^{4} R (R-2 M_*)} \left[105 R^2 (R-2 M_*)^2 (6 M_*^2-14 M_* R+ 7R^2) \ln\left(1-\frac{2 M_*}{R} \right) \right. \nn \\
& & \left. +56 M_*^6+504 M_*^5 R-6720 M_*^4 R^2+12040 M_*^3 R^3-7350 M_*^2 R^4 +1470 M_* R^5  \right]\,, \\
h_{44}^{\ext, \Part} &=& \frac{1}{560 (R-2 M_*)^2 R^8 M_*^{11}} \left[ - 2 J^4 (R- 2 M_*) (768 M_*^{12} + 960 M_*^{11}  R -1264 M_*^{10} R^2 - 264  M_*^{9} R^3 +420 M_*^8 R^4 +770 M_*^7 R^5 \right. \nn \\
& & \left. +1050 M_*^6  R^6 + 791 M_*^5 R^7 -19131 M_*^4 R^8 +180180 M_*^3 R^9 - 319060 M_*^2 R^{10} +194775 M_* R^{11} - 38955 R^{12})  \right. \nn \\
& & \left. - 2 C_{22}^\ext J^2 R^3 (R-2 M_*) M_*^4 (1248 M_*^9 +4544 M_*^8 R +4624 M_*^7 R^2 -528 M_*^6 R^3 - 2292 M_*^5  R^4 - 112404 M_*^4 R^5 \right. \nn \\
& & \left. +1259040 M_*^3 R^6 - 2243740 M_*^2 R^7 + 1369725 M_* R^8 - 273945 R^9) \right. \nn \\
& & \left. + 8 (C_{22}^\ext{})^2 M_*^8 R^6 (240 M_*^7-1872 M_*^6 R -24456 M_*^5 R^2+ 396432 M_*^4 R^3 - 893400 M_*^3 R^4 \right. \nn \\
& & \left.  + 781662 M_*^2 R^5 - 301761 M_* R^6 + 43155 R^7) \right. \nn \\
& & \left. -35 C_{33}^\ext J R^4 (R-2 M_*) M_*^6 (8 M_*^8 + 44 M_*^7 R +60 M_*^6 R^2 +70 M_*^5 R^3 -870 M_*^4 R^4 +7320 M_*^3 R^5 \right. \nn \\
& & \left. -12900 M_*^2 R^6 +7875 M_* R^7 -1575 R^8) \right] \nn \\
& & + \frac{1}{1120 (R-2 M_*) R^4 M_*^{12}} \left[ 210 J^4 (R-2 M_*) R^4 (35 M_*^4-656 M_*^3 R + 1802 M_*^2 R^2-1484 M_* R^3 +371 R^4) \right. \nn \\
& & \left. - 6 C_{22}^\ext J^2 M_*^4 (R-2 M_*) (320 M_*^8 - 96 M_*^7 R - 1008 M_*^6 R^2 + 288 M_*^5  R^3 -3920 M_*^4 R^4 + 158780 M_*^3 R^5 \right. \nn \\
& & \left. - 443530 M_*^2 R^6 + 365260 M_* R^7 - 91315 R^8) \right. \nn \\
& & \left. + 8 (C_{22}^\ext{})^2 R^3 M_*^8 (528 M_*^6 + 1008 M_*^5 R + 134496 M_*^4 R^2 - 477216 M_*^3 R^3 + 548802 M_*^2 R^4 - 258282 M_* R^5 + 43155 R^6) \right. \nn \\
& & \left. + 525 C_{33}^\ext J R^4 (R-2 M_*) M_*^6 (14 M_*^4-188 M_*^3 R + 510 M_*^2  R^2 -420 M_* R^3 + 105 R^4) \right] \ln \left( 1- \frac{2M_*}{R} \right) \nn \\
& & + \frac{27 (C_{22}^\ext)^2 (R-2 M_*) R (13 M_*^2-19 R M_*+3R^2)}{70M_*^4} \left[ \ln \left( 1- \frac{2M_*}{R} \right) \right]^2\,, \\
v_{44}^{\ext, \Hom} &=& - \frac{1}{32  M_*^3 R (R-2 M_*)} \left[ 1470 M_* R^4 - 5880 M_*^2 R^3 + 7210 M_*^3 R^2 - 2660 M_*^4 R  + 84 M_*^5 \right. \nn \\
& & \left. + 105 R (7R^2 - 14 M_* R+ 4 M_*^2) (R-M_*) (R-2 M_*) \ln\left( 1-\frac{2 M_*}{R} \right) \right]\,, \\
v_{44}^{\ext, \Part} &=& - \frac{1}{840 (R-2 M_*)^2 R^8 M_*^{10}} \left[ J^4 (R-2 M_*) (77910 R^{11} - 311640 M_* R^{10} + 377405 M_*^2 R^9 - 129430 M_*^3 R^8 + 777 M_*^4 R^7 \right. \nn \\
& & \left.  - 1400 M_*^5 R^6 - 770 M_*^6 R^5 - 420 M_*^7 R^4 - 1056 M_*^8 R^3 - 192 M_*^9 R^2 + 3168 M_*^{10} R + 2880 M_*^{11})  \right. \nn \\
& & \left. + 2 C_{22}^\ext J^2 (R-2 M_*) M_*^4 R^3 (273945 R^8 - 1095780 M_* R^7 + 1336075 M_*^2 R^6 - 477230 M_*^3 R^5+ 12582 M_*^4 R^4 \right. \nn \\
& & \left. - 3536 M_*^5 R^3 - 6680 M_*^6 R^2 - 8112 M_*^7 R + 3312 M_*^8) \right. \nn \\
& & \left. + 24 (C_{22}^\ext)^2 R^6 M_*^8 (14169 R^6 - 85662 M_* R^5 + 185005 M_*^2 R^4 - 165268 M_*^3 R^3 + 49180 M_*^4 R^2 - 312M_*^6) \right. \nn \\
& & \left. + 35 C_{33}^\ext J (R-2 M_*) M_*^6 R^4 (1575 R^7 - 6300 R^6 M_* + 7590 R^5 M_*^2 \right. \nn \\
& & \left. - 2520 R^4 M_*^3 - 15 R^3 M_*^4 - 40 R^2 M_*^5 - 22 R M_*^6 - 12 M_*^7)  \right] \nn \\
& & + \frac{1}{560 R^4 M_*^{11} (R-2 M_*)} \left[ 35 J^4 (R-2 M_*) R^4  (359 M_*^3 - 1863 R M_*^2 + 2226 R^2 M_* - 742 R^3) \right. \nn \\
& & \left. +2 C_{22}^\ext J^2 M_*^4 (R-2 M_*) (2208 M_*^7 + 688 R M_*^6 - 504 R^2 M_*^5 - 936 R^3 M_*^4 + 48540 R^4 M_*^3 \right. \nn \\
& & \left. - 232290 R^5 M_*^2 + 273945 R^6 M_* - 91315 R^7) \right. \nn \\
& & \left. - 8 (C_{22}^\ext)^2 R^3 M_*^8 (1128 M_*^5 + 11784 M_*^4 R - 78852 M_*^3 R^2 + 122616 M_*^2 R^3 - 71061 M_* R^4 + 13953 R^5) \right. \nn \\ 
& & \left. + 175 C_{33}^\ext J M_*^6 (R-2 M_*) R^4 (47 M_*^3 - 261 R M_*^2 + 315 R^2 M_* - 105 R^3) \right] \ln\left( 1-\frac{2 M_*}{R} \right) \nn \\
& & + \frac{3 (C_{22}^\ext)^2 (220 M_*^4 - 234 R M_*^3 + 45 R^2 M_*^2 - 36 R^3 M_* + 36 R^4)}{70 M_*^4} \left[ \ln\left( 1-\frac{2 M_*}{R} \right) \right]^2\,.
\ea
\ew
$C_{42}^\ext$ and $C_{44}^\ext$ are integration constants that are determined by matching the metric perturbations at the stellar surface. The former is related to the spin correction to the quadrupole moment while the latter is related to the hexadecapole moment.

\section{RNS field equations and sources}
\label{RNSfield}

From the Einstein field equations we have the equations for the metric functions,

\allowdisplaybreaks
\ba \Delta[\tilde{\rho} e^{\gamma/2}]&=&S_{\tilde{\rho}}(r,\mu), \\
       \left(\Delta +\frac{1}{r} \frac{\pd}{\pd r}-\frac{1}{r^2}\mu\frac{\pd}{\pd\mu}\right)\gamma e^{\gamma/2}&=&S_{\gamma}(r,\mu),\\
       \left(\Delta +\frac{2}{r} \frac{\pd}{\pd r}-\frac{2}{r^2}\mu\frac{\pd}{\pd\mu}\right)\omega e^{(\gamma-2\tilde{\rho})/2}&=&S_{\omega}(r,\mu),      
        \ea
where 
\be \Delta=\frac{\pd^2}{\pd r^2}+\frac{2}{r}\frac{\pd}{\pd r}+\frac{1}{r^2}\frac{\pd^2}{\pd\theta^2}+\frac{\cot\theta}{r^2}\frac{\pd}{\pd\theta}+\frac{1}{r^2 \sin^2\theta}\frac{\pd^2}{\pd\phi^2}, \ee
and the sources on the right hand side have the form,
\begin{widetext}
 \begin{align}
 S_{\tilde{\rho}}(r,\mu)=&e^{\gamma/2}\left[8\pi e^{2\alpha}(\rho+p)\frac{1+u^2}{1-u^2}+r^2(1-\mu^2)e^{-2\tilde{\rho}}  \left[\omega_{,r}^2+\frac{1}{r^2}(1-\mu^2)\omega_{,\mu}^2\right] 
                              +\frac{1}{r}\gamma_{,r}-\frac{1}{r^2}\mu\gamma_{,\mu} \right.\nn\\
                            &+\left.\frac{\tilde{\rho}}{2}\left\{16\pi e^{2\alpha}-\gamma_{,r}\left(\frac{1}{2}\gamma_{,r}+\frac{1}{r}\right)-\frac{1}{r^2}\gamma_{,\mu}\left[\frac{1}{2}\gamma_{,\mu}(1-\mu^2)-\mu\right]\right\}\right], \\
  S_{\gamma}(r,\mu)=&e^{\gamma/2}\left\{16\pi e^{2\alpha}p+\frac{\gamma}{2}\left[16\pi e^{2\alpha} p -\frac{1}{2}\gamma_{,r}^2-\frac{1}{2r^2}(1-\mu^2)\gamma_{,\mu}^2\right]\right\},\\
  S_{\omega}(r,\mu)=&e^{(\gamma-2\tilde{\rho})/2} \left[-16\pi e^{2\alpha}\frac{(\Omega-\omega)(\rho+p)}{1-u^2}+\omega\left\{-8\pi e^{2\alpha}\frac{\left[(1+u^2)\rho+2u^2p\right]}{1-u^2}
                                  -\frac{1}{r}\left(2\tilde{\rho}_{,r}+\frac{1}{2}\gamma_{,r}\right)   \right.\right.     \nn\\
                                  &+\frac{1}{r^2}\mu\left(2\tilde{\rho}_{,\mu}+\frac{1}{2}\gamma_{,\mu}\right)+\frac{1}{4}(4\tilde{\rho}_{,r}^2-\gamma_{,r}^2)+\frac{1}{4r^2}(1-\mu^2)(4\tilde{\rho}_{,\mu}^2-\gamma_{,\mu}^2)\nn\\
                                  &-\left.\left.r^2(1-\mu^2)e^{-2\tilde{\rho}}\left[\omega_{,r}^2+\frac{1}{r^2}(1-\mu^2)\omega_{,\mu}^2\right]\right\}\right],                            
 \end{align}
\end{widetext}
where 
\be u=(\Omega-\omega)r \sin\theta e^{-\tilde{\rho}}, \ee
is the proper velocity with respect to the zero angular momentum observers. A useful redefinition of these quantities for use within the RNS code is, $\tilde{S}_{\tilde{\rho}}=r^2 S_{\tilde{\rho}}$, $\tilde{S}_{\gamma}=r^2 S_{\gamma}$, and $\tilde{S}_{\hat{\omega}}=r_\mrm{eq} r^2 S_{\omega}$, where in the last case of the redefined source $\tilde{S}_{\hat{\omega}}$, all the $\omega$'s and $\Omega$'s in $S_{\omega}$ can be substituted by the dimensionless $\hat{\omega}=r_\mrm{eq}\omega$, and $\hat{\Omega}=r_\mrm{eq}\Omega$. We remind that $r_\mrm{eq}$ is a characteristic length scale that gives the coordinate equatorial radius of the star in quasi-isotropic coordinates.

\section{Truncation of the $q_4$ integral in RNS}
\label{q4truncation}

It is difficult to give an estimate of the accuracy with which the RNS code calculates the $q_4$ coefficient. Since though a special truncation scheme has been applied in order to calculate this coefficient, we will attempt to give a theoretical description of the sources of error and based on that an estimate of the accuracy with which the final quantity is calculated. 
 
We have argued in the main text that there is a numerical singularity in the calculation of the multipole related coefficients $q_{2\ell}$ that becomes apparent in higher order coefficients and this is one of the main sources of error in the calculation of $q_4$. The singularity arises in the integral 

\be  \int_0^1 \frac{ds' s'^{2\ell}}{(1-s')^{2\ell+2}}\int_0^1 d\mu' P_{2\ell}(\mu')\tilde{S}_{\tilde{\rho}}(s',\mu'), \ee 
where, in order for the integral to be finite, the quantity $\int_0^1 d\mu' P_{2\ell}(\mu')\tilde{S}_{\tilde{\rho}}(s',\mu'),$ must have the correct asymptotic behaviour, i.e., it should go as a power $(\frac{1-s}{s})^k$ such that the integral, 

 \be  \int_0^1 \frac{ds' s'^{2\ell}}{(1-s')^{2\ell+2}}\left(\frac{1-s'}{s'}\right)^k, \nn\ee 
is finite. This is true when $k\geq 2\ell+2$. Consequently, a fixed error contribution from the angular integration, i.e., an error without an $s$ dependence, causes a numerical singularity at $s=1$. The size of the singular contribution can be estimated as 

 \be \delta \int_0^{1-\varepsilon_1} \frac{ds' s'^{2\ell}}{(1-s')^{2\ell+2}}, \nn\ee 
where $\delta$ is the error contribution of the angular integration and $\varepsilon_1$ is the distance from the singularity at $s=1$. This integral is proportional to $\delta\varepsilon_1^{-5}$ for the $q_4$ ($\ell=2$). Assuming the error that comes only from the numerical integration of the angular integral, and taking into account that the RNS code uses Simpson's integration, then we should have 

\be \delta\varepsilon_1^{-5}\sim \delta_0\left(\frac{1}{MDIV}\right)^5 \varepsilon_1^{-5}=\delta_0\left(\frac{1}{MDIV}\right)^5 \left(\frac{\hat{n}}{SDIV}\right)^{-5}, \nn\ee
where $\hat{n}$ is the number of points in the radial domain that are omitted in the numerical integration and $\delta_0$ is the numerical error in the evaluation of the source function $\tilde{S}_{\tilde{\rho}}$. This function is comprised by numerical derivatives of the metric functions, which are evaluated using a second order accurate central difference formula, and thus its error behaves as, $\delta_0 \sim \max\left[\left(\frac{1}{MDIV}\right)^2,\left(\frac{1}{SDIV}\right)^2\right]$. Therefore, the final error expression will be of the form, 

\be \textrm{error} \sim \left(\frac{SDIV/MDIV}{\hat{n}}\right)^5\left(\frac{1}{MDIV}\right)^2\,. \label{error1}\ee

We can see from this formula that for a number of points such that $\frac{SDIV/MDIV}{\hat{n}}<1$, the singularity can be suppressed in the numerical calculation of $q_4$. 
Therefore the relative error in the calculation of $q_4$ from the singularity will be the error in \ref{error1} divided by $q_4$. This means that the only other variable, apart from $\hat{n}$, that enters the relative error is essentially the value of $q_4$ itself. Consequently the relative error will be larger for small values of $q_4$, i.e., in the case of slow rotation and for less compact objects of low density. The behaviour expected from these theoretical considerations is the behaviour we have observed with the numerical code, by changing the relative size of the grid (the ratio $SDIV/MDIV$) and/or the number $\hat{n}$ of points that are omitted. 
We should also note that increasing the order of the moment that one attempts to calculate increases significantly the number of points that one needs to omit to avoid the corresponding singularity.

\begin{figure}[h]
\includegraphics[width=.50\textwidth]{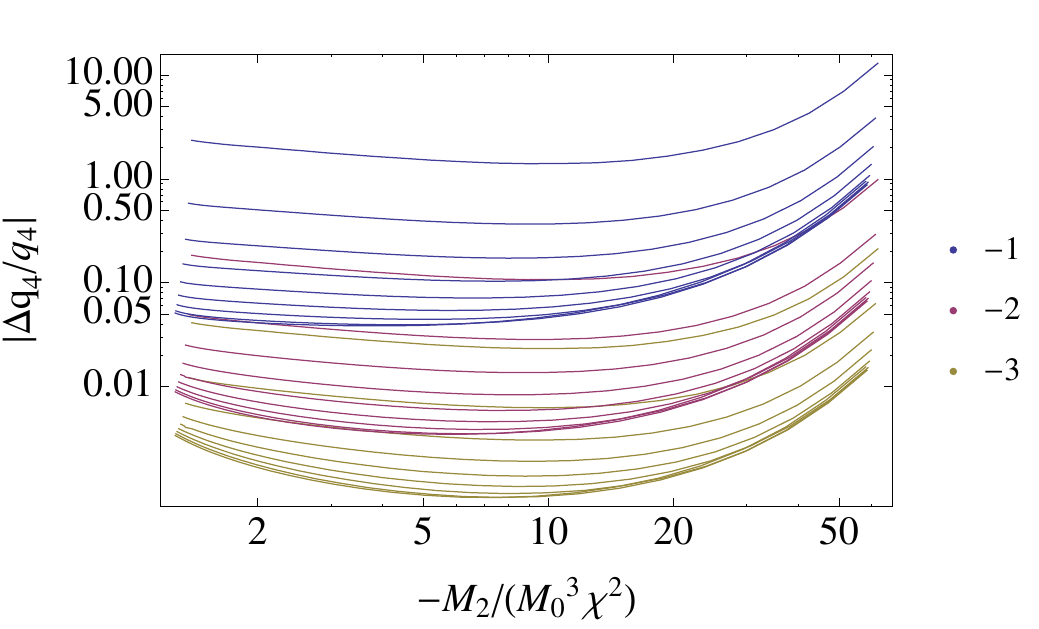}
\caption{
Estimated relative error in the calculation of $q_4$ for the SLy4 EoS, for different number of omitted points and different rotation rates (various curves of same colour). The different colours correspond to the relative differences, $(q_4^{4\#}-q_4^{n\#})/q_4^{4\#}$, with $n\#$ points omitted. The gold lines correspond to the comparison of $q_4$ between 4 points and 3 points being omitted. The figure shows that at the high compactness/low $\bar{M}_2$ end of the plot, the truncation error of $q_4^{4\#}$ dominates, while at the low compactness/high $\bar{M}_2$ end of the plot, the singularity error of $q_4^{3\#}$ dominates.   
}
\protect\label{q4error}
\end{figure}

Apart from the singularity error, there is another source of error in the calculation, which is the error that comes from the truncation of the integral of the coefficient $q_4$. The corresponding deviation of the calculated $q_4$ from the actual value will be, 

\be \delta q_4 \sim \int_{1-\varepsilon_1}^1 \frac{ds' s'^{4}}{(1-s')^{6}}\int_0^1 d\mu' P_{4}(\mu')\tilde{S}_{\tilde{\rho}}(s',\mu'). \ee 
This quantity can be calculated using the asymptotic form of the metric functions to evaluate the asymptotic form of the source $\tilde{S}_{\tilde{\rho}}(s,\mu)$. After some algebra, one can see that the angular integral has the following dependence on $r$,

\be \int_0^1d\mu' P_{4}(\mu')\tilde{S}_{\tilde{\rho}}(s',\mu')\sim \frac{A}{r^6}+O(r^{-7}), \ee
where the numerical coefficient $A$ is an $O(q_4)$ quantity. This means that the deviation will be $\delta q_4 \sim q_4 \varepsilon_1+ O(\varepsilon_1^2)$ and from calculations based on the models evaluated with the RNS code, the relative difference of the calculated from the actual value is, $(\Delta q_4/q_4) \lesssim 1\%$ with up to 4 points excluded from the integral. The relative difference has its maximum value for the more compact objects and it decreases with lower compactness, while it seems to be insensitive to the rotation rate.

Concluding the total relative error will be the sum of these two contributions, with the first one (singularity error) overestimating the value of $q_4$ and the second one (truncation error) underestimating it. From comparing the theoretical estimates of the error against the results calculated with the code after omitting 1,2,3, and 4 points, we can say that the relative difference between the values calculated by omitting 3 and 4 points, i.e., $(q_4^{4\#}-q_4^{3\#})/q_4^{4\#}$, gives an estimate of the accuracy of the calculation. This corresponds to the gold lines in Fig. \ref{q4error}, where we can see that the error is almost everywhere lower than $5\%$ for the models of the lowest rotation rates and it decreases with increasing rotation rate. This picture is typical of all EoSs.


\bibliography{master}
\end{document}